\documentclass[11pt]{article}
\usepackage{a4wide,graphicx}
\usepackage{epsfig,epsf,graphics,psfrag}
\usepackage{amsfonts,amssymb,stmaryrd,latexsym,amsmath}
\usepackage{enumerate}
\usepackage{slashed}

\newcommand{\be}{\begin{equation}}
\newcommand{\ee}{\end{equation}}
\newcommand{\ba}{\begin{eqnarray}}
\newcommand{\ea}{\end{eqnarray}}
\newcommand{\non}{\nonumber}

\newcommand{\lang}{\left\langle}
\newcommand{\rang}{\right\rangle}
\newcommand{\al}{&\!\!\!}

\begin{document}
\title{ \hfill{\tiny FZJ-IKP-TH-2010-08, HISKP-TH-10/09}\\[1.5em]
\bf Effect of charmed meson loops on charmonium transitions}
\author{Feng-Kun Guo$^1$\footnote{{\it E-mail address:}
f.k.guo@fz-juelich.de.  Current address: Helmholtz-Institut f\"ur Strahlen- und
             Kernphysik, Universit\"at Bonn,  D--53115 Bonn, Germany}, %
       ~C.~Hanhart$^{1,2}$\footnote{{\it E-mail address:} c.hanhart@fz-juelich.de},%
       ~Gang Li$^{3,4}$\footnote{{\it E-mail address:} gli@ihep.ac.cn},%
       ~Ulf-G. Mei{\ss}ner$^{1,2,5}$\footnote{{\it E-mail address:} meissner@hiskp.uni-bonn.de},%
       ~and Qiang Zhao$^{3,6}$\footnote{{\it E-mail address:} zhaoq@ihep.ac.cn}%
       \\[2mm]
       {\it\small$\rm ^1$Institut f\"{u}r Kernphysik and J\"ulich Center for Hadron
          Physics,}\\
          {\it\small Forschungszentrum J\"{u}lich, D--52425 J\"{u}lich, Germany}\\
       {\it\small$\rm ^2$Institute for Advanced Simulation,
          Forschungszentrum J\"{u}lich, D--52425 J\"{u}lich, Germany}\\
       {\it\small$\rm ^3$Institute of High Energy Physics, Chinese Academy
          of Sciences, Beijing 100049, China}\\
       {\it\small$\rm ^4$Department of Physics, Qufu Normal University, Qufu,
          273165, China}\\
       {\it\small$\rm ^5$Helmholtz-Institut f\"ur Strahlen- und Kernphysik and
          Bethe Center for Theoretical Physics,}\\
          {\it\small Universit\"at Bonn, D--53115 Bonn, Germany}\\
       {\it\small$\rm ^6$Theoretical Physics Center for Science Facilities,
          CAS, Beijing 100049, China}
}
\date{\today}

\maketitle
\begin{abstract}
The effects of intermediate charmed mesons on charmonium transitions with the
emission of one pion or eta are studied systematically. Based on a
non-relativistic effective field theory we show that charmed meson loops are
enhanced compared to the corresponding tree-level contributions for transitions
between two $S$-wave charmonia as well as for transitions between two $P$-wave
charmonia. On the contrary, for the transitions between one $S$-wave and one
$P$-wave charmonium state, the loops need to be analyzed case by case and often
appear to be suppressed. The relation to and possible implications for an
effective Lagrangian approach are also discussed. This study at the same time
provides a cross check for the numerical evaluations.
\end{abstract}

\vspace{1cm}


\thispagestyle{empty}

\newpage

\section{Introduction}

Since the discovery of the $J/\psi$ more than thirty years ago, the decays of
heavy quarkonia have played an important role in the physics of quarks and
hadrons. During the past decades, experimental studies of the heavy quarkonia at
CLEO, Belle, BaBar, CDF, D0, and BES-II have provided great opportunities for
examining many interesting properties of Quantum Chromodynamics (QCD). At the
present stage, BES-III~\cite{Asner:2008nq} has accumulated the largest data
samples for $J/\psi$ and $\psi'$ decays, and $\rm\overline
P$ANDA~\cite{Lutz:2009ff} plans to accumulate data for charmonia which cannot be
produced directly in electron--positron annihilations. These facilities will
deepen our understanding of the charmonium physics, and hence the
non-perturbative aspects of QCD. Although many theoretical investigations have
been performed in the past thirty years~(for comprehensive reviews, see
Refs.~\cite{Brambilla:2004wf,Asner:2008nq,hqwg}), there remain many mysteries in
charmonium physics to be settled. On the contrary, due to the new experimental
data with unprecedented statistics, many new interesting problems have appeared,
e.g. the nature of many of the new $X,Y,Z$ resonances discovered in the
charmonium mass region has still not been well understood (for recent reviews,
see, e.g.
Refs.~\cite{Swanson:2006st,Eichten:2007qx,Voloshin:2007dx,Godfrey:2008nc,Drenska:2010kg}).

Furthermore, various recent phenomenological calculations suggest
that charmed meson loops may play an important role in the decays of
heavy quarkonia (for an overview, see \cite{Zhao:QNP}). For
instance, using an effective Lagrangian approach (ELA), intermediate
heavy--meson loop contributions are found to be essential for
understanding the puzzling $\psi(3770)$ non-$D\bar{D}$
decays~\cite{Liu:2009dr,Zhang:2009kr}. They are also important in
the $J/\psi$ decays into a vector and a pseudoscalar
meson~\cite{Liu:2006dq} and in the $M1$ radiative transitions
between two charmonia~\cite{Li:2007xr}. Besides, using the on-shell
approximation, the bottom meson loops were suggested to make the
$\Upsilon(5S)$ transitions to the lower $\Upsilon$ states with the
emission of two pions~\cite{Meng:2007tk} or one
$\eta$~\cite{Meng:2008bq} different from those of the
$\Upsilon(4S)$. The inclusion of intermediate heavy mesons in heavy
quarkonium transitions, sometimes called coupled-channel effects,
has been noticed for more than twenty
years~\cite{Lipkin:1986bi,Lipkin:1988tg,Moxhay:1988ri}. Also, the
effect of the mass differences between the neutral and charged
mesons in the intermediate states (i.e. in the meson loops) plays a
role in other isospin breaking processes. This effect, known to be
of particular importance near the continuum thresholds, was already
studied in the $\eta'$ decays~\cite{Borasoy:2006uv}, and in the
decays $\phi\to \omega\pi^0$~\cite{Li:2008xm,Li:2007au}, $J/\psi\to
\phi \eta \pi^0$~\cite{Wu:2007jh,Hanhart:2007bd}, and
$D_{s0}(2317)\to
D_s\pi^0$~\cite{Faessler:2007gv,Lutz:2007sk,Guo:2008gp}.

\begin{figure}[t]
\begin{center}
\vglue-0mm
\includegraphics[width=0.8\textwidth]{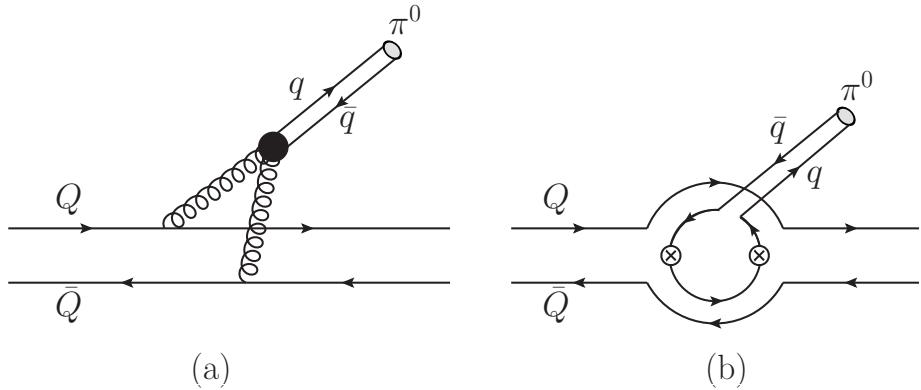}
\vglue-0mm \caption{Schematic diagrams of the QCDME mechanism (a)
and the non-multipole (b) effects of the intermediate heavy meson
loops for heavy quarkonium transition
with the emission of one pion. \label{fig:qcdme}}
\end{center}
\end{figure}

An often used formalism dealing with the hadronic transitions between two heavy
quarkonia is the QCD multipole expansion
(QCDME)~\cite{Gottfried:1977gp,Voloshin:1978hc,Kuang:2006me}. The QCDME is based
on the assumption that the emitted gluons are soft so that their wavelengths are
much larger than the size of a heavy quarkonium. As a result, a multipole
expansion similar to that in classical electrodynamics can be performed. The
soft gluons then hadronize into light meson(s), for instance the pion(s) or eta,
and the matrix elements may be worked out using soft pion theorems. A schematic
diagram for the multipole transition from a heavy quarkonium to another one with
the emission of one pion is plotted in Fig.~\ref{fig:qcdme}(a). However, this
ansatz clearly misses the contribution from intermediate mesons. This can be
understood as a heavy quarkonium can couple to a heavy meson and heavy
anti-meson pair through the non-perturbative production of a light quark and
anti-quark pair, see Fig.~\ref{fig:qcdme}(b). In
Refs.~\cite{Moxhay:1988ri,Zhou:1990ik}, within the framework of QCDME, the
intermediate heavy meson effects were considered to account for non-multipole
effects in the di-pion transitions between two charmonia, the $\psi$ states, or
bottomonia, the $\Upsilon$ states. Better agreement with the experimental data
was obtained.

In light of these phenomenological indications, it is important to have a
theoretical formalism which has a controlled uncertainty to study the effects of
heavy meson loops in the transitions of heavy quarkonia. In
Ref.~\cite{Guo:2009wr}, such a non-relativistic effective field theory (NREFT)
formalism was constructed to investigate the charmed meson loops in the decays
$\psi'\to J/\psi\pi^0(\eta)$. Because the $\psi'$ and $J/\psi$ are isospin and
SU(3) flavor singlets, the decay process $\psi'\to J/\psi\pi^0$ violates isospin
symmetry and  $\psi'\to J/\psi\eta$ violates SU(3) symmetry. Isospin
symmetry can be violated by both  electromagnetic (e.m.) effects and the mass
difference between the $u$ and $d$ quarks.  Because the e.m. effects are
small~\cite{Donoghue:1985vp,Maltman:1990mp} (it can be easily shown in the
framework of chiral perturbation theory (CHPT) with virtual
photons~\cite{Ecker:1988te,Urech:1994hd}, see Section~\ref{sec:em}), these two
decays were  used to extract the light quark mass ratio
$m_u/m_d$~\cite{Ioffe:1979rv,Ioffe:1980mx,Donoghue:1992ac,Donoghue:1993ha,Leutwyler:1996qg}.
The relation between the ratio of the decay widths
$$R_{\pi^0/\eta} \equiv \frac{{\cal B}(\psi'\to J/\psi\pi^0)}{{\cal
B}(\psi'\to J/\psi\eta)}$$
and the light quark masses is given by~\cite{Ioffe:1980mx,Voloshin:2007dx}
\be%
R_{\pi^0/\eta} = 3 \left(\frac{m_d-m_u}{m_d+m_u}\right)^2
\frac{F_\pi^2}{F_\eta^2} \frac{M_\pi^4}{M_\eta^4}
\left|{\vec{q}_\pi\over\vec{q}_\eta}\right|^3, \label{eq:udratio}
\ee%
where $F_{\pi(\eta)}$ and $M_{\pi(\eta)}$ are the decay constant and mass of the
pion (eta), respectively. The extracted quark mass ratio using
Eq.~(\ref{eq:udratio}) and the recent measurements of the decay widths from the
CLEO Collaboration~\cite{Mendez:2008kb},  the BES
Collaboration~\cite{Bai:2004cg} and the Particle Data Group (PDG)
fit~\cite{PDG2010} are listed in Table~\ref{tab:udratio}.
\begin{table}[th]
\begin{center}
\begin{tabular}{| l | l  l |} \hline\hline
            & $R_{\pi^0/\eta}$ & ${m_u/m_d}$ \\ \hline
 CLEO~\cite{Mendez:2008kb} & $(3.88 \pm 0.23 \pm 0.05)\%$ & $0.40\pm0.01$\\%
 BES~\cite{Bai:2004cg} & $(4.8 \pm 0.5)\%$ & $0.35\pm0.02$\\%
 PDG fit~\cite{PDG2010} & $(4.0 \pm 0.3)\%$ & $0.39\pm0.02$\\%
\hline \hline
\end{tabular}
\caption{\label{tab:udratio}The light quark mass ratio $m_u/m_d$ extracted using
Eq.~(\ref{eq:udratio}) from the recent experimental measurements by different
collaborations.}
\end{center}
\end{table}
Comparing with the result obtained using the Goldstone boson masses from
leading order (LO) CHPT~\cite{Weinberg:1977hb,Gasser:1982ap} \be%
\frac{m_u}{m_d} = \frac{M_{K^+}^2-M_{K^0}^2+2M_{\pi^0}^2-M_{\pi^+}^2}
     {M_{K^0}^2-M_{K^+}^2+M_{\pi^+}^2} = 0.56, \label{eq:rudmesonmass} \ee%
     the discrepancy is striking. We remark that there might be sizable higher
     order corrections to this LO result. The up-to-date knowledge of the
     light quark mass ratio from various determinations including lattice
     calculations (but excluding $\psi'$ decays) was summarized by Leutwyler
     as $m_u/m_d=0.47\pm0.08$~\cite{Leutwyler:Stern}. The relatively large
     uncertainty given here thus provides an overlap with the results quoted
     in Table~\ref{tab:udratio}, however, only at the very low end.  In
     Ref.~\cite{Guo:2009wr} it was stressed that for the mentioned transitions
     the effects of charmed meson loops, ignored in the previous analyses,
     should be sizable.  The charmonia $\psi'$ and $J/\psi$ couple to charmed
     and anti-charmed mesons, and the pion is emitted from one intermediate
     charmed meson.  The proper expansion parameter is the velocity of the intermediate meson, which for
     below threshold decays is defined via the analytic continuation of the
     standard definition, namely $v=\sqrt{-E/M_D}$, with $E$ measured relative
     to the open charm threshold.  We find $v\approx0.5$ for most of the
     decays studied.  It is found that loops are enhanced by a factor of $1/v$
     compared to the tree-level contribution where the pion is emitted
     directly from the charmonium. Therefore the dominant (LO) contributions
     to these decays come from the loops instead of from the tree graphs which
     are proportional to the quark mass differences directly, and hence the
     extraction of quark mass differences from $\psi'$ decays mentioned above
     is not reliable. Stated differently: quarkonium decays could only be used
     to extract the light quark mass ratio, if either the loop contributions
     could be controlled quantitatively (at present their uncertainty is quite
     sizable --- see discussion below), or if loop contributions are
     suppressed. %
It turned out that the
     enhancement of the loops in case of the $\psi'$ to $J/\psi$ transitions
     emerges only because the transitions at hand violate isospin or SU(3)
     symmetry.  The power counting is discussed in detail in
     Sec.~\ref{sec:pcloops}.

In Ref.~\cite{Guo:2010zk}, the same NREFT is applied to the decays $\psi'\to
h_c\pi^0$ and $\eta_c'\to\chi_{c0}\pi^0$. As a consequence of the quantum
numbers of the charmonia involved, in these two decays the loop contributions
are highly suppressed, and hence the tree-level terms, i.e. the quark mass
difference terms, dominate the decay amplitudes. Unfortunately there is no
charmonium transition, where one can exploit this observation in order to
extract the light quark mass ratio, since typically the phase space available is
insufficient for an $\eta$ in the final state. However, in the bottomonium
system analogous transitions appear to exist~\cite{Guo:2010ca}, and will allow
for the mentioned analysis. This illustrates that the effective field theory at
hand predicts a highly non-trivial pattern for the loop contributions in
different decays that can be tested experimentally.

\begin{figure}[t]
\begin{center}
\vglue-0mm
\includegraphics[width=\textwidth]{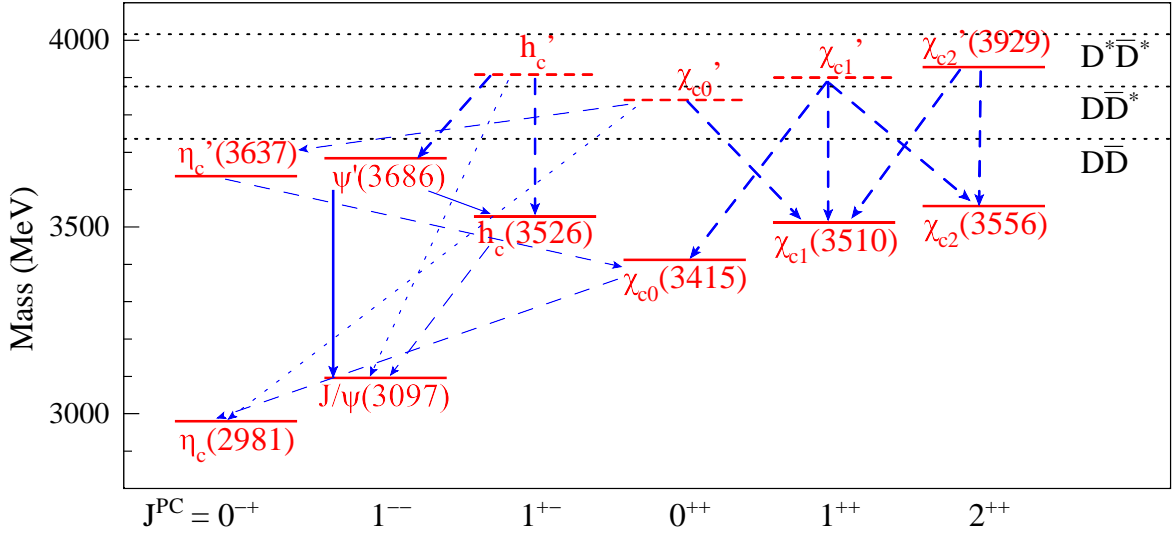}
\vglue-0mm \caption{All possible $S$ and $P$-wave transitions among the ground state
and the first radial excited $S$ and $P$-wave charmonia with the
emission of one light pseudoscalar meson ($\pi^0,\eta$). The unobserved resonances
and decay modes are plotted in dashed. The masses for the unobserved
resonances are taken from predictions in Ref.~\cite{Li:2009zu}. The thresholds for
the $D\bar D$, $D{\bar D}^*$ and $D^*{\bar D}^*$ are represented by
the dotted horizontal lines. The solid lines represent the measured decays.
The dashed and dotted lines are not yet measured with only the former discussed
in the paper and the latter are beyond the range of the applicability of the
chiral EFT (but can be considered in the model-dependent ELA also
discussed here).
Thick lines indicate transitions with enhanced charmed meson loops.
\label{fig:list}}
\end{center}
\end{figure}

In this paper, we will systematically investigate the charmed meson loop
contributions to the transitions between two charmonia with the emission of one
light pseudoscalar meson. We restrict the charmonia to $S$ and $P$-wave states
with radial quantum number $n$ less than or equal to 2. In other words, we will
consider the transitions between or within the following charmonia spin
multiplets: $\{J/\psi,\eta_c\}$, $\{\chi_{c0},\chi_{c1},\chi_{c2},h_c\}$ with
$n=1$, and $\{\psi',\eta_c'\}$, $\{\chi_{c0}',\chi_{c1}',\chi_{c2}',h_c'\}$ with
$n=2$. Charge conjugation  allows for the emission of one light pseudoscalar
meson between two charmonia with the same value of $C$. Considering further the
constraints from  parity conservation, all the allowed transitions are plotted
in Fig.~\ref{fig:list},~\footnote{In the figure, the masses of the so far
unobserved (or unidentified) $2P$ charmonia are taken from a quark model
calculation considering the color-screening effect due to the light quark and
anti-quark pair creation~\cite{Li:2009zu}. Note that these values are only used
for illustration.} and the following will be considered in the paper:
\begin{enumerate}[1)]
\item Transitions between two $S$-wave charmonia: $\psi'\to J/\psi\pi^0$,
    and $\psi'\to J/\psi\eta$.
\item Transitions between one $S$-wave and one $P$-wave charmonium: $\psi'\to
    h_c\pi^0$, $h_c\to J/\psi\pi^0$, $h_c'\to \psi'\pi^0$,
    $\eta_c'\to\chi_{c0}\pi^0$, $\chi_{c0}\to\eta_c\pi^0$, and
    $\chi_{c0}'\to\eta_c'\pi^0$.
\item Transitions between two $P$-wave charmonia:
    $\chi_{c0}'\to\chi_{c1}\pi^0$,
    $\chi_{c1}'\to\chi_{cJ}\pi^0$~($J=0,1,2$),
    $\chi_{c2}'\to\chi_{c1(2)}\pi^0$, and $h_c'\to h_c\pi^0$.
\end{enumerate}
In fact, as plotted in Fig.~\ref{fig:list}, the decays $h_c'\to J/\psi\pi^0
(\eta)$ and $\chi_{c0}'\to\eta_c\pi^0 (\eta)$ can also occur. However, the mass
difference between the initial and final charmonium exceeds 800~MeV, of order
${\cal O}(\Lambda_\chi)$, with $\Lambda_\chi\approx1$~GeV denoting the typical
hadronic scale. Since the chiral expansion is an expansion in $p/\Lambda_\chi$,
with $p$ denoting a typical momentum or mass, for those energies the chiral
expansion is not expected to converge any more. We therefore do not consider
these transitions. There could also be $D$-wave transitions
$\chi_{c2}\to\eta_c\pi^0(\eta)$ and $\chi_{c2}'\to\eta_c'\pi^0$ (not shown in
the figure). However, their partial decay widths would be too small to be
detected in the near future because of the $D$-wave suppression, the isospin or
SU(3) breaking, and small phase space. They will also not be considered here.

The paper is organized as follows. In Section~\ref{sec:tree}, the tree-level
chiral effective Lagrangians and the resulting amplitudes are given for all the
decays discussed in the paper. The isospin and SU(3) breaking are given by quark
mass differences.  Consistent with earlier analyses, e.m. contributions are
found to be small and can be neglected. Various aspects of the charmed meson
loops will be discussed in Section~\ref{sec:loops}. Section~\ref{sec:results} is
devoted to the results of the meson loops for various transitions in the NREFT.
In Section~\ref{ELA-formulation}, a detailed parallel study of the meson loop
transitions in the framework of the ELA is presented and the results are
compared to those from the NREFT. A brief summary is given in the last section.
Various technicalities such as the utilized loop functions, the decay amplitudes
from the meson loops and the ingredients of the ELA are given in the Appendices.

\section{Effective Lagrangians for tree-level diagrams}
\label{sec:tree}

Isospin breaking has two sources. One is the mass difference between the up and
down quarks, and the other one is of e.m. origin because photons do not have
definite isospin. This section is devoted to the construction of the LO chiral
Lagrangians for the tree-level diagrams of the transitions considered in this
paper. Both the quark mass difference and the e.m. effects will be taken into
account.

The chiral effective Lagrangians are of the most general form which
is invariant under the transformations of SU(3)$_L\times$SU(3)$_R$,
parity and charge conjugation. The charmonia are treated as matter
fields, and the pion and eta are the Goldstone bosons of the
spontaneous breaking of SU(3)$_L\times$SU(3)$_R$ down to its vector
subgroup SU(3)$_V$. The charmonia are SU(3) singlets, so they do not
change under the chiral transformation. The Goldstone boson fields
\begin{eqnarray}
\label{eq:phi}
 \phi =
  \left(
    \begin{array}{c c c}
 \frac{1}{\sqrt{2}}\pi^0 + \frac{1}{\sqrt{6}}\eta & \pi^+ & K^+\\
\pi^- & - \frac{1}{\sqrt{2}}\pi^0 + \frac{1}{\sqrt{6}}\eta & K^0 \\
K^- & {\bar K}^0 & - \frac{2}{\sqrt{6}}\eta
    \end{array}
\right) ,
\end{eqnarray}
where we have approximated the $\eta$ as one element of the octet SU(3)
representation, are collected in $u=\exp \left( {i\phi / \sqrt{2}F}\right)$
with $F$ being the pion decay constant in the chiral limit.  Under the
transformation of SU(3)$_L\times$SU(3)$_R$, we have
\be%
u \to R u h^{\dag} = h u L^{\dag},
\ee%
where $h$ is the compensator field. It is convenient to construct chiral
Lagrangians using operators whose chiral transformation is $O\to h O h^\dag$.
The following such building blocks will be used
\ba%
u_{\mu} &\!\!\!=&\!\!\! i\left(u^{\dag}\partial_{\mu}u + \partial_{\mu}uu^{\dag}\right), \non\\
\chi_\pm &\!\!\!=&\!\!\! u^\dagger \chi u^\dagger \pm u\chi^\dag u,\non\\
Q_\pm &\!\!\!=&\!\!\! \frac12\left( u^\dagger Q u \pm uQu^\dagger
\right),
\ea%
where the diagonal quark mass matrix and the charge matrix are
\ba%
\chi &\!\!\!=&\!\!\! 2B_0\cdot {\rm diag}\left(m_u,m_d,
m_s\right), \non\\
Q &\!\!\!=&\!\!\! e \cdot{\rm diag}\left(2/3,-1/3, -1/3\right),
\ea%
in terms of $B_0= |\langle 0 |\bar q q |0\rangle|/F^2 $ and the elementary
electric charge $e$ ($e>0$).

In the heavy quark limit with $m_Q\to\infty$, the coupling of a heavy
quark to a gluon is spin-independent~\cite{Neubert:1993mb}. As a
result, there is a spin symmetry in that limit, and the heavy
quarkonia, which differ from each other only in the total spin of the
heavy quark and  anti-quark, can be grouped into the same spin
multiplet. It is then convenient to introduce a single field for a spin
multiplet of heavy quarkonia~\cite{Jenkins:1992nb,Casalbuoni:1992yd}
using the trace formalism proposed for single-heavy
mesons~\cite{Falk:1990yz,Bjorken:1990hs}. In this way, the consequence
of the heavy quark spin symmetry can be obtained automatically by
evaluating a trace in spinor space. The construction of charmonium
fields with arbitrary orbital angular momentum $l$ in the trace
formalism can be found in Ref.~\cite{Casalbuoni:1992yd}. Since we are
dealing with the transitions between two charmonia, the heavy quark
four-velocity is conserved up to higher order corrections. In this
case, it is convenient to use the two-component notation as introduced
in Ref.~\cite{Hu:2005gf}. For doing that, the four-velocity is chosen
to be $v^\mu=(1,\vec{0})$. In the two-component notation, the field for
the $S$-wave charmonia reads
\be%
J = \vec{\psi}\cdot\vec{\sigma}+\eta_c,
\ee%
with $\vec{\psi}$ and $\eta_c$ annihilating the $J/\psi$ and
$\eta_c$ states. The field for the $P$-wave charmonia is
\footnote{The sign convention for the $\chi_{c2}$ and $\chi_{c1}$ fields are
different from those in Ref.~\cite{Fleming:2008yn}. }
\be%
\chi^i = \sigma^j
\left(-\chi_{c2}^{ij}-\frac{1}{\sqrt{2}}\epsilon^{ijk}\chi_{c1}^k +
\frac{1}{\sqrt{3}}\delta^{ij}\chi_{c0} \right) + h_c^i,
\ee%
where $\chi_{c2}^{ij}$, $\chi_{c1}^i$, $\chi_{c0}$ and $h_c$
annihilate the $\chi_{c2}$, $\chi_{c1}$, $\chi_{c0}$ and $h_c$
states, respectively. $\chi_{c2}^{ij}$ is a symmetric and traceless
tensor.

The quantum numbers of the charmonia determine their parity and charge
conjugation transformation properties.
The parity transformations for the charmonia fields are given by
\ba%
J \overset{\cal P}{\to} - J, \quad \chi^i \overset{\cal P}{\to} \chi^i,
\ea%
and the charge conjugation transformations are given by
\ba%
J \al\overset{\cal C}{\to}\al \sigma_2 J^T \sigma_2 = -\vec{\psi}\cdot\vec{\sigma}+\eta_c, \non\\
\chi^i \al\overset{\cal C}{\to}\al -\sigma_2\chi^i\sigma_2 = \sigma^j
\left(-\chi_{c2}^{ij}-\frac{1}{\sqrt{2}}\epsilon^{ijk}\chi_{c1}^k +
\frac{1}{\sqrt{3}}\delta^{ij}\chi_{c0} \right) - h_c^i,
\ea%
where $J^T$ is the transpose of $J$. Denoting the rotation in the SU(2) spin
space of the heavy quark (anti-quark) by $S$ ($\bar S$), the transformation of
the charmonium fields reads
\be%
J \overset{\cal S}{\to} SJ{\bar S}^\dag, \quad \chi^i \overset{\cal S}{\to} S\chi^i{\bar S}^\dag.
\ee%
The transformation for a heavy quarkonium field with arbitrary orbital angular
momentum is given in~\cite{Casalbuoni:1992yd} in four-component notation. In
two-component notation, the transformation properties for the $\chi_{cJ}$ fields
can be found in Ref.~\cite{Fleming:2008yn}.

\subsection{Quark mass difference}
\label{sec:treeLag}

In the transitions considered in this paper,  one pion or eta is emitted.
Therefore we need to construct the chiral Lagrangian using the external field $\chi_-$,
which is proportional to the light quark mass matrix and contains an odd number of
the Goldstone bosons. Under parity and charge conjugation, $\chi_-$ transforms as
\be%
\chi_- \overset{\cal P}{\to} -\chi_- = \chi_-^\dag, \quad \chi_- \overset{\cal C}{\to} \chi_-^T,
\ee%
respectively.

The LO Lagrangian for the transitions between two $S$-wave charmonia can be
constructed considering parity conservation, which requires the presence of a
derivative, charge conjugation, chiral symmetry and Galilean invariance,
\be%
\label{eq:LSStree} {\cal L}_{SS} = \frac{A}{4} \left[\lang J'\sigma^i J^{\dag}\rang - \lang J^{\dag}\sigma^iJ'\rang\right]
\partial^i \left(\chi_-\right)_{aa},
\ee%
where $\lang \ldots\rang$ denotes the trace in spinor space, the subscript
$a=u,d,s$ is a flavor index, and $aa$ as a sum over it denotes the trace in
flavor space. Similarly, the Lagrangian for the transitions between one $S$- and
one $P$-wave charmonium states is
\be%
\label{eq:LSPtree} {\cal L}_{SP} = \frac{i}{4}C \left[
\lang\vec{\chi}^\dag\cdot\vec{\sigma}J'\rang + \lang
J'\vec{\sigma}\cdot\vec{\chi}^\dag\rang \right] \left(\chi_-\right)_{aa},
\ee%
and that for the transitions between two $P$-wave charmonia is
\be%
\label{eq:LPPtree} {\cal L}_{PP} = i\frac{\gamma}{2}\epsilon^{ijk} \lang
\chi^{i\prime}\chi^{j\dag}\rang \partial^k \left(\chi_-\right)_{aa},
\ee%
These Lagrangians were first proposed in Ref.~\cite{Casalbuoni:1992fd} in
four-component notation. Note that due to the presence of a Pauli matrix between
the two heavy quarkonium fields in Eqs.~(\ref{eq:LSStree},\ref{eq:LSPtree}),
the heavy quark spin symmetry is violated. On the contrary,
the Lagrangian ${\cal L}_{PP}$ preserves the spin symmetry.

The goal of the present work is to set up an effective field theory that allows
one to systematically study both loop as well as tree level transition
amplitudes. To prepare for this we need to assign an order of magnitude estimate
to the coupling constants given above. They may be determined in principle as
the result of some matching procedure between the hadronic matrix elements and
the more fundamental quark--gluon dynamics calculated within (potential)
nonrelativistic QCD ((p)NRQCD) --- for recent reviews, see
Ref.~\cite{Brambilla:2004jw,Brambilla:2004wf,Asner:2008nq,hqwg}. For instance,
the coupling constant $A$ in the Lagrangian Eq.~(\ref{eq:LSStree}) has a mass
dimension $-2$. There are several different scales in the physics related to
heavy quarkonia. They are the heavy quark mass $m_Q$, the momentum $m_Qv_Q$, the
inverse of which sets the length scale of a heavy quarkonium, and the energy
scale $m_Qv_Q^2$~\cite{Caswell:1985ui}, where $v_Q$ denotes the velocity of the
heavy quark within a heavy quarkonium to be distinguished from the velocity $v$
of the heavy mesons in the loops to be introduced in the next section --- for an
estimate of the values of various scales in heavy quarkonia, one may refer to
\cite{Braaten:1996ix}. In addition, there is the nonperturbative QCD scale
$\Lambda_{\rm QCD}$. In this paper, we are considering the low-lying heavy
quarkonia. For these states, it is believed that $m_Qv_Q^2\gtrsim \Lambda_{\rm
QCD}$, which defines the weak-coupling regime~\cite{Brambilla:2004jw} --- e.g.
with $v_c^2\simeq 0.3$ and $m_c=1.5$ GeV we find $m_Qv_Q^2\simeq 450$ MeV.
 Since most of the coupling constants introduced
in the Lagrangians are dimensionful, they should have certain scaling properties
expressed by the above mentioned scales.
The tree-level Lagrangian describes a process with the emission of soft gluons,
which then hadronize into a pion or an eta. As mentioned in the Introduction,
the applicable regime of our effective field theory is limited to the
transitions with the pion (eta) energy much smaller than $\Lambda_{\chi}$, i.e.
$E_{\pi(\eta)}\lesssim 600$~MeV. Hence the energy of the emitted gluons should
also be $\lesssim 600$~MeV. Therefore, the proper dimensionful parameter that
sets the scale for this nonperturbative process should be either $m_cv_c^2$,
which is sometimes called ultrasoft, or $\Lambda_{\rm QCD}$. As mentioned above,
the charmonia considered in this paper are weakly-coupled, i.e. $m_cv_c^2\gtrsim
\Lambda_{\rm QCD}$. So conservatively, one may take $\Lambda_{\rm QCD}$ to set
the soft scale. Furthermore, since the transition violates spin symmetry we have
to put in a factor ${\Lambda_{\rm QCD}}/{m_c}$ to finally get
\be%
\label{eq:Ascaling} A \sim \frac1{\Lambda_{\rm QCD} 2m_c} \frac{\Lambda_{\rm
QCD}}{m_c} =\frac1{2m_c^2},
\ee%
where a factor of $1/(2m_c)$ was introduced to make $A$ have the correct
dimension. In addition, assigning the pion decay constant as $F\sim\Lambda_{\rm
QCD}$, and the quark condensate as $|\langle 0 |\bar q q |0\rangle|\sim
\Lambda_{\rm QCD}^3$, one has
\be%
B_{du} = \frac{B_0}{F}(m_d-m_u) \sim \delta,
\ee%
with $\delta$ denoting the quark mass difference. So using the expressions given
in Table~\ref{tab:treeamp}, the tree-level amplitude for a pionic transition
scales as
\be%
\label{eq:scalingtree} {\cal M}_{\rm tree}^{SS} \sim \frac1{m_c} q\delta,
\ee%
where $1/(2m_c)$ has been canceled by the factor $\sqrt{M_iM_f}$ due to
nonrelativistic normalization.

The dimension of the coupling constant $\gamma$ in the Lagrangian for the
transitions between two $P$-wave quarkonia is the same as that of $A$. However,
for these transitions, the spin symmetry is preserved as can be seen from
Eq.~(\ref{eq:LPPtree}). So in the scaling of $\gamma$ the suppression factor
$\Lambda_{\rm QCD}/m_c$ should not be present. This is the only difference from
that of $A$. The dimension of $C$ in Eq.~(\ref{eq:LSPtree}) is higher than that
of $A$ or $\gamma$ by one unit. Therefore, analogous to
Eq.~(\ref{eq:scalingtree}), the scaling of the tree-level amplitude for a pionic
transition between two $P$-wave charmonia and that for a transition between one
$S$- and one $P$-wave states should be given by
\ba%
\label{eq:scalingtreeSPPP}{\cal M}_{\rm tree}^{PP} \sim \frac1{\Lambda_{\rm
QCD}} q\delta, \qquad {\cal M}_{\rm tree}^{SP} \sim \delta.
\ea%

\subsection{Virtual photons}
\label{sec:em}

The e.m. effects come from virtual photons exchanged in the processes. The
inclusion of  virtual photons has been first considered systematically for
three-flavor CHPT in Ref.~\cite{Urech:1994hd}.

Since the photons are virtual, we need to consider operators with at least two
powers of electric charge. At ${\cal O}(m_q^0\alpha)$, with $\alpha\equiv
e^2/(4\pi) \simeq 1/137$  the fine structure constant, one virtual photon is
exchanged. There are three types of operators, and we will discuss them one by
one:
\begin{enumerate}[1)]
\item The virtual photon is exchanged between light quarks, which is the
    standard case in CHPT with virtual photons. One needs quadratic
    combinations of the spurions $Q_+$ and $Q_-$ which act on the light
    quarks. The  parity and charge conjugation properties of the $Q_\pm$ are given
    by~\cite{Meissner:1997ii}
\be%
Q_\pm \overset{\cal P}{\to} \pm Q_\pm^\dag, \quad Q_\pm \overset{\cal
C}{\to} \pm Q_\pm^T.
\ee%
    There is only one light pseudoscalar meson in the final states of all
    the transitions considered in the paper.  $Q_+$ and $Q_-$ contain even
    and odd number of the Goldstone fields, respectively, and their
    expansion reads
\ba%
Q_+ \al=\al Q + {\cal O}\left(\phi^2\right), \non\\
Q_- \al=\al \frac{i}{\sqrt{2}F}(Q\phi-\phi Q) + {\cal O}\left(\phi^3\right).
\ea%
Therefore, at ${\cal O}(\alpha)$, the possible virtual photon operators for
the one pion (eta) emission transitions between charmonia are
$(Q_+Q_-)_{aa}$ and $(Q_+)_{aa} ( Q_-)_{aa}$. The traces come from the fact
that the charmonia are SU(3) singlets. However, one can easily show that
\ba%
\label{eq:Qtraces}
(Q_-)_{aa}  \al=\al 0, \non\\
(Q_+Q_-)_{aa}  \al=\al 0.
\ea%
Thus, there is no electromagnetic contribution to the one pion emission
transitions at order ${\cal O}(\alpha)$. Actually, there is a more general
relation
\be%
(Q_+^nQ_-)_{aa} = 0 + {\cal O}\left(\phi^3\right).
\ee%
That means, for any transition with the emission of one soft pion
between two iso-singlets, the contribution from virtual photons
exchanged between light quarks vanishes at tree-level.

\item The virtual photon is exchanged inside the heavy quarkonia. In this
    case, no operator containing light mesons without derivative or quark
    mass can be constructed. This may be understood as virtual photons
    exchanged inside the heavy quarkonia cannot contribute to the isospin
    breaking transitions.

\item  The virtual photon is exchanged between a heavy (anti-)quark and a
    light (anti-)quark. This kind of virtual photon is important in
    understanding the isospin mass splitting of heavy
    hadrons~\cite{Jenkins:1992hx,Guo:2008ns}. In principle, this will give a
    non-vanishing contribution to isospin breaking transitions. For the
    transitions considered here, however, only one Goldstone boson is
    emitted. Therefore, the operator for the light flavor part should be
    $(Q_-)_{aa}$. However, the trace of $Q_-$ vanishes (see
    Eq.~(\ref{eq:Qtraces})).
\end{enumerate}
Therefore, there is no e.m. contribution to the isospin breaking heavy
quarkonium transitions at order ${\cal O}(\alpha)$ and thus they can be
neglected compared to the quark mass difference terms. This conclusion agrees
with those of earlier studies in Refs.~\cite{Donoghue:1985vp,Maltman:1990mp}.

\subsection{Tree-level amplitudes}\label{tree-level}

Before working out the tree-level amplitudes using the Lagrangians given in
Section~\ref{sec:treeLag}, one subtlety needs to be addressed. In
Eq.~(\ref{eq:phi}), the $\pi^0$ and $\eta$ are SU(3) flavor eigen-states.
However, they are not exactly the same as the physical pion and eta which are
mass eigen-states. Denoting the physical states by ${\tilde \pi}^0$ and $\tilde
\eta$, the $\pi^0-\eta$ mixing is given as
\begin{eqnarray}\label{eq:pietamixing}
\pi^0 \al=\al
\tilde{\pi}^0\cos\epsilon_{\pi^0\eta} - \tilde{\eta}\sin\epsilon_{\pi^0\eta} =
\tilde{\pi}^0 -\epsilon_{\pi^0\eta}\tilde{\eta} + {\cal O}\left(\epsilon_{\pi^0\eta}^2\right),
\nonumber\\
\eta \al=\al \tilde{\eta}\cos\epsilon_{\pi^0\eta} + \tilde{\pi}^0\sin\epsilon_{\pi^0\eta}
= \tilde{\eta}+\epsilon_{\pi^0\eta}\tilde{\pi}^0 + {\cal O}\left(\epsilon_{\pi^0\eta}^2\right),
\end{eqnarray}
where $\epsilon_{\pi^0\eta}$ is the well-known $\pi^0-\eta$
mixing angle, which reads to LO in the chiral expansion
\begin{equation}%
\epsilon_{\pi^0\eta}
= \frac{\sqrt{3}}{4}\frac{m_d-m_u}{m_s-\hat m}
\end{equation}%
with $\hat m=(m_u+m_d)/2$ the average mass of the $u$ and $d$ quarks. Using
Dashen's theorem~\cite{Dashen:1969eg}, one may express the mixing angle in terms
of the masses of the Goldstone bosons at  LO in CHPT
\be%
\epsilon_{\pi^0\eta} =
\frac{1}{\sqrt{3}}\frac{M_{K^0}^2-M_{K^+}^2+M_{\pi^+}^2-M_{\pi^0}^2}{M_\eta^2-M_{\pi^0}^2}
= 0.01.
\ee%
The mixing of the $\pi^0$ or $\eta$ with the $\eta'$ is not  considered since it
is of higher order. The reason is that the $\eta'$ is not a Goldstone boson of
the spontaneous breaking of SU(3)$_L\times$SU(3)$_R$ to the vector subgroup
SU(3)$_V$ and its mass as a large scale provides a suppression.

The tree-level amplitudes for the charmonium transitions with the emission of
one pion or eta are listed in Table~\ref{tab:treeamp},\footnote{We have checked
that the ratios among the spin-averaged absolute square of the transition
amplitudes for the transitions between $P$-wave charmonia given in
Ref.~\cite{Casalbuoni:1992fd} can be reproduced.} where we have defined
$B_{du}=B_0(m_d-m_u)/F$ and $B_{sl}=B_0(m_s-{\hat m})/F$.
\begin{table*}[t]
\begin{center}
\renewcommand{\arraystretch}{1.3}
\begin{tabular}{|l|l|}\hline\hline
$\psi'\to J/\psi\pi^0$ & $i6A\epsilon^{ijk}\varepsilon^i(\psi')\varepsilon^j(J/\psi)q^kB_{du}$\\
$\psi'\to J/\psi\eta$ & $i({8}/{\sqrt{3}})A\epsilon^{ijk}\varepsilon^i(\psi')\varepsilon^j(J/\psi)q^kB_{sl}$\\
$\psi'\to h_c\pi^0$ & $6C\vec{\varepsilon}(\psi')\cdot\vec{\varepsilon}(h_c)B_{du}$\\
$\eta_c'\to \chi_{c0}\pi^0$ & $6\sqrt{3}CB_{du}$\\
$\chi_{c0}'\to \chi_{c1}\pi^0$ & $-2\sqrt{6}i\gamma\vec{\varepsilon}(\chi_{c1})\cdot\vec{q}B_{du}$\\
$\chi_{c1}'\to \chi_{c1}\pi^0$ & $-i3\gamma\epsilon^{ijk}\varepsilon^i(\chi_{c1}')\varepsilon^j(\chi_{c1})q^kB_{du}$\\
$\chi_{c1}'\to \chi_{c2}\pi^0$ & $3\sqrt{2}i\gamma\varepsilon^i(\chi_{c1}')\varepsilon^{ij}(\chi_{c2})q^jB_{du}$\\
$\chi_{c2}'\to \chi_{c2}\pi^0$ & $-i6\gamma\epsilon^{ijk}\varepsilon^{il}(\chi_{c2}')\varepsilon^{jl}(\chi_{c2})q^kB_{du}$\\
$h_c'\to h_c\pi^0$ & $-i6\gamma\epsilon^{ijk}\varepsilon^i(h_c')\varepsilon^j(h_c)q^kB_{du}$\\
\hline\hline%
\end{tabular}
\caption{\label{tab:treeamp}Tree-level amplitudes for the charmonium transitions
with the emission of one pion or eta. A factor of $\sqrt{M_iM_f}$, with
$M_{i(f)}$ denoting the mass of the initial (final) charmonium, should be
multiplied to all the expressions to account for the non-relativistic
normalization of the charmonium fields.}
\end{center}
\end{table*}
A factor of $\sqrt{M_iM_f}$, with $M_{i(f)}$ denoting the mass of the initial
(final) charmonium, should be multiplied to all the expressions to account for
the non-relativistic normalization of the charmonium fields used in the
effective Lagrangians.

\section{Decay amplitudes from charmed meson loops}

\label{sec:loops}

\subsection{Charmed meson loops}

\begin{figure}[t]
\begin{center}
\vglue-0mm
\includegraphics[width=\textwidth]{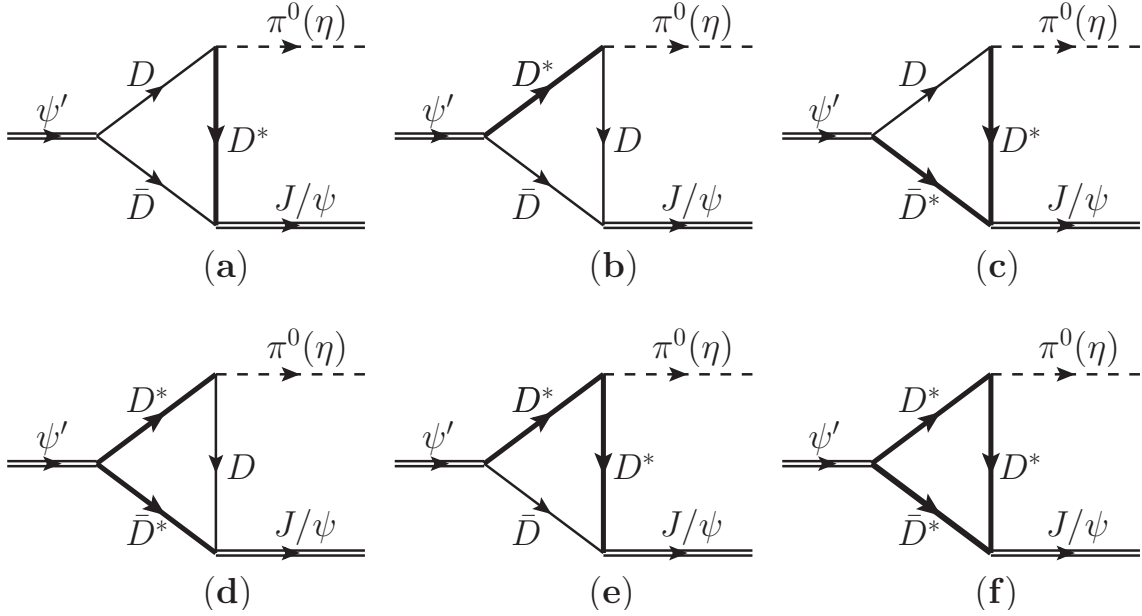}
\vglue-0mm \caption{The decays $\psi'\to J/\psi\pi^0(\eta)$ through
triangle charmed-meson loops. Charmonia, light mesons, pseudoscalar
and vector charmed mesons, are denoted by double, dashed, thin and
thick solid lines, respectively. \label{fig:loops}}
\end{center}
\end{figure}

In this section, we list all the possible loops (i.e. the triangle graphs)
with the lowest-lying
pseudoscalar and vector charmed mesons for each transition. There are three
charmed mesons in each loop. To be specific, we denote the one connecting the
initial charmonium and the light meson as $M1$, the one connecting two charmonia
as $M2$, and the one connecting the final charmonium and the light meson as
$M3$. The meson $Mi$ has a mass $m_i$. For instance, in Fig.~\ref{fig:loops}
which shows all loops contributing to the $\psi'\to J/\psi\pi^0(\eta)$, $M1$,
$M2$ and $M3$ are the $D$, $\bar D$ and $D^*$, respectively in diagram (a). All
the loops contributing to each decay are listed in Table~\ref{tab:loops}.
\begin{table*}[t]
\begin{center}
\renewcommand{\arraystretch}{1.3}
\begin{tabular}{l|l}\hline\hline
$\psi'\to J/\psi\pi^0(\eta)$ & $[D,{\bar D},D^*]$, $[D^*,{\bar D},D]$, $[D,{\bar D}^*,D^*]$, %
                               $[D^*,{\bar D}^*,D]$, $[D^*,{\bar D},D^*]$, $[D^*,{\bar D}^*,D^*]$\\
$\psi'\to h_c\pi^0$ & $[D,{\bar D},D^*]$, $[D^*,{\bar D},D^*]$,  $[D^*,{\bar D}^*,D]$, %
                      $[D,{\bar D}^*,D^*]$, $[D^*,{\bar D}^*,D^*]$\\
$h_c\to J/\psi\pi^0$ & $[D^*,{\bar D},D]$, $[D^*,{\bar D},D^*]$, $[D,{\bar D}^*,D^*]$, %
                       $[D^*,{\bar D}^*,D]$, $[D^*,{\bar D}^*,D^*]$\\
$\eta_c'\to \chi_{c0}\pi^0$ & $[D^*,{\bar D},D]$, $[D,{\bar D}^*,D^*]$, $[D^*,{\bar D}^*,D^*]$\\
$\chi_{c0}\to \eta_c\pi^0$ & $[D,{\bar D},D^*]$, $[D^*,{\bar D}^*,D]$, $[D^*,{\bar D}^*,D^*]$\\
$\chi_{c0}'\to \chi_{c1}\pi^0$ & $[D,{\bar D},D^*]$, $[D^*,{\bar D}^*,D]$\\
$\chi_{c1}'\to \chi_{c0}\pi^0$ & $[D^*,{\bar D},D]$, $[D,{\bar D}^*,D^*]$\\
$\chi_{c1}'\to \chi_{c1}\pi^0$ & $[D^*,{\bar D},D^*]$\\
$\chi_{c1}'\to \chi_{c2}\pi^0$ & $[D,{\bar D}^*,D^*]$\\
$\chi_{c2}'\to \chi_{c1}\pi^0$ & $[D^*,{\bar D}^*,D]$\\
$\chi_{c2}'\to \chi_{c2}\pi^0$ & $[D^*,{\bar D}^*,D^*]$\\
$h_c'\to h_c\pi^0$ & $[D^*,{\bar D},D^*]$, $[D,{\bar D}^*,D^*]$, $[D^*,{\bar D}^*,D]$, %
                     $[D^*,{\bar D}^*,D^*]$\\
\hline\hline%
\end{tabular}
\caption{\label{tab:loops}All the loops contributing to each
transition. The mesons are listed as $[M1,M2,M3]$. Flavor labels are
dropped for simplicity.}
\end{center}
\end{table*}

\subsection{Leading order effective Lagrangians}

In order to calculate the leading contributions from the charmed meson loops, we
need the LO effective Lagrangians for the couplings. Because the pion and eta
are pseudo-Goldstone bosons of the spontaneous chiral symmetry breaking of QCD,
their coupling to the charmed mesons in the low-energy limit is constrained by
chiral symmetry. The effective Lagrangians were constructed considering both the
heavy quark symmetry and chiral symmetry in
Refs.~\cite{Burdman:1992gh,Wise:1992hn,Yan:1992gz} (for a review, we refer to
Ref.~\cite{Casalbuoni:1996pg}). In the two-component notation of
Ref.~\cite{Hu:2005gf}, the charmed mesons are represented by
$H_a=\vec{V}_a\cdot\vec{\sigma}+P_a$, with $V_a$ and $P_a$ denoting the vector
and pseudoscalar charmed mesons, respectively, $\vec{\sigma}$ are the Pauli
matrices, and $a$ is the light flavor index. Explicitly, one can write
$P_a(V_a)=\left(D^{(*)0},D^{(*)+},D_s^{(*)+}\right)$. The lowest order chiral
effective Lagrangian for the axial coupling is~\cite{Hu:2005gf}
\be%
{\cal L_\phi} = -{g\over 2}\lang
H_a^{\dag}H_b^{\,}\vec{\sigma}\cdot\vec{u}_{ba}\rang, \label{eq:Lphi}
\ee%
where the axial current is $\vec{u}=-\sqrt{2}\vec{\partial}\phi/F+{\cal
O}(\phi^3)$.

The LO Lagrangian for the coupling of the $S$- or $P$-wave charmonium
fields to the charmed and anti-charmed mesons can be constructed
considering parity, charge conjugation and spin symmetry. In
two-component notation, the one for the $S$-wave charmonia $J/\psi$
and $\eta_c$ reads~\cite{Guo:2009wr}
\be%
{\cal L}_\psi = i \frac{g_2}{2} \lang J^\dag H_a \vec{\sigma}\cdot
\!\overleftrightarrow{\partial}\!{\bar H}_a\rang + {\rm H.c.}, \label{eq:Lpsi0}
\ee%
where $A\overleftrightarrow{\partial}\!B\equiv
A(\vec{\partial}B)-(\vec{\partial}A)B$, and ${\bar H}_a=-\vec{{\bar
V}}_a\cdot\vec{\sigma}+{\bar P}_a$ is the field for anti-charmed
mesons~\cite{Fleming:2008yn}. The Lagrangian for the $P$-wave charmonia at LO
is~\cite{Fleming:2008yn}
\be%
{\cal L}_\chi = i \frac{g_1}{2} \lang \chi^{\dag i} H_a \sigma^i {\bar H}_a\rang
+ {\rm H.c.} \label{eq:Lchi0}
\ee%
These Lagrangians were introduced in Ref.~\cite{Colangelo:2003sa} in
four-component notation. The values of $g_1$ and $g_2$ are twice of those in
Ref.~\cite{Colangelo:2003sa}.~\footnote{In the definition of the Lagrangians of
Ref.~\cite{Colangelo:2003sa}, each term is doubled for heavy quarkonia with the
same flavor of quark and anti-quark. Hence the values of the coupling constants
there should be half of those in our paper.} The Lagrangians for the coupling of
the radial excited charmonia to the charmed and anti-charmed mesons have the
same form as Eqs.~(\ref{eq:Lpsi0},\ref{eq:Lchi0}) with the coupling constants
changed to those for the excited states $g_2'$ and $g_1'$. For later use, we
evaluate the traces in Eqs.~(\ref{eq:Lpsi0},\ref{eq:Lchi0}), and rewrite the
Lagrangians in terms of the meson fields. The Lagrangian for the $J/\psi$ and
$\eta_c$ is
\ba%
{\cal L}_\psi \al=\al ig_2 \psi^{\dag i} \left( V_a^j
\overleftrightarrow{\partial}\!^i {\bar V}_a^j - V_a^i
\overleftrightarrow{\partial}\!^j {\bar V}_a^j - V_a^j
\overleftrightarrow{\partial}\!^j {\bar V}_a^i \right) + g_2
\epsilon^{ijk}\psi^{\dag i} \left( P_a \overleftrightarrow{\partial}\!^j {\bar
V}_a^k - V_a^j \overleftrightarrow{\partial}\!^k {\bar P}_a \right)
\non\\
\al\al + i g_2 \psi^{\dag i} P_a \overleftrightarrow{\partial}\!^j {\bar P}_a +
g_2\eta_c^\dag \epsilon^{ijk} V_a^i \overleftrightarrow{\partial}\!^j {\bar
V}_a^k + ig_2\eta_c^\dag \left( V_a^i \overleftrightarrow{\partial}\!^i {\bar
P}_a - P_a \overleftrightarrow{\partial}\!^i {\bar V}_a^i \right) + {\rm H.c.}
\ea%
The Lagrangian for the $\chi_{cJ}$ and $h_c$ reads
\ba%
{\cal L}_\chi \al=\al ig_1\chi_{c2}^{\dag ij} \left( V_a^i{\bar
V}_a^j + V_a^j{\bar V}_a^i \right) + \sqrt{2} g_1 \chi_{c1}^{\dag i}
\left( V_a^i{\bar P}_a + P_a{\bar V}_a^i \right) +
\frac{i}{\sqrt{3}}g_1 \chi_{c0}^\dag \left( \vec{V}_a\cdot\vec{{\bar
V}}_a + 3 P_a{\bar P}_a \right) \non\\
\al\al - g_1\epsilon^{ijk}h_c^{\dag i}V_a^j{\bar V}_a^k +
ig_1h_c^{\dag i} \left( V_a^i{\bar P}_a - P_a{\bar V}_a^i \right) +
{\rm H.c.},
\ea%
where the trace and symmetry properties $\chi_{c2}^{ij}\delta^{ij}=0$ and
$\chi_{c2}^{ij}\epsilon^{ijk}=0$ have been used in the derivations.

\subsection{Decay amplitudes of the pion or eta emission transitions}
\label{sec:loopamp}

The amplitudes for all the transitions with charged charmed meson
loops are listed in Appendix~\ref{app:amp}. The charmonia are
isospin and SU(3) flavor singlets, the pions form an isospin triplet
and the $\eta$ is an element of the SU(3) octet. Therefore, the
transitions between two charmonia with the emission of one pion or
eta break isospin or SU(3) symmetry. The leading contributions to
the eta transition amplitudes are given by the differences
between the non-strange and strange charmed meson loops.
The decay amplitude for the $\psi'\to J/\psi\eta$ is
\be%
\label{eq:mpsieta} {\cal M}\left(\psi'\to J/\psi \eta\right) =
\frac{1}{\sqrt{3}} \left[{\cal M}\left(\psi'\to J/\psi\phi\right)_{0} + {\cal
M}\left(\psi'\to J/\psi\phi\right)_{\pm} - 2{\cal M}\left(\psi'\to
J/\psi\phi\right)_{s}\right],
\ee%
where $\phi$ denotes the light pseudoscalar meson, and the
expression of ${\cal M}\left(\psi'\to J/\psi\phi\right)_{\pm}$ is
given as Eq.~(\ref{eq:amppsipsi}) in Appendix~\ref{app:amp}. The
amplitude with lower index $0$ can be obtained by simply replacing
the charged charmed mesons in Eq.~(\ref{eq:amppsipsi}) by the
neutral ones, and the lower index $s$ denotes the charmed-strange
meson loops. The leading contributions to
the pionic transition amplitudes are given by the differences
between the neutral and charged charmed meson loops, and also from
the the $\pi^0-\eta$ mixing through the loops contributing to the eta transition.
From Eq.~(\ref{eq:pietamixing}), the physical pion is
\be%
\tilde{\pi}^0 =
\pi^0+\epsilon_{\pi^0\eta}\eta + {\cal O}\left(\epsilon_{\pi^0\eta}^2\right).
\ee%
Hence, the amplitude for the $\psi'\to J/\psi\pi^0$ reads
\ba%
\label{eq:Api0} {\cal M}\left(\psi'\to J/\psi \pi^0\right) \al=\al
 \frac{\epsilon_{\pi^0\eta}}{\sqrt{3}}
\left[{\cal M}\left(\psi'\to J/\psi\phi\right)_{0} + {\cal
M}\left(\psi'\to J/\psi\phi\right)_{\pm} - 2{\cal M}\left(\psi'\to
J/\psi\phi\right)_{s}\right]\non\\
\al\al + {\cal
M}\left(\psi'\to J/\psi\phi\right)_{0} - {\cal M}\left(\psi'\to
J/\psi\phi\right)_{\pm}.
\ea%
Note that although the loop amplitudes ${\cal M}\left(\psi'\to
J/\psi\phi\right)_{0,\pm,s}$ take the same form as those in
Eq.~(\ref{eq:mpsieta}),  the momentum $q$ is different for the
different decays, it is given by  the three-momentum of the light
meson in the final state.
For all the other transitions considered in this
paper, the available phase spaces only allow the emission of one pion, and the
amplitudes can be obtained in a similar way to Eq.~(\ref{eq:Api0}) using the
equations given in Appendix~\ref{app:amp}.

As can be seen from the presence of a Pauli matrix between two charmonium fields
in the Lagrangians ${\cal L}_{SS}$ and ${\cal L}_{SP}$, the transitions between
the $S$-wave charmonia violate the heavy quark spin symmetry, so do those
between one $S$ and one $P$-wave charmonia. Were the heavy quark spin symmetry
exact, the vector and pseudoscalar charmed mesons would have the same mass. In
the heavy quark spin symmetric world, all the meson-loop amplitudes for the
transitions between two $S$-wave charmonia, and those between one $S$ and one
$P$-wave charmonia would vanish. This is because the contributions from
different loops would cancel with each other completely as one may easily see
from the amplitudes listed in Appendix~\ref{app:amp} by putting $M_{D}=M_{D^*}$.
This means the vector and pseudoscalar heavy mesons have to be considered
simultaneously to keep the structure of the spin symmetry which is used in
constructing the Lagrangians and relating the coupling constants for different
transitions. The transitions between the $P$-wave charmonia respect the spin
symmetry, so the resulting amplitudes do not vanish in the given heavy quark
spin symmetric world as shown in Appendix~\ref{app:Pamp}. Keeping spin symmetry
structure of the tree-level amplitudes might be a general feature of the heavy
hadron loops,
so that the
analysis of the spin partner of heavy hadron molecules in
Ref.~\cite{Guo:2009id}, which is based on the spin symmetry without considering
loops, would not be affected.

\subsection{Power counting of the loops}
\label{sec:pcloops}

Before proceeding to numerical calculations, one must analyze the power counting
of the loops in the NREFT formalism. Any loop diagram in the paper is composed
of two vertices for the coupling of a charmonium to charmed and anti-charmed
mesons --- both characterized by a vertex structure and a coupling constant, one
vertex for the coupling of a pion or eta to charmed mesons, and three
propagators for charmed mesons. The power counting of a given diagram is
obtained via estimating each individual ingredient by a typical value. Based on
this analysis, each diagram can be assigned a definite order $n$ in the given
expansion parameter, which for the transitions at hand turn out to be the
velocity $v$ of the intermediate heavy mesons. Once a complete calculation up to
order $n$ is performed, the uncertainty of the calculation may be estimated as
$v^{(n+1)}$ --- we will show below that additional scales introduced by the
dimensionful coupling constants do not distort this picture.

Since effective field theories are in general non-renormalizable, they have to
be regularized and renormalized order by order. A consistent power counting thus
has to guarantee that at each order where there are divergences there are also
appropriate counterterms available. In this section we check this for the
transitions and the power counting at hand.

Let us first focus on the propagators, and take the scalar loop integral as an
example. The scalar loop integral in $d$ dimensions for the triangle graphs
under consideration is defined as
\be%
\label{eq:scalarloop}
I(q){=} \int\!\frac{d^dl}{(2\pi)^d} \frac{i}{D}
{=}\int\!\frac{d^dl}{(2\pi)^d}
\frac{i}{\left(l^2-m_1^2+i\epsilon\right)
\left[(P-l)^2-m_2^2+i\epsilon\right]
\left[(l-q)^2-m_3^2+i\epsilon\right] },
\ee%
where $P$ is the momentum of the initial charmonium and $q$ is the momentum of
the light meson in the final state. Non-relativistically, in the rest-frame of
the initial charmonium, the loop can be written as
\begin{eqnarray}\nonumber
I(q) \!\!\! &=&  \!\!\!  \frac{i}{8m_1m_2m_3}
 \int\!\frac{d^4l}{(2\pi)^4} \frac{1}{
\left(l^0{-}T_1(|\vec l|)\right) \left(P^0{-}l^0{-}T_2(|\vec
l|)\right)
\left(l^0{-}q^0{-}T_3(|\vec l{-}\vec q|)\right) } \\
 &=&  \!\!\!  \frac{1}{8m_1m_2m_3}
 \int\!\frac{d^3l}{(2\pi)^3} \frac{1}{
\left(E_i{-}T_2(|\vec l|){-}T_1(|\vec l|)\right)
\left(E_f{-}T_2(|\vec l|){-}T_3(|\vec l{-}\vec q|)\right) }
\label{eq:loop}
\end{eqnarray}
where $T_i(p)=p^2/2m_i=m_i v^2/2$, with $v$ being the charmed meson velocity,
denotes the kinetic energy for the charmed mesons with masses $m_1,m_2$ and
$m_3$, $E_i=M_i-m_1-m_2$ and $E_f=M_f-m_2-m_3-E_\pi$ denote the energies
available for the first (before the pion emission) and second (after the pion
emission) two--heavy--meson intermediate state. One may assign the charmed meson
momentum as $M_Dv$. Here, we will only count the power of $v$ since the
dimension of the loops can be simply implemented by multiplying proper power of
$M_D$. Thus, the scalar loop scales as $1/(16\pi) \, [v^3/(v^2)^2]=1/(16\pi v)$,
since in the last line of Eq.~(\ref{eq:loop}) each of the non-relativistic
propagators is counted as $1/v^2$ and the integral measure is counted as
$v^3/(16\pi)$, where it was used that the loops are dominated by the unitarity
cut, which produces a factor of $\pi$ to be combined with the standard factor
$1/(4\pi)^2$ from the integral measure. This factor is common to all loop
contributions as given in Table~\ref{tab:pcloops}.

As indicated in the Eq.~(\ref{eq:Lphi}) for the axial-coupling of the pion or
eta to the charmed mesons, the corresponding vertex is proportional to the
external momentum of the pion or eta, denoted by $q$ --- this gives one power of
$q$ in the expressions for all the loop contributions in
Table~\ref{tab:pcloops}. Further, we have to account for the isospin or SU(3)
violation as well as the momentum dependence of the charmonium--charmed meson
vertices --- the scaling of the coupling constants will be discussed below. To
account for the corresponding symmetry breaking, in each power counting estimate
for the loops listed in Table~\ref{tab:pcloops}, we have pulled out the meson
mass difference, denoted as $\Delta$,\footnote{These meson mass differences are,
of course, generated by quark mass differences and e.m. effects. For the charmed
mesons, $\Delta$ is of similar size as the quark mass differences $\delta$.}
which is a small energy scale, and divided $\Delta$ by a factor which
characterizes the intrinsic energy, $v^2$, for balance.
 The vertices are more
complicated since their scaling behavior depends on the quantum numbers of the
charmonia in both the initial and final states. We therefore classify the vertices into three groups:
\begin{enumerate}[1)]
\item Both the vertices for the initial and final charmonia are in $S$-wave.
    This corresponds to the transitions between two $P$-wave charmonia
    (denoted by $PP$ in Table~\ref{tab:pcloops}). In this case, the vertices
    do not give any non-trivial contribution to the power counting, and they
    scale as ${\cal O}(v^0q^0)={\cal O}(1)$.
\item One vertex is in $S$-wave, and the other one is in $P$-wave. This
    corresponds to the transitions between one $S$-wave and one $P$-wave
    charmonia (denoted by $SP$ in Table~\ref{tab:pcloops}). In this case,
    the loop momentum must be contracted with the external momentum $q$, and
    hence render the scale of the $P$-wave vertex to be ${\cal O}(q)$. In
    this case a factor  $1/M_D^2$ needs to be introduced to match
    dimensions.
\item Both the vertices are in $P$-wave. This corresponds to the transitions
    between two $S$-wave charmonia (denoted by $SS$ in
    Table~\ref{tab:pcloops}). In this case the loops are tensor loops, which
    can be split into two parts as given in Eq.~(\ref{Aeq:tensorloop}). The
    part $q^iq^jI_0^{(2)}(q)$ scales as two powers of the eternal momentum,
    i.e., ${\cal O}(q^2)$. In the other part, the two momenta in the
    numerator of the loop integrand contract with each other, and as a
    result, the Kronecker delta appears. Here all momenta appearing are
    internal momenta, which, by assumption, scale as $v$. Thus, for this
    piece of the integral the two vertex functions together scale as $v^2$.
  In the transitions considered in the paper, one
    always has $q\lesssim M_Dv$. Hence, the product of the two vertices in this
    case can be counted as ${\cal O}(v^2)$.
\end{enumerate}

The last ingredients to be discussed are the coupling constants $g$, $g_1$ and
$g_2$.  $g$ is the axial coupling constant for the heavy mesons. It is
dimensionless, and should be of order unity.  Based on the underlying
Lagrangians, e.g., the dimension of $g_2$, the coupling of the $S$--wave
charmonia to the open charm ground states, is $-3/2$.  In
Ref.~\cite{Colangelo:2003sa}, using vector meson dominance arguments, the
authors obtain $g_2 = \sqrt{M_{J/\psi}}/(M_Df_{J/\psi})$, where $f_{J/\psi}$ is
the decay constant of the $J/\psi$. On the quark level it scales with the
$J/\psi$ wave function at the origin that, on dimensional grounds, should be
$f_{J/\psi}\sim m_cv_c^{3/2}$ --- the quark mass and velocity $m_c$ and $v_c$
were introduced at the end of Section~\ref{sec:treeLag}. Hence, we have
\be%
g_2 \sim \frac{\sqrt{2}}{(m_cv_c)^{3/2}} \ .
 \ee%
Using $m_c=1.5$ GeV and $v_c^2=0.3$ we get $g_2=1.9$~GeV$^{-3/2}$, which is
close to independent model estimates in a range of 2.1...2.9~GeV$^{-3/2}$ for
this quantity existing in the
literature~\cite{Matinyan:1998cb,Deandrea:2003pv,Matheus:2002nq}.

The expression for $g_1$ derived from vector meson dominance is $g_1 =
-2\sqrt{M_{\chi_{c0}}/3} / f_{\chi_{c0}}$~\cite{Colangelo:2003sa}. From
dimensional analysis, the decay constant of the $P$-wave charmonium $\chi_{c0}$
should scale with the first derivative of the wave function at the origin.
Hence,
\be%
\label{eq:g1scaling}
 g_1 \sim - \sqrt{\frac{2}{3}} \frac{2}{\sqrt{m_c}v_c^{5/2}}.
\ee%

Using the above scaling, we can work out the power counting of the loops for
different processes.For the $SS$ transitions, as mentioned at the end of
    Section~\ref{sec:loopamp}, the loop amplitudes vanish in case of
    $M_D=M_{D^*}$, i.e., the loop amplitude violates spin symmetry as the
    tree level amplitude does. So a factor of ${\Lambda_{\rm QCD}}/{m_c}$
    should also be considered in the scaling of the loop amplitude as well
    as the estimate for the loop integral itself, $1/(16\pi v)$, and a factor
    $v^2$, which originates from the two decay vertices. Furthermore, the
    nonrelativistic normalization factor $\sqrt{M_iM_f}m_1m_2m_3$ gives a
    factor $2m_c^4$. As explained in the above, a factor of
    $\Delta/(m_Dv^2)\sim\Delta/(m_cv^2)$ should also be introduced to
    account for the isospin breaking. Collecting all factors, we get for the
    scaling of the loop amplitude for the $SS$ transitions
    \ba%
    \label{eq:scalingloop} %
    {\cal M}_{\rm loop}^{SS} \al\sim\al \frac{g}{F}g_2g_2'
    q 2m_c^4\frac{v^2}{16\pi v} \frac{\Lambda_{\rm QCD}}{m_c} \frac{\Delta}{m_cv^2} \non\\
    \al\sim\al \frac1{4\pi v_c^3} \frac1{m_c}
    \frac{q\Delta}{v}.
    \ea%
    As will be shown later, these rules for power counting of the loops are
    satisfied by explicit non-relativistic calculations.

    Similarly, for the $SP$ and $PP$ transitions, using the scaling of $g_1$
    given in Eq.(\ref{eq:g1scaling}) we have
    \ba%
    \label{eq:scalingloopSP} %
    {\cal M}_{\rm loop}^{SP} \al\sim\al \frac{g}{F}g_1g_2
    q 2m_c^4\frac1{16\pi v} \frac{q}{M_D^2} \frac{\Lambda_{\rm QCD}}{m_c} \frac{\Delta}{m_cv^2} \non\\
    \al\sim\al \frac1{2\sqrt{3}\pi v_c^4}
    \frac{q^2\Delta}{M_D^2v^3},
    \ea%
    and
    \ba%
    \label{eq:scalingloopPP} %
    {\cal M}_{\rm loop}^{PP} \al\sim\al \frac{g}{F}g_1g_1'
    q 2m_c^4\frac1{16\pi m_c^2v}  \frac{\Delta}{m_cv^2} \non\\
    \al\sim\al \frac1{3\pi v_c^5} \frac1{\Lambda_{\rm QCD}}
    \frac{q\Delta}{v^3}.
    \ea%
     In the last equation, a factor of $1/M_D^2\sim1/m_c^2$ was taken into
     account to give the correct dimension of the scalar loop. This can be done
     as mentioned below Eq.~(\ref{eq:loop}).

\begin{table}[t]
\begin{center}
\begin{tabular}{| l | c | c | } \hline
     &  Tree-level   & Loops  \\ \hline
$SS$ &  \parbox{1.6cm}{$$\frac1{m_c}q\delta$$}
     & \parbox{5.5cm}{$$\frac{\cal N}{m_c}\left(\frac{v^3}{v^4}\right)v^2q\left(\frac{\Delta}{v^2}\right) = \frac{\cal N}{m_c}\frac{q\Delta}{v}$$}\\
$SP$ &  \parbox{1.6cm}{$$\delta$$}
     & \parbox{5.5cm}{$${\cal N}\left(\frac{v^3}{v^4}\right)\frac{qq}{M_D^2}\left(\frac{\Delta}{v^2}\right) = {\cal N}\frac{q^2}{v^3M_D^2}\Delta $$} \\
$PP$ &  \parbox{1.6cm}{$$\frac1{\Lambda_{\rm QCD}}q\delta$$}
     & \parbox{5.5cm}{$$\frac{\cal N}{\Lambda_{\rm QCD}}\left(\frac{v^3}{v^4}\right)q\left(\frac{\Delta}{v^2}\right) = \frac{\cal N}{\Lambda_{\rm QCD}}\frac{q\Delta}{v^3}$$}\\
\hline
\end{tabular}
\caption{\label{tab:pcloops}Power counting of the tree-level amplitudes and the
(leading) loops. Here $SS$, $SP$ and $PP$ represent transitions between two
$S$-wave, one $S$-wave and one $P$-wave, and two $P$-wave charmonia,
respectively. The parameter $\delta$ denotes the quark mass differences, and
$\Delta$ the charmed meson mass differences. They are the strength parameters
for isospin or SU(3) symmetry violation. $v$ is the heavy-meson velocity in the
intermediate loops, $q$ the momentum of the outgoing pseudoscalar meson, and
$M_D$ the mass of the heavy mesons in the loop. For the origin of the individual
factors, see the text. ${\cal N}=1/(4\pi v_c^3)$, $1/(2\sqrt{3}\pi v_c^4)$ and
$1/(3\pi v_c^5)$ for the $SS$, $SP$ and $PP$ transitions, respectively, where
$v_c$ denotes the charm quark velocity inside the charmonia.}
\end{center}
\end{table}

The scaling for the loop amplitudes in Eqs.~(\ref{eq:scalingloop}),
(\ref{eq:scalingloopSP}) and (\ref{eq:scalingloopPP}) need to be compared to
the corresponding estimates for the tree-level amplitudes. The relevant
estimate is given explicitly in Eq.~(\ref{eq:scalingtree}) for the $SS$
transitions and Eq.~(\ref{eq:scalingtreeSPPP}) for the $PP$ and $SP$
transitions --- the final results for all transitions are given in
Table~\ref{tab:pcloops}.  For charmed mesons, $\Delta$ is of similar size as
$\delta$ as may be checked numerically, i.e. $M_{D^+}-M_{D^0}\simeq m_d-m_u$
and $M_{D_s^+}-M_{D^+}\simeq m_s-m_d$. At this stage, let us make a remark
about the factor containing the charm quark velocity scaling of the coupling
constants, i.e. ${\cal N}=1/(4\pi v_c^3)$, $1/(2\sqrt{3}\pi v_c^4)$ and
$1/(3\pi v_c^5)$ for the $SS$, $SP$ and $PP$ transitions, respectively. Taking
$v_c=\sqrt{0.3}$, we get ${\cal N}=0.5, 1$ and 2 for the SS, SP and PP
transitions, respectively. All these numbers are of order unity. Although
$\cal N$ scales differently in powers of $v_c$ for different processes, the
numerical values are not very different. This is because the charm quark
velocity $v_c$ in such a charmonium is not so small.  Hence the numerical
values of the presumably well-separated scales may be similar in
practice. Sometimes, the order of the scales is even reversed. For instance,
purely from the scaling, $f_{J/\psi}\sim m_cv_c^{3/2}$ should be larger than
$f_{\chi_{c0}}\sim m_cv_c^{5/2}$. However, using the experimental value
$\Gamma(J/\psi\to e^+e^-)=5.55\pm0.14$~keV~\cite{PDG2010}, and the relation
between decay constant and the leptonic width of the $J/\psi$ \be%
\Gamma(J/\psi\to e^+e^-) =
\frac{16\pi}{27}\frac{\alpha^2}{M_{J/\psi}}f_{J/\psi}^2, \ee
$f_{J/\psi}=416$~MeV, while the numerical result from QCD sum rules gives
$f_{\chi_{c0}}=510\pm40$~MeV~\cite{Colangelo:2002mj}. Furthermore, in view
that there must be unknown numerical factors in the coupling constants
$g_1, \ g_2$ and also the the tree-level ones, and the $v_c$ scaling of
tree-level couplings is not known yet as mentioned before
Eq.~(\ref{eq:Ascaling}), we will neglect the subtlety caused by $\cal N$ in
the power counting, and just take it to be unity for all the transitions
(numerical calculations in the next section will not be affected). Based on
the underlying power counting, we therefore predict loops to be enhanced by a
power of $1/v$ for $SS$ and $1/v^3$ for $PP$ transitions, while for $SP$
transitions in many cases loops appear to be suppressed --- see detailed
discussion below.  One should also keep in mind that the meson velocity $v$ in
some of the processes to be discussed is as large as 0.5 so that $1/v$ for the
$SS$ transitions might be in practice not a large enhancement. But it should
be sufficient to say that the the loops are important compared to the
tree-level contributions, and should not be neglected in any realistic analysis.

As mentioned above, a consistent power counting needs to ensure that all
appearing divergences can be absorbed into appropriate counterterms. This is of
importance here since the leading counterterms are supplied by the tree-level
amplitudes discussed above, which in some cases appear in higher orders than the
leading loops.  In such a case, they might get renormalized by absorbing the
divergence of the loops. Although the scalar loop, defined in
Eq.~(\ref{eq:scalarloop}) is convergent, for some transitions there are momentum
factors at the vertices, resulting in divergent integrals. In
Appendix~\ref{app:amp} the contributions to all transitions are expressed in
terms of a few fundamental integrals defined in Appendix~\ref{app:loop}.
 Note that, in order to account for the non-relativistic
normalization of the charmonium and charmed meson fields, one needs to multiply
each amplitude by a factor of $\sqrt{M_iM_f}m_1m_2m_3$, where $M_{i(f)}$ is the
mass of the initial (final) charmonium, and $m_i$~$(i=1,2,3)$ are the masses of
the charmed mesons in the loops. The loop functions $I^{(1)}(q)$, $I^{(2)}_0(q)$
and $I^{(2)}_1(q)$ are constructed from the basic scalar loop functions $I(q)$
defined in Eq.~(\ref{eq:scalarloop})
\be%
q^iI^{(1)}(q) = i\int\!\frac{d^dl}{(2\pi)^d} \frac{l^i}{D} \ , \ \
q^iq^jI^{(2)}_0(q)+ \delta^{ij}\vec q\ ^2I^{(2)}_1(q)
 = i\int\!\frac{d^dl}{(2\pi)^d} \frac{l^il^j}{D} \ ,
\ee%
 and $B(c)$, which in $d$ space-time
dimensions is
\be%
B(c) = 4\pi N
\int\!\frac{d^{d-1}l}{(2\pi)^{d-1}}
\frac{1}{\vec{ l}\ ^2+c-i\epsilon},
\ee%
where $N=\mu_{12}\mu_{23}/(16\pi m_1m_2m_3)$ with $\mu_{ij}=m_im_j/(m_i+m_j)$
for the reduced masses. For later use, we define the following quantities
\ba%
\label{eq:abc} a = \left(\frac{\mu_{23}}{m_3}\right)^2 \vec{ q}\ ^2,
\quad c = 2\mu_{12}b_{12}, \quad
c'=2\mu_{23}b_{23}+\frac{\mu_{23}}{m_3}\vec{ q}\ ^2,
\ea%
where $b_{12}=m_1+m_2-M_i$, and $b_{23}=m_2+m_3+q^0-M_i$.

In our full calculation all integrals are evaluated using dimensional
regularization. In this section, however, in order to make all divergences
explicit, we investigate the divergence structure by introducing a sharp
momentum cut-off in three dimensions. With this one finds
\ba%
B(c)=4\pi N\int^\Lambda\!\frac{d^{3}l}{(2\pi)^{3}} \frac{1}{\vec{ l}\ ^2+c-i\epsilon}
\al=\al N\left\{\frac{2\Lambda}{\pi} - {\sqrt{c-i\epsilon}} +
{\cal O}\left(\frac{1}{\Lambda}\right)\right\}.
\ea%
The non-analytic part is finite. It is determined by unitarity, and
hence does not depend on the choice of regularization method.
In the integral $I^{(1)}$ only the linear combination $B(c'-a)-B(c)$
appears --- c.f. Eq.~(\ref{eq:I1}) --- and thus  the divergence of $B(c)$
cancels.
 The UV divergent part of $\vec{q}\ ^2 I^{(2)}_{0,1}(q)$
is
\be%
\vec{q}\ ^2 I^{(2)}_0(q)^{\rm UV} = \vec{q}\ ^2 I^{(2)}_1(q)^{\rm
UV} = \frac{N}{\pi}\Lambda \ .
\ee%
For the loop amplitudes given in Appendix~\ref{app:amp}, only the transitions
between two $S$-wave charmonia, i.e. $\psi'\to J/\psi\pi^0(\eta)$, are
divergent.

In order to estimate the finite parts of the pertinent integrals we may use
$a\ll c \approx c' \approx \sqrt{2\mu b} \approx M_Dv$ and expand the expressions in
a series around $a=0$. With this we find
\ba%
\label{eq:loopsLO}
 B(c'-a) \al=\al B(c) + {\cal O}(a) = -N\sqrt{c} + {\cal
O}(a), \non \\
I(q) \al\approx\al N \frac{1}{\sqrt{c}} + {\cal O}(a), \non\\
I^{(1)}(q) \al\approx\al N \frac{\mu_{23}}{m_3} \frac{1}{2\sqrt{c}}
+ {\cal O}(a), \non\\
\vec{q}\ ^2I^{(2)}_1(q)^{\rm finite} \al\approx\al -N {\sqrt{c}}
+ {\cal O}(a), \non\\
\vec{q}\ ^2I^{(2)}_0(q)^{\rm finite} \al=\al {\cal O}(a) \ .
\ea%
The exact expressions are given in Appendix~\ref{app:loop}.

As an example we now focus on the analysis of diagram (b) in
Fig.~\ref{fig:loops} to the amplitude of the $\psi'\to J/\psi\pi^0$ --- the
discussion is easily generalized to the other diagrams.  The amplitude reads
\be%
{\cal M}(\psi'\to J/\psi\pi^0)_{\rm (b)} = N_{\rm (b)} \left[
\vec{q}\ ^2 I_1^{(2)}(q,D^{*0},D^0,D^0) - \vec{q}\ ^2
I_1^{(2)}(q,D^{*\pm},D^\pm,D^\pm) \right],
\ee%
with $N_{\rm (b)}=-4(g/F)g_2g_2'
\epsilon^{ijk}q^i\varepsilon^j(\psi')\varepsilon^k(J/\psi)$.
The contribution of the finite part of the loop function
$I^{(2)}_1(q)$ to diagram (b) behaves as
\be%
{\cal M}(\psi'\to J/\psi\pi^0)^{\rm finite}_{\rm (b)} \sim -N_{\rm
(b)} \left(N_n\sqrt{2\mu_nb_n} - N_c\sqrt{2\mu_cb_c}\right),
\ee%
where the lower index $n$ means neutral, and $c$ charged. Denoting the mass
difference between the charged and neutral charmed mesons by
$\Delta$,\footnote{We neglect the difference between the mass difference for the
pseudoscalar mesons and that for the vector ones, which is of higher order in
heavy quark expansion. Empirically, one finds
$\Delta_P=m_{D^{\pm}}-m_{D^0}=4.77\pm0.10$~MeV, and
$\Delta_V=m_{D^{*\pm}}-m_{D^{*0}}=3.29\pm0.21$~MeV. Their difference is about
30\% of $\Delta_P$, which can be understood as ${\cal O}(\Lambda_{\rm QCD}/m_c)$
effects.} we have $\mu_c=\mu_n+\Delta/2$ and $b_c=b_n+2\Delta$. Thus, we have
\ba%
\label{eq:pcbfinite} {\cal M}(\psi'\to J/\psi\pi^0)^{\rm finite}_{\rm (b)}
\al\sim\al N_{\rm (b)} N \Delta
\frac{2\mu_n+b_n/2}{\sqrt{2\mu_n b_n}} + {\cal O}(\Delta^2) \non\\
\al\sim\al N_{\rm (b)} N \frac{\Delta}{v},
\ea%
which is consistent with the power counting analysis for $SS$ transitions
given above. Here, we neglect the difference between $N_n$ and
$N_c$ since it is of higher order. On the other hand, the UV
divergence of diagram (b) is
\ba%
{\cal M}(\psi'\to J/\psi\pi^0)^{\rm UV}_{\rm (b)} \al=\al
N_{\rm (b)}\frac{\Lambda}{\pi} \left(N_n-N_c\right) \non\\
\al\sim\al N_{\rm (b)}N \frac{\Lambda}{\mu_n} \frac{\Delta}{\pi}.
\ea%
Therefore, for diagram (b), the finite part is of order ${\cal O}\left(\Delta
v^{-1}\right)$, while the UV divergence is of order ${\cal O}\left(\Delta
v^{0}\right)$. Hence the UV divergence is one order higher in the expansion of
$v$. Recalling the tree-level contribution to the $\psi'\to J/\psi\pi^0$ starts
from the same order as ${\cal O}\left(\Delta v^{0}\right)$ --- see the column
for $SS$ transitions in Table~\ref{tab:pcloops} --- such a divergence can be
renormalized by a counterterm in the Lagrangian for the tree-level contribution.

As a result of this analysis we summarize, comparing the loop contributions with
the tree-level decay amplitudes given in the last section and in
Table~\ref{tab:pcloops}, where we assume the same scale for the light quark mass
and the heavy meson mass differences, $\delta\simeq\Delta$~\cite{Guo:2008gp},
that the loop contributions for the $SS$ transitions are enhanced by a factor of
$1/v\approx 2$, and for the $PP$ transitions even by a factor of $1/v^3\approx
10$. The situation for the $SP$ transitions should be analyzed case by case
since an enhancement factor $1/v^3$ competes with a suppression factor
$q^2/M_D^2$. For the $SP$ transitions with small phase space, the external
momentum $q$ might be small enough to make $q^2/(v^3M_D^2)$ much smaller than 1,
which is satisfied for the decays $\psi'\to h_c\pi^0$ and $\eta_c'\to
\chi_{c0}\pi^0$~\cite{Guo:2010zk}. In this case, the decay is dominated by the
tree-level contributions. However, for the decays with external momentum
$q\gtrsim M_D v^{3/2}$, the factor $q^2/(v^3M_D^2)\gtrsim1$, so is no more a
suppression, and hence the tree-level contributions are at least as important as
the loop contributions. In summary, our  power counting was shown to be
consistent with the divergence structure of the pertinent integrals.



\section{Results for the decay widths}

\label{sec:results}

In this section, we give the results from explicit calculations of all the mentioned
transitions with emphasis on the contributions from charmed meson loops.
In the numerical evaluations
 we use the following values for the meson masses~\cite{PDG2010}
\ba%
\label{eq:masses} M_{\pi^0} \al=\al 134.98~{\rm MeV}, \qquad M_{\eta} = 547.85~{\rm MeV}, \non\\
M_{D^+} \al=\al 1869.60~{\rm MeV},  \quad M_{D^0} = 1864.83~{\rm MeV}, \quad M_{D_s} = 1968.47~{\rm MeV}, \non\\
M_{D^{*+}} \al=\al 2010.25~{\rm MeV}, \quad M_{D^{*0}} = 2006.96~{\rm MeV},
\quad M_{D_s^*} = 2112.3~{\rm MeV}, \non\\
M_{J/\psi} \al=\al 3096.92~{\rm MeV}, \quad M_{\eta_c} = 2980.3~{\rm MeV}, \qquad
M_{\psi'} = 3686.09~{\rm MeV}, \non\\
M_{\eta_c'} \al=\al 3637~{\rm MeV}, \qquad ~ M_{h_c} = 3525.42~{\rm MeV}, \quad
M_{\chi_{c0}} = 3414.75~{\rm MeV}, \non\\
M_{\chi_{c1}} \al=\al 3510.66~{\rm MeV}, \quad M_{\chi_{c2}} = 3556.20~{\rm
MeV}, \quad  M_{\chi_{c2}'} = 3929~{\rm MeV}.
\ea%
Here we have identified the $\chi_{c2}'$ with the $Z(3930)$ as
done by the Particle Data Group, see e.g.
Refs.~\cite{Uehara:2005qd,Eichten:2005ga,Eichten:2007qx,Li:2009zu}.  The
$Z(3930)$ was observed in the $D{\bar D}$ mass distribution in photon-photon
collisions by the Belle Collaboration~\cite{Uehara:2005qd} with a mass of
$3929\pm5\pm2$~MeV and width of $29\pm10\pm2$~MeV. The observed angular
distribution suggests that its quantum numbers are $J^{PC}=2^{++}$. The widths
of the $\chi_{c2}'$ calculated in quark models are consistent with the observed
values for the $Z(3930)$~\cite{Eichten:2005ga,Li:2009zu}.

There is no unambiguous candidate for either of the other three excited $P$-wave
charmonia $\chi_{c0}'$, $\chi_{c1}'$ and $h_c'$. Very recently, the Belle
Collaboration observed an enhancement, called the $X(3915)$, in the
$J/\psi\omega$ mass distribution in photon-photon
collisions~\cite{Uehara:2009tx}. The mass and width of the $X(3915)$ were
reported to be $3915\pm3\pm2$ and $17\pm10\pm3$,
respectively~\cite{Uehara:2009tx}. In Ref.~\cite{Liu:2009fe}, the authors
suggest the $X(3915)$ as the $\chi_{c0}'$. However, this assignment may be
criticized from the following points: First, since the $\chi_{c0}'$ is above the
$D{\bar D}$ threshold, and it couples to $D{\bar D}$ in an $S$-wave, one would
expect it to have a larger width than the 17 MeV of the $X(3915)$. Second, were
the $Z(3930)$ the $\chi_{c2}'$, the mass difference
$M_{Z(3930)}-M_{X(3915)}=14\pm6$~MeV is too small for the hyperfine splitting,
since the mass splitting between the ground state $\chi_{c2}$ and $\chi_{c0}$ is
one order of magnitude larger,
$M_{\chi_{c2}}-M_{\chi_{c0}}=141$~MeV~\cite{PDG2010}. In addition, one expects
the hyperfine splittings for the bottomonia are smaller than the corresponding
ones for the charmonia, as can be checked from all the measured cases. The mass
splitting between the excited $P$-wave bottomonia $\chi_{b2}'$ and $\chi_{b0}'$
is $36.2\pm0.8$~MeV~\cite{PDG2010}. It is larger than $M_{Z(3930)}-M_{X(3915)}$,
and therefore does not support the assignment of the $X(3915)$ and $Z(3930)$ as
the $\chi_{c0}'$ and $\chi_{c2}'$ simultaneously. Based on the above arguments,
we shall let the mass of the $\chi_{c0}'$ run in a range from 3800~MeV to
3930~MeV which covers the predicted values from quark
models~\cite{Barnes:2005pb,Li:2009zu}.

In the observed spectrum of the charmonia, the only candidate of the
$\chi_{c1}'$ with well-established quantum numbers is the $X(3872)$ discovered
by the Belle Collaboration~\cite{Choi:2003ue}. However, due to the proximity to
the $D\bar D^*$ threshold, the interpretation of the $X(3872)$ as a molecular
state~\cite{Tornqvist:2004qy,Braaten:2007dw} or virtual
state~\cite{Hanhart:2007yq} (for an update of the latter analysis see
Ref.~\cite{Kalashnikova:2009gt}) is very intriguing~(for reviews, see, e.g.
Refs.~\cite{Swanson:2006st,Eichten:2007qx,Voloshin:2007dx,Godfrey:2008nc}).
Therefore, although a molecular interpretation of the $X(3872)$ was questioned
in Ref.~\cite{Bignamini:2009sk} (see also Ref.~\cite{Artoisenet:2009wk} for a
critical re-evaluation), we shall not identify the $X(3872)$ as the
$\chi_{c1}'$, and the mass of the $\chi_{c1}'$ will also be allowed to vary.
To be specific, a range from 3.83~GeV to 3.93~GeV will be chosen which
covers the the predicted value from some quark models, e.g.
Ref.~\cite{Li:2009zu}, and the mass of the $X(3872)$.

\subsection{Transitions between the $S$-wave charmonia}

\label{sec:S-S}

The decays studied in this class are $\psi'\to J/\psi \pi^0$ and  $\psi'\to
J/\psi \eta$. As shown in Sec.~\ref{sec:pcloops} (see especially
Table~\ref{tab:pcloops}), for these transitions the charmed meson loops are
enhanced by $1/v$. The velocity can be estimated as $v\sim
\sqrt{(2M_{\hat{D}}-M_{\hat{\psi}})/M_{\hat{D}}}\simeq0.53$ with $M_{\hat{D}}$
being the averaged charmed-meson mass, and
$M_{\hat{\psi}}=(M_{J/\psi}+M_{\psi'})/2$. Hence, the LO result for the width is
provided by the loops~\cite{Guo:2009wr}, although the relatively large expansion
parameter leads to a sizable uncertainty.

Taking into account only the loop contributions, we give the
numerical results for the $\psi'\to J/\psi\pi^0$ and $\psi'\to
J/\psi\eta$. To account for the non-relativistic normalization, one
needs to multiply the amplitudes by a proper factor.

One way to do this is to multiply the amplitude for each loop by the factor
$\sqrt{M_{\psi'}M_{J/\psi}}\times m_1m_2m_3$, with $m_{i}$ being the masses of
the charmed mesons in the loop. In this way, the factor may differ for different
loops. From the decay  $D^*\to D\pi$ the coupling $g$, defined in
Eq.~(\ref{eq:Lphi}), can be fixed to $g=0.6$ using $F=92.4$~MeV.  We thus get
for the decay widths
\ba%
\nonumber
 \Gamma(\psi'\to J/\psi\pi^0) \al=\al (0.048\pm0.025)
g_2^2{g_2'}^2~{\rm keV},
\\
\label{eq:widthpsi}
\Gamma(\psi'\to J/\psi\eta) \al=\al (0.43\pm0.23) g_2^2{g_2'}^2~{\rm keV},
\ea%
where an uncertainty of 53\%, which is the value of $v$, has been taken into
account. Here and in the following values for $g_2$ and $g_2'$ are given in unit
of GeV$^{-3/2}$. The resulting ratio, which is parameter-free, reads
\be%
\label{eq:Rpi0eta} R_{\pi^0/\eta} = 0.11\pm0.06,
\ee%
with an uncertainty of 53\%. Comparing with the measured values listed in
Table~\ref{tab:udratio}, it is within two sigma of the CLEO and PDG-fit data,
and even overlaps within uncertainties with the result given by the BES
Collaboration. A more conservative estimate of the uncertainty of the ratio may
be given by assuming the uncertainties in Eq.~(\ref{eq:widthpsi}) are
uncorrelated, which would give $R_{\pi^0/\eta}$ in a much larger range from 0.03
to 0.36. It is consistent with the data in Table~\ref{tab:udratio}.

Within the effective field theory we can not predict the absolute rates for the
decays, for the couplings are unknown. However, we may use the results given
above to extract some averaged coupling constants that may then be compared to
the corresponding values in the literature. Using the experimental data for two
decay widths $\Gamma(\psi'\to J/\psi\pi^0)=0.40\pm0.03$~keV and $\Gamma(\psi'\to
J/\psi\eta)=10.0\pm0.4$~keV~\cite{PDG2010}, the value of $G\equiv
\sqrt{g_2g_2'}$ can be deduced. Since with the more conservative uncertainty
estimate a ratio consistent with the data is obtained, a combined fit to both
widths is possible. The widths for both decays agree with the data within
uncertainties giving $ G \approx 2.0~{\rm GeV}^{-3/2}. $ We can define
dimensionless coupling constants $g_{\psi D_{(s)}^{(*)}D_{(s)}^{(*)}} =
g_2\sqrt{M_{J/\psi}M_{D_{(s)}^{(*)}}M_{D_{(s)}^{(*)}}}$ and similar quantities
for the $\psi'$. Assuming the coupling constants are SU(6) symmetric, i.e.
$g_{\psi DD} = g_2\sqrt{M_{J/\psi}{\bar m}^2}$ with ${\bar m}$ being the average
mass of the charmed meson spin-flavor SU(6) multiplet, we obtain $G_{\psi
DD}\equiv\sqrt{g_{\psi DD}g_{\psi' DD}} \approx 7.3$.

Another way to account for the non-relativistic normalization is to
use an overall normalization factor, i.e.,
$\sqrt{M_{\psi'}M_{J/\psi}}{\bar m}^3$. In this way, the results are
collected as follows
 \ba%
 \Gamma(\psi'\to J/\psi\pi^0) \al=\al
(0.098\pm0.052) g_2^2{g_2'}^2~{\rm keV},
\\
\Gamma(\psi'\to J/\psi\eta) \al=\al (0.84\pm0.45) g_2^2{g_2'}^2~{\rm keV},
\ea%
and their ratio  is almost the same as that given in Eq.~(\ref{eq:Rpi0eta}). The
combined coupling constant is extracted from a combined fit as $ G \approx
1.7~{\rm GeV}^{-3/2} $ and the dimensionless one is $G_{\psi DD}\approx 6.0$.

Note that because the $\psi'$ and $J/\psi$ are below the open charm threshold,
their coupling to the charmed mesons cannot be extracted directly using the
decay widths. Nevertheless, there are several theoretical estimates. Using
vector meson dominance, the value for the $J/\psi DD$ coupling was estimated to
be 7.7 in Ref.~\cite{Matinyan:1998cb} and $8.0\pm0.5$ in
Ref.~\cite{Deandrea:2003pv}. Using QCD sum rules, Ref.~\cite{Matheus:2002nq} obtained
$8.2\pm1.3$, and using an SU(4) chiral model, it was estimated to be 4.93 in
Ref.~\cite{Haglin:2000ar}. Assuming $g_{\psi DD}=g_{\psi' DD}$, the $J/\psi DD$
coupling extracted from the $\psi'\to J/\psi\pi^0(\eta)$ considering only the
charmed meson loops 
is of similar size as these phenomenological estimates.

Since the results for the widths considering these two ways to account for the
non-relativistic normalization are consistent within uncertainties, in the
following, we shall only use the former one, i.e. using the physical masses of
the intermediate mesons in the factor $\sqrt{M_{i}M_{f}}m_1m_2m_3$.

\subsection{Transitions between the $S$- and $P$-wave
charmonia}\label{sp-nreft}

In these transitions, the situation is different. As analyzed in
Section~\ref{sec:pcloops}, the loops do not necessarily dominate the transitions. Especially
for those decays which have a very small phase space, the contribution from the
charmed-meson loops is highly suppressed~\cite{Guo:2010zk}. For these decays
the momentum of the emitted pion is so small, that $q^2/(v^3M_D^2)\ll 1$.
 Two examples are given by  $\psi'\to
h_c\pi^0$, with $q=86$~MeV, and  $\eta_c'\to\chi_{c0}\pi^0$, with $q=171$~MeV.
Considering that the velocity $v$ is about $\sqrt{[2M_{\hat{D}}-(M_{\psi'}+M_{h_c})/2]/M_{\hat{D}}}\simeq0.4$,
in case of the former transition
the dimensionless factor is
\be%
\frac{1}{v^3} \frac{\vec{q}\ ^2}{m_D^2} \approx 0.03 \ .
\ee%
Thus, if all couplings are natural, the heavy-meson loop contributions are
estimated to only give a few percent correction to the tree-level contribution.
In case only loops are considered we get for the width of the $\psi'\to
h_c\pi^0$ the  very small prediction,
\be%
\label{eq:psi'hc_loop} \Gamma(\psi'\to h_c\pi^0)_{\rm loop} = 2.1\times10^{-7}
g_1^2 g_2^{\prime\ 2}~{\rm keV},
\ee%
where the value of $g_1$ is given in unit of GeV$^{-1/2}$. The decay width has
been measured by the BES-III Collaboration very recently~\cite{BESIIIhc}, and
the absolute value of the branching ratio for the $\psi'\to h_c\pi^0$ is ${\cal
B}(\psi^\prime\to\pi^0 h_c) = (8.4\pm 1.3\pm 1.0)\times 10^{-4}$. Using the PDG
value for the width of the $\psi'$, $\Gamma(\psi') =
304\pm9$~keV~\cite{PDG2010}, the measured width is $\Gamma(\psi'\to
h_c\pi^0)=0.26\pm0.05$~keV. Taking into account that $g_2'$ is about
2~GeV$^{-3/2}$ as extracted in Section~\ref{sec:S-S} and Ref.~\cite{Guo:2009wr},
and $g_1$ is about $-4$~GeV$^{1/2}$ from an estimate using the vector meson
dominance~\cite{Colangelo:2003sa}, the numerical value $\Gamma(\psi'\to
h_c\pi^0)_{\rm loop}\sim10^{-5}$~keV is orders of magnitude smaller than the
measured value. This confirms our very rough estimate for the loop contributions
in the above.

The same situation happens to the $\eta_c'\to \chi_{c0}\pi^0$. For
this decay, the momentum of the pion is 171~MeV, and the suppression
factor is
\be%
\frac{1}{v^3} \frac{\vec{q}\ ^2}{m_D^2} \approx 0.1 \ .
\ee%
Thus, here, for natural couplings, the loops are expected to give a correction
of the order of 10\% to the tree-level amplitude. If again only loops are
considered, the width is again quite small
\be%
\Gamma(\eta_c'\to \chi_{c0}\pi^0)_{\rm loop} = 1.0\times10^{-5} g_1^2
g_2^{\prime\ 2}~{\rm keV}.
\ee%

The pion momentum is 382~MeV for the transitions $h_c\to J/\psi\pi^0$, and
387~MeV for the $\chi_{c0}\to\eta_c\pi^0$. The dimensionless suppression factor
is about 0.3, still a small number although much larger than the previous two
cases. Especially through the interference with the tree-level amplitude meson
loops may give a significant contribution for the mentioned decays and should
not be neglected in any quantitative analysis.
 The widths,
considering only the loop contributions, are given by
 \ba\label{eq-p-s}%
\Gamma(h_c\to J/\psi\pi^0)_{\rm loop} \al=\al 1.9\times10^{-4} g_1^2
g_2^2~{\rm keV}, \non\\
\Gamma(\chi_{c0}\to\eta_c\pi^0)_{\rm loop} \al=\al 3.3\times10^{-4} g_1^2
g_2^2~{\rm keV}.
\ea%

\begin{figure}[t]
\begin{center}
\vglue-0mm
\includegraphics[width=0.6\textwidth]{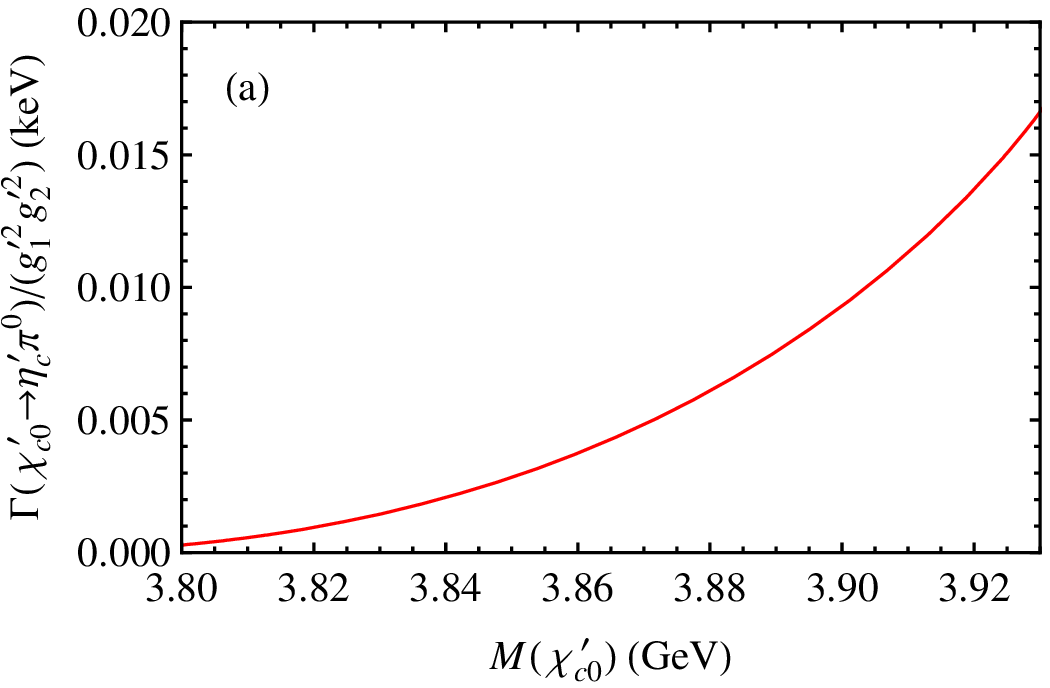}\\[3mm]
\includegraphics[width=0.6\textwidth]{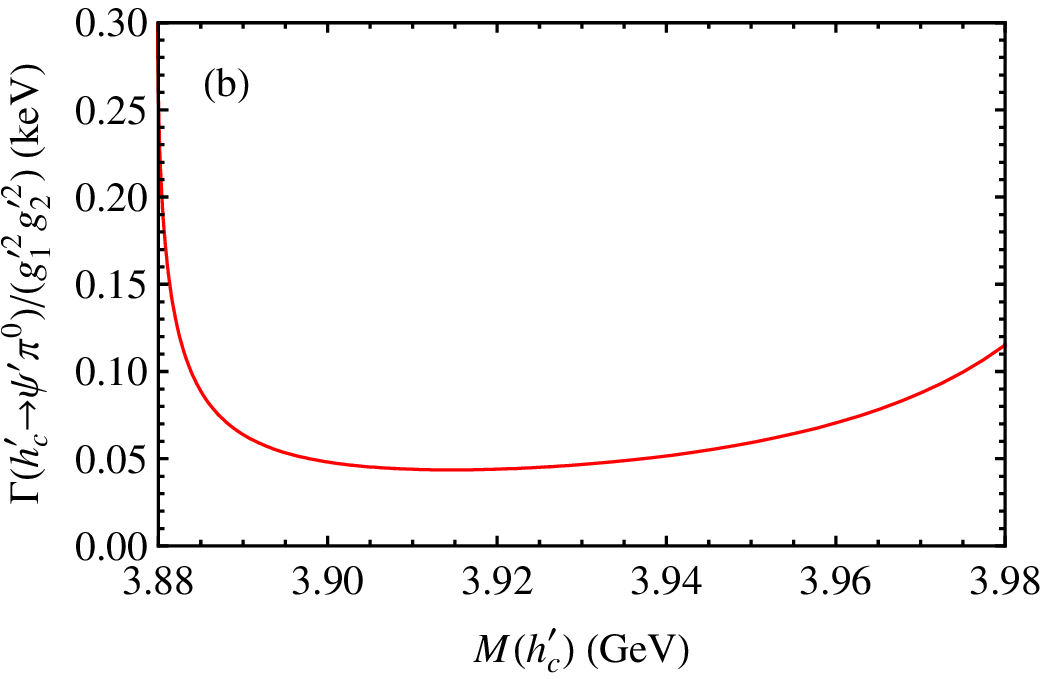}
\vglue-0mm \caption{The widths of the decays
$\chi_{c0}'\to\eta_c'\pi^0$  and $h_c'\to \psi'\pi^0$ considering
only meson loops. \label{fig:GSPchic0p}}
\end{center}
\end{figure}

For the transitions $\chi_{c0}'\to\eta_c'\pi^0$ and $h_c'\to \psi'\pi^0$, the
situation is more complicated. Although the pion momentum is small for these
decays, the dimensionless factor $(q_\pi/M_D)^2/v^3$ does not necessarily gives
a large suppression. For these two transitions, the masses of both the initial
and final charmonia are close to the thresholds of the charmed mesons involved
in the loops. As a result, the velocity as approximated by $\sqrt{\bar{
|b|}/{\bar M}_D}$, where $\bar{ |b|}=(|b_{12}|+|b_{23}|)/2$, is small. As
discussed at the beginning of this section, the $\chi_{c0}'$ and $h_c'$ have not
been unambiguously identified so far. To illustrate the point, one may use
values for their masses from quark model calculations. For the
$\chi_{c0}'\to\eta_c'\pi^0$, considering the loop $[D,{\bar D},D^*]$, the
velocity is about 0.30 if we use $M_{\chi_{c0}'}=3842$~MeV predicted in
Ref.~\cite{Li:2009zu}, and $(q_\pi/M_D)^2/v^3\approx0.23$ is still a suppression
factor. If we use $M_{\chi_{c0}'}=3916$~MeV predicted in the quark
model~\cite{Barnes:2005pb}, the velocity is about 0.33, and
$(q_\pi/M_D)^2/v^3\approx0.43$. The loops are not highly suppressed, and they
may have a contribution comparable to or at least not much smaller than the
tree-level one. Taking into account only the loops, one gets
\ba%
\Gamma(\chi_{c0}'\to\eta_c'\pi^0)_{\rm loop} = 0.002 (0.010) g_1^{\prime2}
g_2^{\prime2}~{\rm keV},
\ea%
where the numbers outside and inside the parentheses are obtained using
$M_{\chi_{c0}'}$ predicted in Ref.~\cite{Li:2009zu} and \cite{Barnes:2005pb},
respectively --- the difference reflects the difference in the
factor $(q_\pi/M_D)^2/v^3$ which enters squared in the expression for the width. In
Fig.~\ref{fig:GSPchic0p}~(a), the width of the decay $\chi_{c0}'\to\eta_c'\pi^0$
as a function of $M_{\chi_{c0}'}$ is shown.
For the decay $h_c'\to \psi'\pi^0$, considering the loop $[D^*,{\bar D},D]$, the
velocity is about 0.14 using $M_{h_c'}=3908$~MeV given in Ref.~\cite{Li:2009zu}.
As a result, the factor $(q_\pi/M_D)^2/v^3\approx2.9$ even gives an enhancement.
Then the width in this case is induced mainly by the charmed meson loops. The
result for a varying $M_{h_c'}$ is shown in Fig.~\ref{fig:GSPchic0p}~(b). In the
figure, the thresholds of the $D\bar D$ and  $D^*\bar D^*$ are approached at the
lower and higher end, respectively. Consequently, the curve for the width shows
an increasing tendency at both ends. Because the width of the
$\chi_{c0}'\to\eta_c'\pi^0$ is tree-level dominated, and that of the $h_c'\to
\psi'\pi^0$ is dominated by loops, which give an enhancement, we predict their
ratio to be significantly smaller than the one derived from the assumption that
both are tree-level dominated
\be%
\frac{\Gamma\left(\chi_{c0}'\to\eta_c'\pi^0\right)_{\rm
tree}}{\Gamma\left(h_c'\to \psi'\pi^0\right)_{\rm tree}} = 3
\frac{q_1}{q_2}\frac{M_{\eta_c'}M_{h_c'}}{M_{\chi_{c0}'}M_{\psi'}},
\ee%
where $q_1$ and $q_2$ are the pion momenta in the rest frame of the initial
state for the decays $\chi_{c0}'\to\eta_c'\pi^0$ and $h_c'\to \psi'\pi^0$,
respectively. Once the masses of the $\chi_{c0}'$ and $h_c'$ are measured, our
statement on the ratio can be tested.

\subsection{Transitions between two $P$-wave
charmonia}\label{NREFT-p-wave}

As has been shown in Section~\ref{sec:pcloops}, the charmed meson loops scale as
$q\Delta/v^3$. Compared to the tree-level amplitudes, the loops are enhanced by
a factor of $1/v^3$, see  Table~\ref{tab:treeamp}.

Therefore, it is reasonable to neglect the contributions from the tree-level
diagrams for the transitions between two $P$-wave charmonia. There are in total
seven transition processes from the first radial excited $P$-wave charmonia to
the ground state ones. All of the amplitudes are proportional to the product of
the same coupling constants $gg_1g_1'/F$. Therefore, the ratios among these
decay widths can be predicted without any free parameter.

\begin{figure}[t]
\begin{center}
\vglue-0mm
\includegraphics[width=0.6\textwidth]{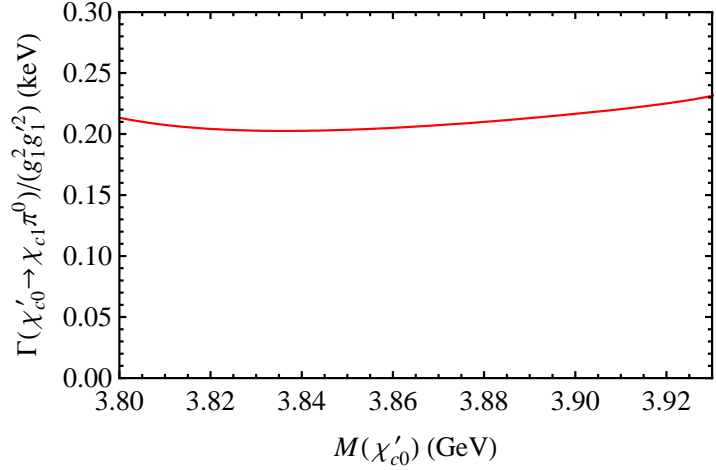}
\vglue-0mm \caption{The width of the decay $\chi_{c0}'\to\chi_{c1}\pi^0$.  \label{fig:Gchic0}}
\end{center}
\end{figure}

Instead of taking some value of the unknown coupling constants $g_1$ and $g_1'$
from model estimates, we choose to show the coupling-constant dependent width
for the $\chi_{c0}'\to\chi_{c1}\pi^0$ in Fig.~\ref{fig:Gchic0}. The
curve is obtained by setting the coupling constants $g_1$ and $g_1'$ to
1~GeV$^{-1/2}$, and the width can be obtained by multiplying the value plotted
in the figure by $g_1^2g_1^{\prime2}$.

\begin{figure}[t]
\begin{center}
\vglue-0mm
\includegraphics[width=0.6\textwidth]{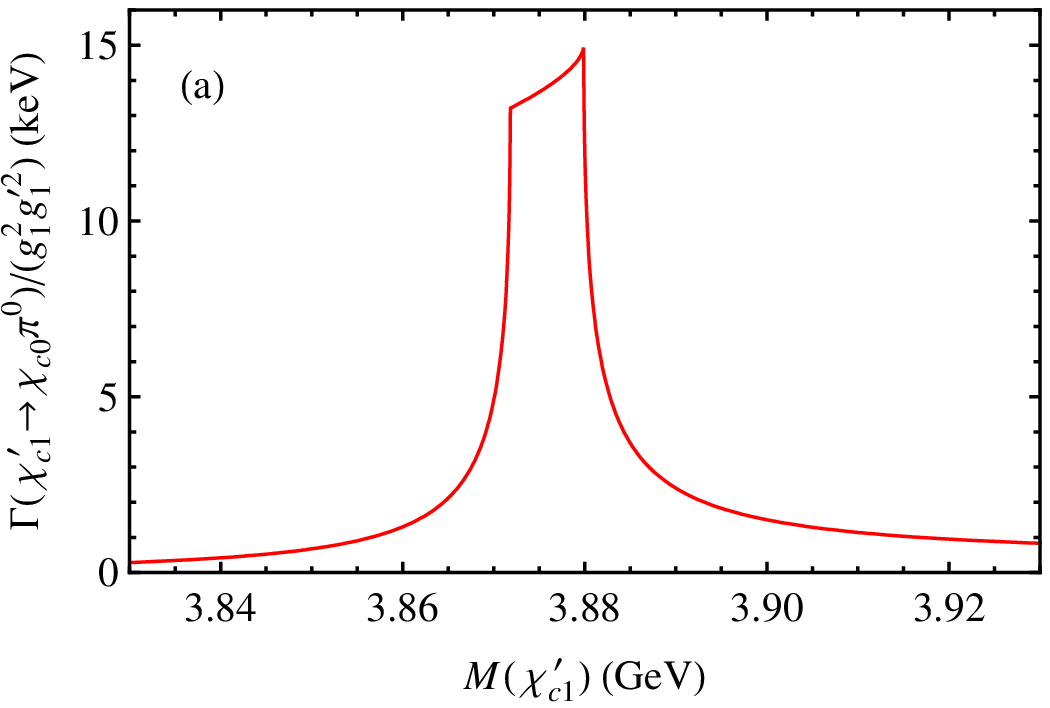}\\[3mm]
\includegraphics[width=0.6\textwidth]{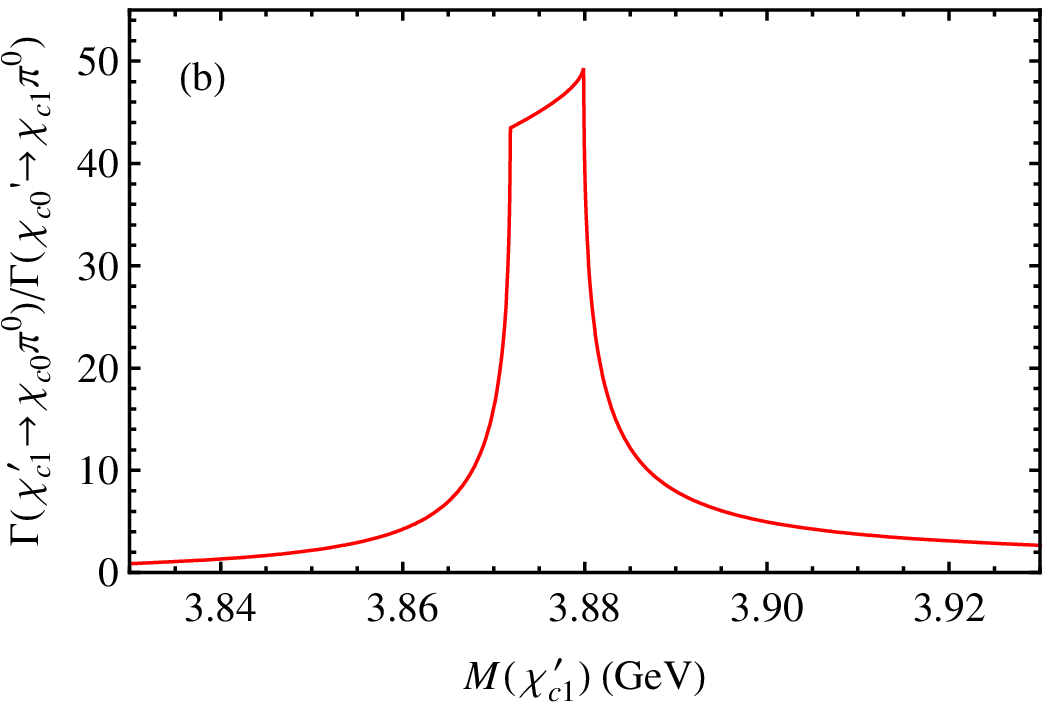}
\vglue-0mm \caption{(a) The width of the decay $\chi_{c1}'\to\chi_{c0}\pi^0$.
(b) The parameter-free ratio between the two decay widths
$\Gamma(\chi_{c1}'\to\chi_{c0}\pi^0)_{\rm loop}/\Gamma(\chi_{c0}'\to\chi_{c1}\pi^0)_{\rm loop}$,
here $M_{\chi_{c0}'}=3842$~MeV as predicted in Ref.~\cite{Li:2009zu} is used. \label{fig:Rchic1001}
}
\end{center}
\end{figure}

The mass of the $\chi_{c1}'$ is also allowed to vary, and the width for the
$\chi_{c1}'\to\chi_{c0}\pi^0$ is shown in Fig.~\ref{fig:Rchic1001}~(a). In
Fig.~\ref{fig:Rchic1001}~(b), the parameter-free ratio of the decay widths
$\Gamma(\chi_{c1}'\to\chi_{c0}\pi^0)_{\rm
loop}/\Gamma(\chi_{c0}'\to\chi_{c1}\pi^0)_{\rm loop}$ is shown.
In the plots, a double-cusp structure is prominent. These cusps correspond to
the thresholds of the neutral and charged $D{\bar D}^*$ mesons. When the
thresholds are approaching, the velocities of the intermediate charmed mesons
decrease, and then the enhancement characterized by $1/v^3$ becomes larger.
A similar double-cusp structure was found in a study of $a_0(980)-f_0(980)$
mixing in the decays of charmonia~\cite{Wu:2007jh,Hanhart:2007bd,Wu:2008hx},
where the cusps are due to the thresholds of the charged and neutral kaons~\cite{Achasov:1979xc}.
\begin{figure}[t]
\begin{center}
\vglue-0mm
\includegraphics[width=0.6\textwidth]{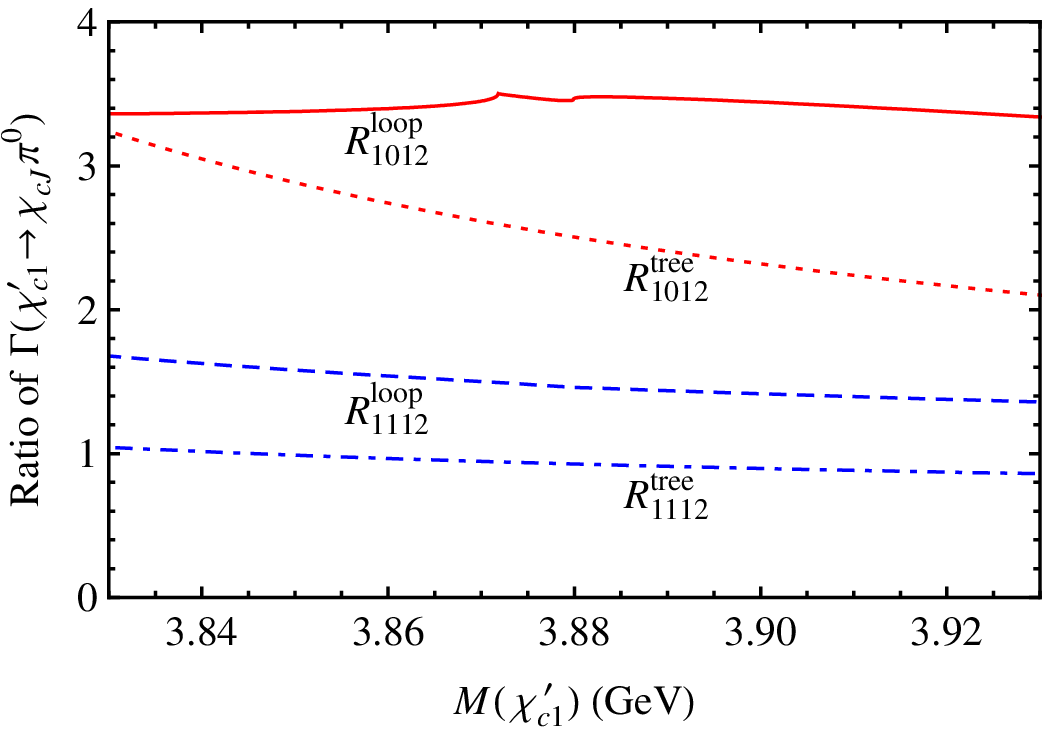}
\vglue-0mm \caption{ The parameter-free ratios among the decay widths of the $\chi_{c1}'\to\chi_{cJ}\pi^0$
considering only the meson loops. The corresponding tree-level predictions are
also shown.
\label{fig:Rchic1J}}
\end{center}
\end{figure}
For the decays of the $\chi_{c1}'\to\chi_{cJ}\pi^0$, the ratios defined as
\ba%
\label{ratio-chicj} R_{1012}^{\rm loop}  \equiv
\frac{\Gamma(\chi_{c1}'\to\chi_{c0}\pi^0)_{\rm
loop}}{\Gamma(\chi_{c1}'\to\chi_{c2}\pi^0)_{\rm loop}},
\quad R_{1112}^{\rm loop}  \equiv \frac{\Gamma(\chi_{c1}'\to\chi_{c1}\pi^0)_{\rm
loop}}{\Gamma(\chi_{c1}'\to\chi_{c2}\pi^0)_{\rm loop}},
\ea%
are plotted in Fig.~\ref{fig:Rchic1J} as the solid and dashed lines,
respectively, in a parameter-free way. One may also construct similar ratios
taking into account only the tree-level contributions, which are
\ba%
R_{1012}^{\rm tree} = \frac{4q_{\pi0}^3}{5q_{\pi2}^3}
\frac{M_{\chi_{c0}}}{M_{\chi_{c2}}}, \quad R_{1112}^{\rm tree} =
\frac{3q_{\pi1}^3}{5q_{\pi2}^3} \frac{M_{\chi_{c1}}}{M_{\chi_{c2}}},
\ea%
where $q_{\pi J}$ is the pion momentum for the decays
$\chi_{c1}'\to\chi_{cJ}\pi^0$. The dotted and dot-dashed curves in
Fig.~\ref{fig:Rchic1J} represent these ratios, respectively. Comparing the
curves considering only the meson loops with the tree-level ones, the difference
is not tremendous. Since the uncertainty is large in our loop calculations, c.f.
Eq.~(\ref{eq:Rpi0eta}) and the discussion below, it may be difficult to draw a
definite conclusion from the comparison without a refined uncertainty analysis.

The ratios for the tree-level contributions agree with those derived in
Ref.~\cite{Casalbuoni:1992fd} using chiral Lagrangians, but differ from those
given for the $c{\bar c}$ option of the $X(3872)$ in
Ref.~\cite{Dubynskiy:2007tj} using the QCDME. However, as shown above, even if
the $X(3872)$ is a $c{\bar c}$ state, its decays into the $\chi_{cJ}\pi^0$ are
mainly given by the intermediate charmed meson loops, i.e. by non-mulitpole
contributions.

\begin{figure}[htb]
\begin{center}
\vglue-0mm
\includegraphics[width=0.6\textwidth]{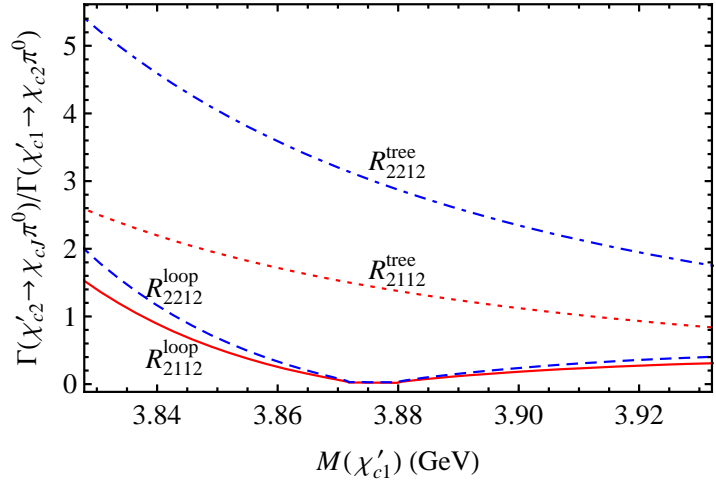}
\vglue-0mm \caption{Ratio of the decay widths $\chi_{c2}'\to\chi_{c1}\pi^0$ and
$\chi_{c2}'\to\chi_{c2}\pi^0$ to $\Gamma(\chi_{c1}'\to\chi_{c2}\pi^0)$. \label{fig:Gchic2}}
\end{center}
\end{figure}
\begin{figure}[t]
\begin{center}
\vglue-0mm
\includegraphics[width=0.6\textwidth]{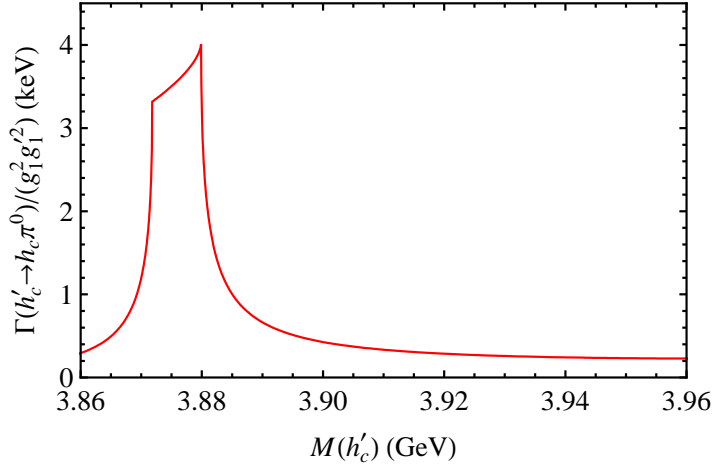}
\vglue-0mm \caption{The width of the decay $h_c'\to h_c\pi^0$ considering only the meson loops. \label{fig:Ghcp}}
\end{center}
\end{figure}

The widths of the $\chi_{c2}'$ decays into the $\chi_{c1}\pi^0$ and
$\chi_{c2}\pi^0$ considering only the meson loops are
\ba%
\label{widths-p-p}\Gamma(\chi_{c2}'\to\chi_{c1}\pi^0)_{\rm loop} \al=\al
(0.08\pm0.03) g_1^2
g_1^{\prime2}~{\rm keV}, \non\\
\Gamma(\chi_{c2}'\to\chi_{c2}\pi^0)_{\rm loop} \al=\al (0.10\pm0.04) g_1^2
g_1^{\prime2}~{\rm keV},
\ea%
where the uncertainties are 33\% and 36\%, which are the velocities of the
intermediate charmed mesons for these two processes, respectively. The ratios of
decay widths defined as
\ba%
\label{ratio-p-p-2} R_{2112}^{\rm loop}  \equiv
\frac{\Gamma(\chi_{c2}'\to\chi_{c1}\pi^0)_{\rm
loop}}{\Gamma(\chi_{c1}'\to\chi_{c2}\pi^0)_{\rm loop}},
\quad R_{2212}^{\rm loop}  \equiv \frac{\Gamma(\chi_{c2}'\to\chi_{c2}\pi^0)_{\rm
loop}}{\Gamma(\chi_{c1}'\to\chi_{c2}\pi^0)_{\rm loop}},
\ea%
for the loop contributions only are shown as the solid and dashed lines in
Fig.~\ref{fig:Gchic2}. The same ratios considering only the tree-level
contributions
\ba%
R_{2112}^{\rm tree} = \frac{3q_{1}^3}{5q_{2}^3}
\frac{M_{\chi_{c1}}M_{\chi_{c1}'}}{M_{\chi_{c2}'}M_{\chi_{c2}}}, \quad
R_{2212}^{\rm tree} = \frac{9q_{1}^3}{5q_{2}^3}
\frac{M_{\chi_{c1}'}}{M_{\chi_{c2}'}},
\ea%
where $q_i$ is the pion momentum for the corresponding decay, are given by the
dotted and dot-dashed lines in Fig.~\ref{fig:Gchic2}, respectively. The
difference is obvious, and may be tested by future experiments. The width of the
decay $h_c'\to h_c\pi^0$ considering only the meson loops is shown in
Fig.~\ref{fig:Ghcp}. The double-cusp structure is again due to the coupling to
the intermediate $D\bar D^*$ mesons.

\section{Comparison with the effective Lagrangian
approach}\label{ELA-formulation}

 As emphasized before, the power counting of NREFT provides a control of the
 theoretical uncertainties. However, for transitions where the mass difference
 between the initial and final charmonium exceeds $\sim 800$ MeV, i.e.
 ${\cal O}(\Lambda_\chi)$, the NREFT is not applicable any more.  The ELA, on the
 other hand, which  deals with the non-local effects by introducing
 empirical form factors to cut off the divergences, can be applied to
 broader kinematic regions. But the disadvantage is that, due to a lack of
 knowledge about the behavior of the counterterms, model-dependence will be
 present in association with the cut-off energies and different forms for the
 form factors. By comparing these two approaches with each other, we expect
 that the model-dependent aspects of the ELA can be identified and reduced. In
 this way, one may have more confidence in the ELA when it is applied in
 kinematical regions where the NREFT is not applicable.

In this section, we first present the general formulae for the ELA.
The calculation results will be summarized and compared with the
NREFT for those transitions studied in the previous section. The
detailed formulation is given in Appendix~\ref{app:ELA}.

The loop transition amplitudes for the transitions in  Fig.~\ref{fig:list} can
be expressed in a general form in the ELA as follows:
 \begin{eqnarray}
 M_{fi}=\int \frac {d^4 q_2} {(2\pi)^4} \sum_{D^\ast \ \mbox{pol.}}
 \frac {V_1V_2V_3} {a_1 a_2 a_3}\prod_i{\cal F}_i(m_i,q_i^2)
 \end{eqnarray}
 where $V_i \ (i=1,2,3)$ are the vertex functions;
 $a_i = q_i^2-m_i^2 \ (i=1,2,3)$ are the denominators of the intermediate meson propagators.
We adopt the form factor, $\prod_i{\cal F}_i(m_i,q_i^2)$, which is a
product of  monopole form factors for each internal meson,
i.e.
\begin{equation}\label{ELA-form-factor}
\prod_i{\cal F}_i(m_i,q_i^2)\equiv {\cal F}_1(m_1,q_1^2){\cal
F}_2(m_2,q_2^2){\cal F}_3(m_3,q_3^2) \ ,
\end{equation}
with
\begin{equation}{\cal F}_i(m_{i}, q_i^2) \equiv \left(\frac
{\Lambda_i^2-m_{i}^2} {\Lambda_i^2-q_i^2}\right),
\label{ffpara}
\end{equation}
 where
$\Lambda_i\equiv m_i+\alpha\Lambda_{\rm QCD}$~\cite{Cheng:2004ru} and the QCD
energy scale is $\Lambda_{\rm QCD} = 220$ MeV. The numerator is introduced in
order not to modify the expression at the on-shell point. This form factor is
supposed to parameterize the non-local effects of the vertex functions and to
remove the loop divergence in the integrals. In this approach the local
couplings for a charmonium to charmed mesons, or a light meson to charmed mesons
are the same as used in the NREFT, while the form factor parameter $\alpha$ will
be determined by comparison to experimental information. Thus, it is assumed
here that all --- or at least the dominant part --- of the short-range physics
related to meson loops can be parameterized in the form of Eq.~(\ref{ffpara})
with a reaction independent parameter $\alpha$.

Although used widely and very convenient for the actual evaluation of the
pertinent integrals, the form factor parameterization given in
Eq.~(\ref{ELA-form-factor}) also has its drawbacks. Especially, unphysical
thresholds, which are located at $\Lambda_i+m_j=m_i+m_j+\alpha \Lambda_{\rm
QCD}$ and $\Lambda_i+\Lambda_j=m_i+m_j+2\alpha \Lambda_{\rm QCD}$, are
introduced into the integrals. Thus, for sufficiently heavy decaying charmonia
and small values of $\alpha$, these singularities are located nearby or even
inside the physical regime and additionally introduce unphysical contributions.
Clearly, those are not part of the NREFT. We therefore expect, and indeed
observe, that there exist significant quantitative differences between these two
approaches in the decays of heavy charmonia. It suggests that a different form
factor parameterization or larger values of $\alpha$ should be considered. On the
other hand, for those cases, where the form factor singularities do not
contribute, we expect interesting insights from the comparison of the ELA and
NREFT results.

\begin{table*}[t]
\begin{center}
\renewcommand{\arraystretch}{1.3}
\begin{tabular}{l|l|c|c}\hline\hline
& Decays & Qualitative & Quantitative  \\ \hline
$SS$ & $\psi^\prime\to J/\psi\pi^0$ &   $\surd$  & $\surd$           \\
& $\psi^\prime\to J/\psi\eta$ &     $\surd$  & $\surd$        \\
\hline 
 & $\psi^\prime\to h_c\pi^0$ &     $\surd$ & $\times$      \\
 & $h_c\to J/\psi\pi^0$ &          $\surd$  & $\surd$    \\
$SP$ & $\eta_c^\prime\to \chi_{c0}\pi^0$ &     $\surd$ & $\times$     \\
 & $\chi_{c0}\to \eta_c\pi^0$ &       $\surd$ & $\times$       \\
 & $h_c^\prime\to \psi^\prime\pi^0$ &   $\surd$  & $\times$  \\
 & $\chi_{c0}^\prime\to \eta_c^\prime\pi^0$ &     $\surd$ & $\times$  \\  \hline
 &
 $\chi_{c1}^\prime\to\chi_{c0}\pi^0/\chi_{c0}^\prime\to\chi_{c1}\pi^0$
 &    $\surd$  & $\times$  \\
 &
 $\chi_{c1}^\prime\to\chi_{c0}\pi^0/\chi_{c1}^\prime\to\chi_{c2}\pi^0$
 &    $\surd$  & $\times$ \\
  $PP$ & $\chi_{c1}^\prime\to\chi_{c1}\pi^0/\chi_{c1}^\prime\to\chi_{c2}\pi^0$
 &    $\surd$  & $\surd$  \\
 &
 $\chi_{c2}^\prime\to\chi_{c1}\pi^0/\chi_{c1}^\prime\to\chi_{c2}\pi^0$
 &    $\surd$  & $\surd$ \\
 &
 $\chi_{c2}^\prime\to\chi_{c2}\pi^0/\chi_{c1}^\prime\to\chi_{c2}\pi^0$
 &   $\surd$  & $\surd$   \\  \hline\hline
\end{tabular}
\caption{\label{tab:comparison} Summary of the {\em qualitative} and {\em
quantitative} features in the comparison between the ELA and NREFT results for
various decay transitions. For $PP$ transitions, we take the width ratios in
order to compare the ELA with the NREFT results. The appearance of the
crosses can be understood, as explained in the text.}
\end{center}
\end{table*}

\subsection{Transitions between the $S$-wave charmonia}

\begin{figure}[t]
\begin{center}
\vglue-0mm
\includegraphics[width=0.7\textwidth]{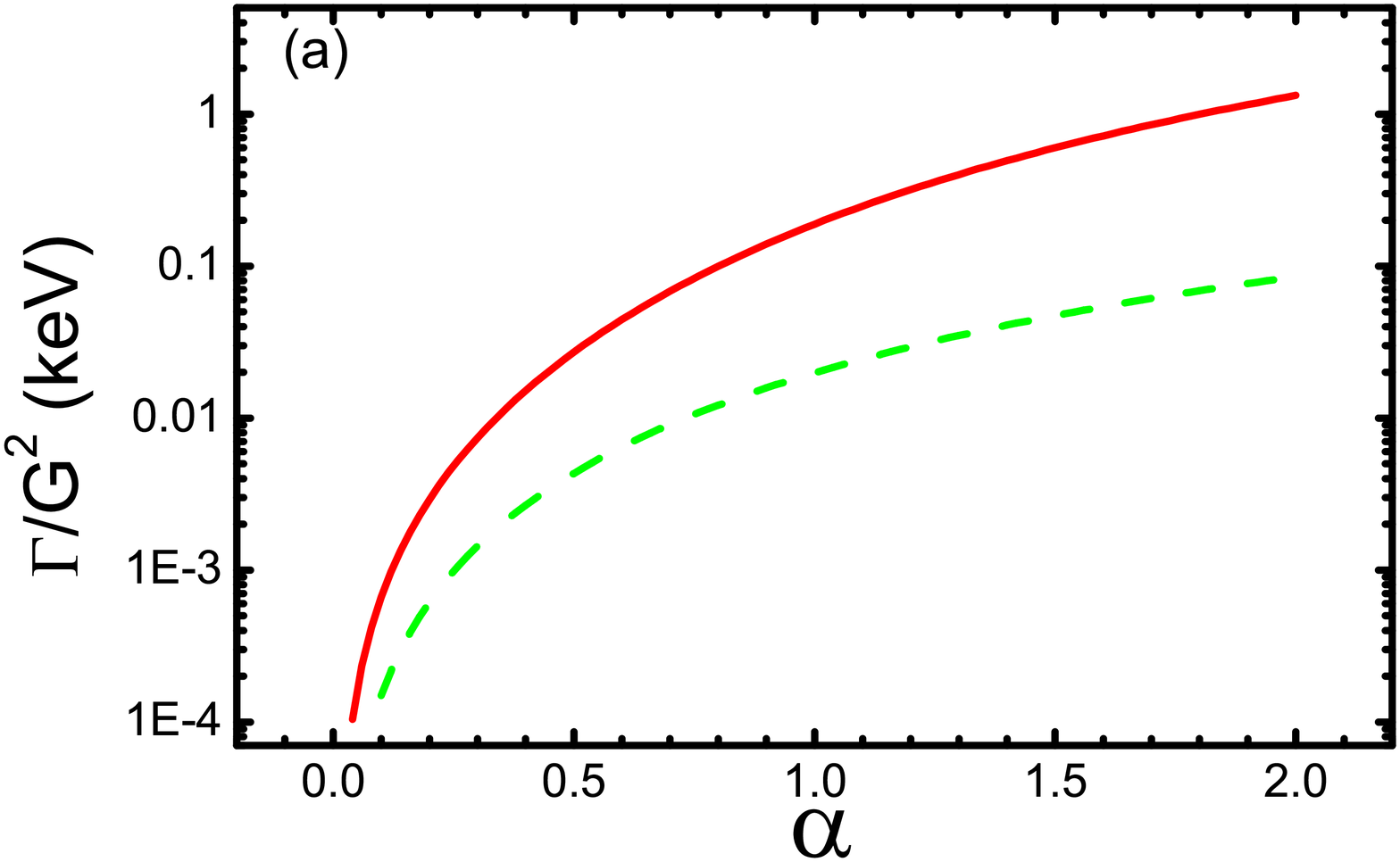}\\ \vglue-5mm
\includegraphics[width=0.7\textwidth]{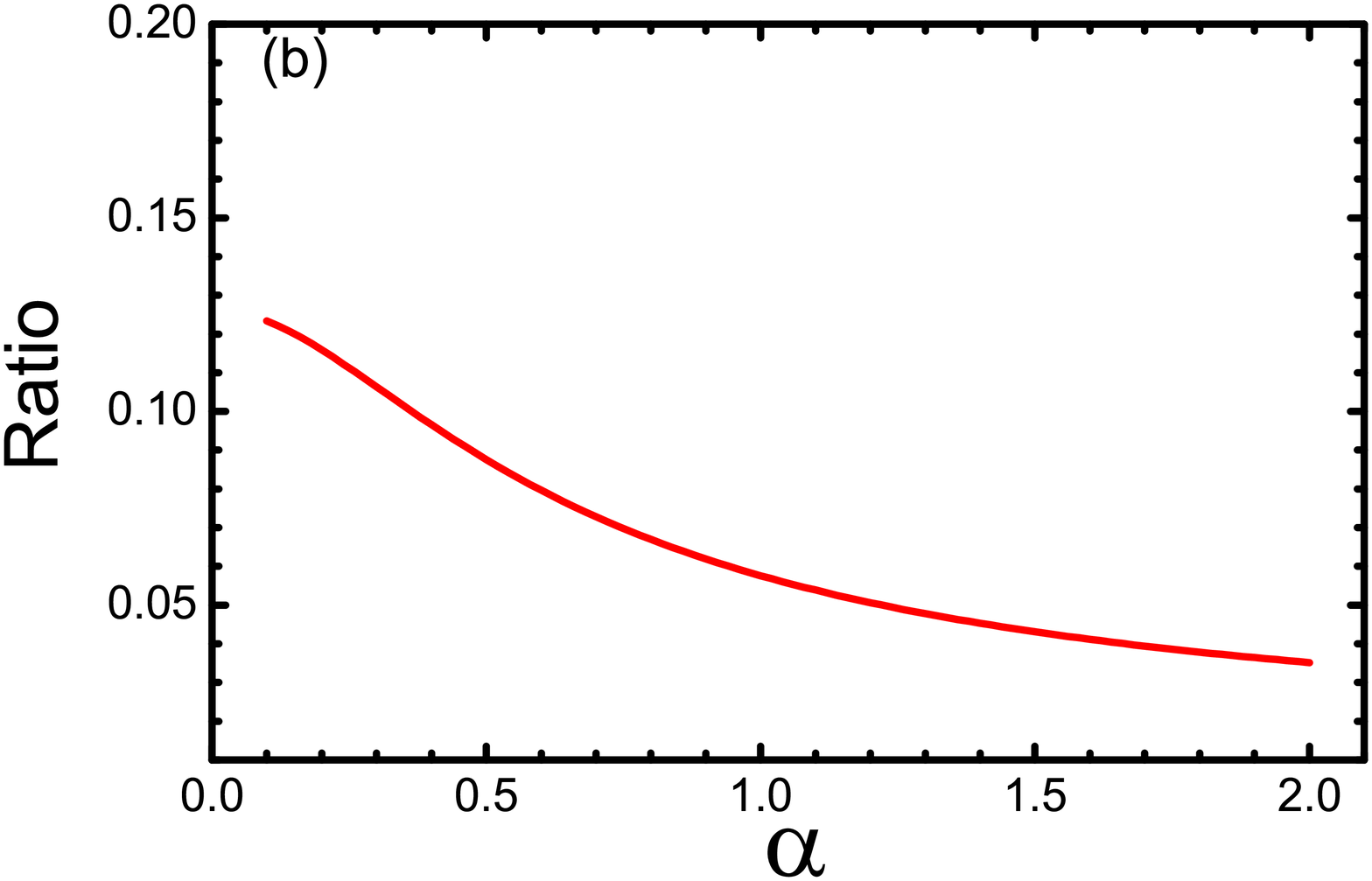}
\vglue-0mm \caption{(a) $\alpha$-dependence for the decay widths of
the decay $\psi'\to J/\psi \pi^0$ (dashed line) and $J/\psi \eta$
(solid line). (b) $\alpha$-dependence of the ratio $R_{\pi^0/\eta}$.
The coupling is
defined as $G^2\equiv g_2^2{g'}_2^2$. \label{fig:psipi_eta}}
\end{center}
\end{figure}

For transitions between the $S$-wave charmonia, i.e. $\psi'\to J/\psi
\pi^0(\eta)$, we plot the $\alpha$-dependence of the partial decay widths of
$\psi'\to J/\psi \pi^0(\eta)$ in Fig.~\ref{fig:psipi_eta} (a) as shown by the
solid and dashed lines, respectively. The $\pi^0-\eta$ mixing has been taken
into account. Using $\alpha$ in a range of $\alpha\simeq 1...2$, the measured
ratio $R_{\pi^0/\eta}$ by both the CLEO and BES Collaborations can be
reproduced, and the central value of the PDG fit as listed in
Table~\ref{tab:udratio} may be obtained with $\alpha=1.64$,
\begin{eqnarray}
\Gamma(\psi^\prime\to J/\psi\pi^0)& =& 0.031 g_2^2 g_2^{'2} \
\mbox{keV}, \\
\Gamma(\psi^\prime\to J/\psi\eta) &=& 0.77 g_2^2 g_2^{'2} \
\mbox{keV},
\end{eqnarray}
where the vertex couplings are taken the same as those in the NREFT. These
results are consistent with the NREFT.

In Fig.~\ref{fig:psipi_eta} (b), the ratio $R_{\pi^0/\eta}$ calculated in the
ELA is given in terms of $\alpha$.  The predominant feature is that the
$\alpha$-dependence of the ratios is quite stable. It is because the loop
amplitudes for the transitions play a dominant role in the transition and have
the same divergence structure which is regularized by the form factors. With
$\alpha=1.64$, the resulting ratio $R_{\pi^0/\eta}^{\rm loop}$ is consistent
with the result from the NREFT~\cite{Guo:2009wr}.

\begin{figure}[t]
\begin{center}
\vglue-0mm
\includegraphics[width=0.7\textwidth]{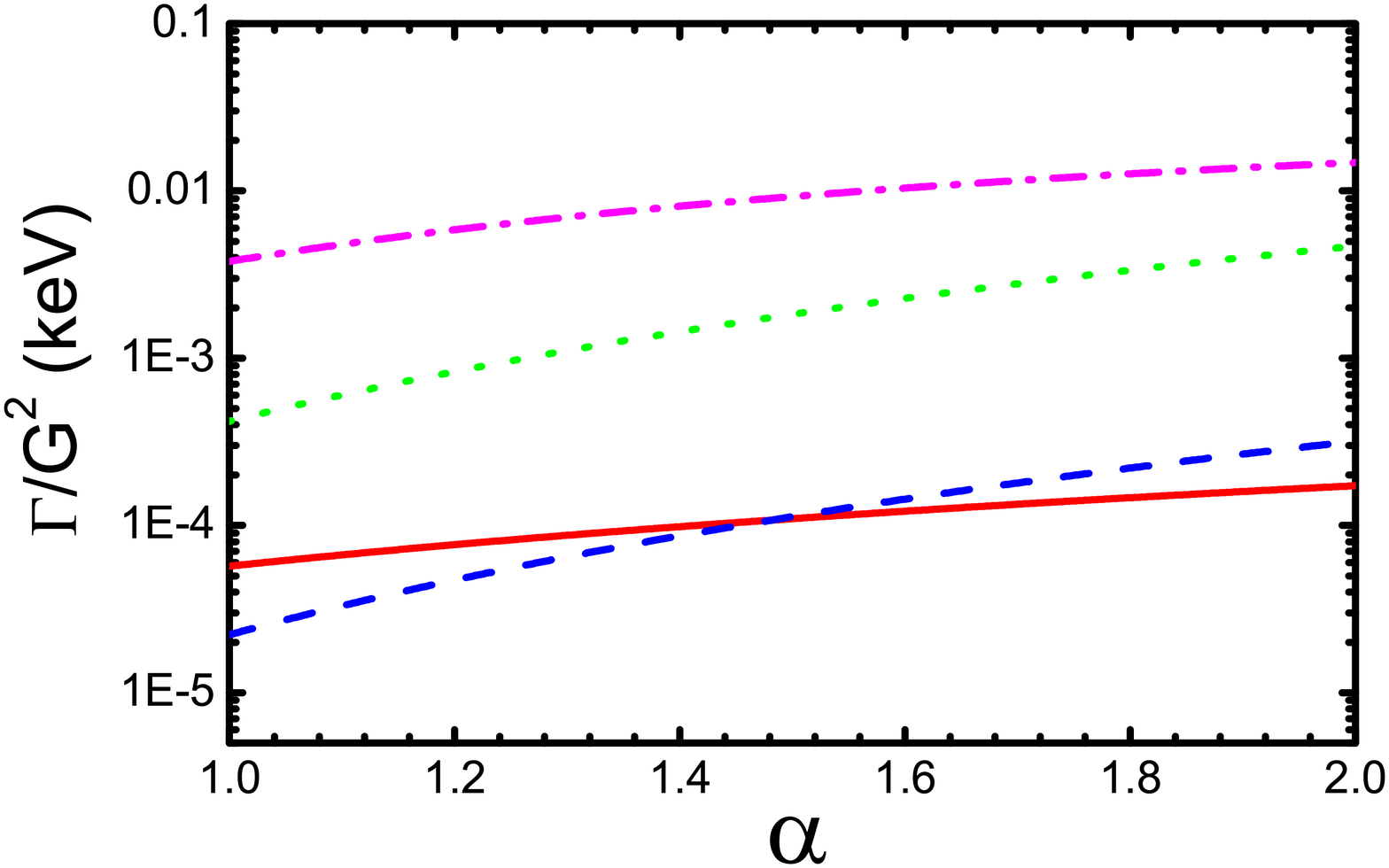}
\vglue-0mm \caption{$\alpha$-dependence of the decay widths of
$\psi'\to h_c \pi^0$ (solid line), $h_c\to J/\psi \pi^0$ (dashed
line), $\chi_{c0}\to \eta_c\pi^0$ (dotted line), and $\eta_c'\to
\chi_{c0}\pi^0$ (dot-dashed line). The coupling is $G^2\equiv
g_1^2{g'}_2^{2}$ for $\psi'\to h_c \pi^0$ and
$\eta_c'\to \chi_{c0}\pi^0$, while $G^2\equiv g_1^2{g}_2^{2}$ for
$h_c\to J/\psi \pi^0$ and $\chi_{c0}\to \eta_c\pi^0$.
\label{fig:chic1pchic0pi0}}
\end{center}
\end{figure}

\subsection{Transitions between the $S$- and $P$-wave charmonia}

Recall that the power counting suggests suppressions on most of the transitions
between the $S$ and $P$-wave charmonia via intermediate meson loops, such as,
$\psi'\to h_c\pi^0$, while the loops in the transition $h_c'\to \psi'\pi^0$ are
enhanced. The dominance of tree-level contribution means that the physics is
described by contact interactions -- sometimes loosely called short-distance
physics. On the contrary, in the ELA, since form factors are adopted, part of the
short-range physics is already included in the loops. Therefore, one may expect
 large differences between the NREFT and the ELA in this sector,
and the ELA results would typically be larger than those in the NREFT,
especially for the tree-level dominant transitions.

With the form factor parameter determined in the previous subsection and the
same couplings for the vertices as in the NREFT, we now examine the predictions
from the ELA for those transitions. Instead of going to all the channels, we
shall concentrate on two sets of decays, i) $\psi'\to h_c\pi^0$ and $h_c\to
J/\psi \pi^0$, and ii) $\eta_c^\prime\to \chi_{c0}\pi^0$ and $\chi_{c0}\to
\eta_c\pi^0$, and summarize the others in Table~\ref{tab:comparison}.

The two transitions in each set involve similar coupling vertices, but different
kinematics. Their ratios will again eliminate the uncertainties arising from the
unknown coupling constants. In Fig.~\ref{fig:chic1pchic0pi0}, the decay widths
with the charmonium--$D$-meson couplings normalized to unity are plotted for
these two pairs of decay channels, i.e. $\psi'\to h_c\pi^0$ (solid line),
$h_c\to J/\psi \pi^0$ (dashed line), $\eta_c^\prime\to \chi_{c0}\pi^0$
(dot-dashed line) and $\chi_{c0}\to \eta_c\pi^0$ (dotted line) in a range of
$\alpha=1...2$. Note that the BES-III measurement gives a central value
$\Gamma^{\rm exp}(\psi'\to h_c\pi^0)=0.61$~keV~\cite{BESIIIhc}. Within
$\alpha=1...2$ which is found in the transitions $\psi^\prime\to J/\psi
\pi^0(\eta)$, the partial decay width of $\psi'\to h_c\pi^0$ from the meson
loops is much smaller than the experimental data~\cite{BESIIIhc}, which confirms
the suppression of the meson loops found in the NREFT, although these two
approaches give quite different values for the meson loop magnitudes. This
comparison is listed in Table~\ref{tab:comparison} as qualitative agreement.

Such a suppression also occurs to $h_c\to J/\psi \pi^0$ in the ELA, and exhibits
some peculiar features in comparison with the NREFT expectation. Since the decay
$h_c\to J/\psi \pi^0$ has a relatively large phase space, the power counting
suppression in the NREFT is not as much as for  $\psi'\to h_c\pi^0$ as pointed
out in the previous Section. In the ELA, however, as shown in
Fig.~\ref{fig:chic1pchic0pi0}, these two decay widths are of the same order in
the range of $\alpha=1...2$. Nevertheless, the suppression of the charmed meson
loops as analyzed in the NREFT in both decays is confirmed in the ELA. The
partial decay width of the transition $h_c\to J/\psi \pi^0$ considering only
loops is even quantitatively consistent with the NREFT expectation.

As shown by the dotted and dot-dashed lines in Fig.~\ref{fig:chic1pchic0pi0}, we
also observe a suppression of the meson loops in $\eta_c'\to \chi_{c0}\pi^0$ and
$\chi_{c0}\to \eta_c\pi^0$ which, however, is weaker as in the NREFT
calculation. As already mentioned, this may be understood as part of the
short-range physics is already mimicked by the form factors in the ELA. Such an
uncertainty seems inevitable in the ELA and a measurement of the ratios of
branching fractions would be less sensitive to it. The normalized widths of
$\eta_c'\to \chi_{c0}\pi^0$ and $\chi_{c0}\to \eta_c\pi^0$ have a different
ordering compared with the NREFT results, and the $\eta_c'\to \chi_{c0}\pi^0$
appears to be one order of magnitude larger. This is because the mass of the
$\eta_c'$ is much closer to the $D\bar{D^*}$ threshold than $\eta_c$, and the
form factor parameterization adopted in the ELA makes the effect of the
proximity more prominent than the NREFT.

The other decay channels of interest are between the radial excitation $S$- and
$P$-wave charmonia, such as $h_c^\prime\to \psi^\prime \pi^0$ and
$\chi_{c0}^\prime\to \eta_c^\prime\pi^0$. The NREFT predicts that the width of
$\chi_{c0}^\prime\to \eta_c^\prime\pi^0$ is tree-level dominant, but the charmed
meson loops may give a significant contribution, while the transition
$h_c^\prime\to \psi^\prime \pi^0$ is dominated by the meson loops. We find that
the ELA result for $h_c^\prime\to \psi^\prime \pi^0$ has a much larger
normalized decay width considering loops only, which indicates the dominance of
the loops for this transitions. It agrees qualitatively with the NREFT one quite
well though has a larger value. The ELA result of the transition
$\chi_{c0}^\prime\to \eta_c^\prime\pi^0$ is smaller than that of $h_c^\prime\to
\psi^\prime \pi^0$ while much larger than that of the other $SP$ transitions. We
consider this as a qualitative agreement with the NREFT, since in the NREFT the
pattern is similar.

This is consistent with our prospect that these two approaches would agree with
each other given the dominance of the meson loops in the transitions. It should
also be pointed out that in the ELA, singularities would appear when the mass of
the initial charmonium is larger than the artificial threshold
$\Lambda_1+\Lambda_2$ introduced by the form factor. When the ``threshold'' is
approaching, one typically gets larger results. The comparison of these two
approaches turns out to be a useful tool for our understanding the properties of
the ELA.

\subsection{Transitions between two $P$-wave charmonia}

The transitions between $P$-wave charmonia, especially for the $\chi_{cJ}$
decays, allow for an especially sensitive test of the meson loop transitions, as
the power counting analysis suggests that the meson loop contributions relative
to the tree-level transitions will be enhanced significantly (c.f. table
\ref{tab:pcloops}). Thus, the comparison between NREFT and ELA will help
restrict the possible parameter ranges in the ELA as well as study further the
implications of the adopted form factor parameterization.

\begin{figure}[thb]
\begin{center}
\vglue-0mm
\includegraphics[width=0.7\textwidth]{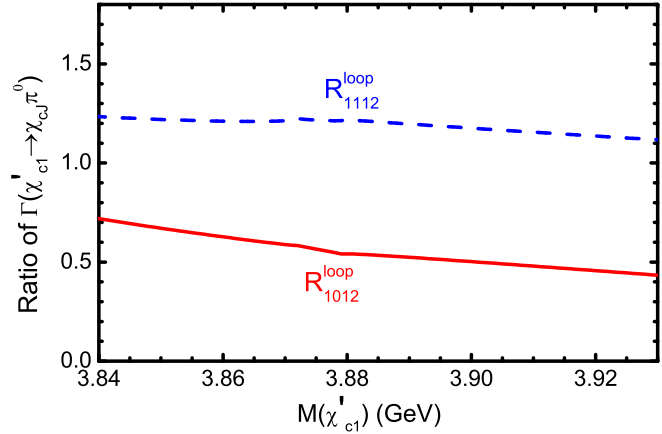}
\vglue-0mm \caption{Parameter-free ratios among the decay widths of
$\chi_{c1}'\to \chi_{cJ}\pi^0$ considering only the meson loops in
the ELA. \label{fig:Rchic1pchicJ_M}}
\end{center}
\end{figure}

\begin{figure}[thb]
\begin{center}
\vglue-0mm
\includegraphics[width=0.7\textwidth]{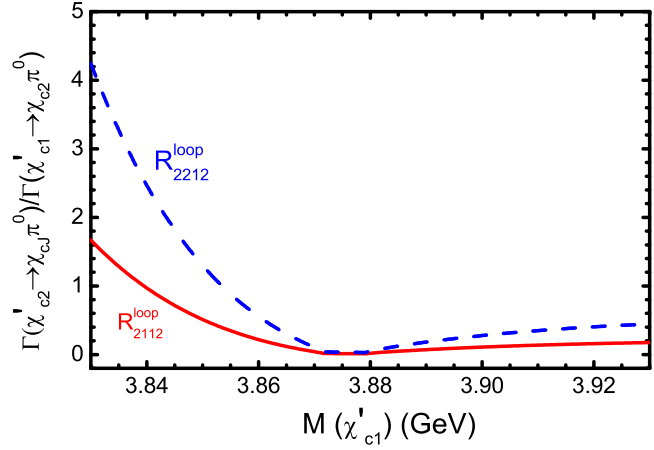}
\vglue-0mm \caption{Ratios of the decay widths
$\chi_{c2}'\to\chi_{c1}\pi^0$ and $\chi_{c2}'\to\chi_{c2}\pi^0$ to
$\Gamma(\chi_{c1}'\to\chi_{c2}\pi^0)$ in the ELA.
\label{fig_Rchic2J_M}}
\end{center}
\end{figure}

In the case of $\chi_{c0}'\to \chi_{c1}\pi^0$ and $\chi_{c1}'\to
\chi_{c0}\pi^0$, our calculation shows that the normalized partial widths,
$\Gamma(\chi_{c0}'\to \chi_{c1}\pi^0)/(g_1^2 g_2^2)$, and the mass-dependence of
$\Gamma(\chi_{c1}'\to \chi_{c0}\pi^0)/(g_1^2 g_1^{\prime 2})$ have similar
structures as those given by the NREFT, but the relative magnitude of the decay
widths is different. The ELA predicts larger results for the widths of the
transitions $\chi_{c0}'\to \chi_{c1}\pi^0$ and $\chi_{c1}'\to \chi_{cJ}\pi^0$
than the corresponding values in the NREFT, except for the $\chi_{c1}'\to
\chi_{c0}\pi^0$ which is opposite.

A comparison is also made for these two methods in the predictions of the
ratios, $R_{1012}^{\rm loop}$ and $R_{1112}^{\rm loop}$, as defined by
Eq.~(\ref{ratio-chicj}). In Fig.~\ref{fig:Rchic1pchicJ_M}, these two ratios are
plotted in terms of $\alpha$ values. It is sensible to observe the stability of
the ratios within the varying $\chi_{c1}^\prime$ mass region. In comparison with
the NREFT results in Fig.~\ref{fig:Rchic1J}, ratio $R_{1112}^{\rm loop}$ is
consistent with the dashed line there even in magnitude, while ratio
$R_{1012}^{\rm loop}$ exhibits an inverse relative magnitude between
$\Gamma(\chi_{c1}^\prime\to \chi_{c0}\pi^0)$ and $\Gamma(\chi_{c1}^\prime\to
\chi_{c2}\pi^0)$. Although there exist significant discrepancies between these
two methods here, we emphasize that the flat behavior of the ratios again
suggests some systematic model-independent features of the meson loop
contributions. Especially, there are two kinks in $R_{1012}^{\rm loop}$ in both
methods, which indicate the opening of the $D^0{\bar D}^{*0}$ and $D^+D^{*-}$
thresholds. In this sense, it is also listed as in qualitative agreement in
Table~\ref{tab:comparison}.

We further compare the ELA results for the $P$-wave decays,
$\chi_{c2}'\to\chi_{c1}\pi^0$ and $\chi_{c2}'\to\chi_{c2}\pi^0$, with the NREFT
ones. For these two transitions, the contributing loops have $M_1=M_2=D^*$. The
mass of the initial state $M_{\chi_{c2}}$ is smaller than the threshold
$2M_{D^*}$, and much smaller than the artificial threshold $2(M_{D^*}+\alpha
\Lambda_{\rm QCD})$, and hence the unphysical singularities would have little
effect. So one may expect the ELA would give similar results compared with the
NREFT. The meson loops in the ELA give,
\begin{eqnarray}
\Gamma(\chi_{c2}' \to \chi_{c1} \pi^0)_{\rm loop}=0.06 g_1^2 g_1^{\prime^2} \ \mbox{keV} \ , \nonumber \\
\Gamma(\chi_{c2}' \to \chi_{c2} \pi^0)_{\rm loop}=0.19 g_1^2
g_1^{\prime^2} \ \mbox{keV} \ ,
\end{eqnarray}
with $\alpha=1.64$ as determined previously. They are consistent with those
given by the NREFT in Eq.~(\ref{widths-p-p}). In Fig.~\ref{fig_Rchic2J_M}, we
present the ratios of the decay widths, $R^{\rm loop}_{2112}$ and $R^{\rm
loop}_{2212}$, as defined by Eq.~(\ref{ratio-p-p-2}), to compare with the solid
and dashed lines in Fig.~\ref{fig:Gchic2}. It is interesting to see the
consistency of these two methods here in both mass evolution and magnitude, as
expected.

\subsection{Summary of the ELA results}

As can be seen from Table~\ref{tab:comparison}, the NREFT and ELA results for
all the transitions are in qualitative agreement. For both the $SS$ and
$PP$ transitions 
the results for the two approaches are in quantitative agreement as long as the
decaying charmonium state is not sufficiently heavy to allow the form factor
singularities to matter numerically. The observed level of agreement between
these two approaches provides a further support for the power counting of the
NREFT. It demonstrates nicely two important aspects of this effective field
theory, namely, the applicability of the non-relativistic treatment in these
charmonia transitions (note that the ELA formulations are relativistic), and the
dominance of the long-range pieces of the meson loops which is driven by the
unitary cuts for the $SS$ and $PP$ transitions. At the same time, it provides
additional confidence for applying the ELA to reactions where the NREFT is no
longer applicable, such as in case of charmonium decays into light final states.

The situation is different for the $SP$ transitions. Most of these transitions
are expected to be dominated by ``short-range'' physics, i.e. the tree-level
contributions, in the NREFT. From the effective field theory point of view, the
tree-level amplitudes serve two important duties: On the one hand, they provide
a way to include the short-ranged quark dynamics. On the other hand, they may
absorb the ultra-violet behavior of the meson loops and, if necessary, absorb
their divergences. In the ELA, part of the short-range physics, which is
completely omitted so far in the NREFT, is parameterized by the form factors.
Thus, for the $SP$ transitions we expect, and indeed find, that there are
quantitative differences between these two approaches. However, it is important
to note that the qualitative role of the loops is the same in both treatments.

\section{Summary}

\label{sec:summary}

In this paper, the effects of  charmed meson loops in the
transitions between two charmonia with the emission of one pion or
eta have been systematically investigated. The power counting for
the loop amplitudes within the framework of a non-relativistic
effective field theory is given.  The difference among the power
counting estimates for the ratios of tree--level and loop
contributions for the various transitions considered comes from the
quantum number difference of the involved charmonia. It is found
that the loops are enhanced in the transitions between two $S$-wave
charmonia by a factor of $1/v$ and in the transitions between two
$P$-wave charmonia by a factor of $1/v^3$. As a result, even if the
$X(3872)$ is a $c\bar c$ charmonium with $J^{PC}=1^{++}$, the
dominant contribution to its decays into $\chi_{cJ}\pi^0$ is given
by the intermediate charmed meson loops rather than the tree-level
ones. For the transitions between one $S$-wave and one $P$-wave
charmonia, because of the competition between $1/v^3$ and $q^2/M_D^2$
one should analyze case by case. The loops are highly suppressed for
the decays $\psi'\to h_c\pi^0$ and $\eta_c'\to\chi_{c0}\pi^0$ which
have a small phase space, while the suppression for the decay
$\chi_{c0}'\to\eta_c'\pi^0$ is more moderate, and they are even
enhanced in the decay $h_c'\to \psi'\pi^0$.

Among the loop dominated transitions, predictions for the ratios among the decay
widths are given. In these ratios, the dependence of the unknown coupling
constants is canceled. A detailed calculation in the framework of an effective
Lagrangian approach is also given in comparison with the NREFT results. We find
that the results from these two methods are qualitatively consistent with each
other. Significant deviations appeared only when the results became sensitive to
the particular form of the form factors used in the ELA, pointing at possible
improvements for this approach, or when the considered transition is dominated
by short-range physics.
 Although the lack of constraints on the structure of the
counterterms would lead to uncertainties with the absolute magnitudes of the
partial decay widths, the ratios between related channels are less sensitive to
such uncertainties, and model-independent aspects of the meson loop
contributions can be highlighted. Those loop-dominant channels should be
testable experimentally in the future, for instance, with Belle, $\rm\overline
P$ANDA, BES-III, LHC-b and Super $B$ factory.

\medskip

\section*{Acknowledgments}
We would like to thank Yu Jia, Thomas Mannel and Joan Soto for useful
discussions. F.-K.G., C.H. and U.-G.M. would like to thank from the Helm\-holtz
Association through funds provided to the virtual institute ``Spin and strong
QCD'' (VH-VI-231) and by the DFG (SFB/TR 16, ``Subnuclear Structure of
Matter''), and the European Community-Research Infrastructure Integrating
Activity ``Study of Strongly Interacting Matter'' (acronym HadronPhysics2, Grant
Agreement n. 227431) under the Seventh Framework Programme of EU. U.-G.M. also
thanks the BMBF for support (grant 06BN9006). Q.Z. and G.L. acknowledge the
supports, in part, from the National Natural Science Foundation of China (Grants
No. 10675131,  10491306 and 10947007), Chinese Academy of Sciences
(KJCX3-SYW-N2), and Ministry of Science and Technology of China (2009CB825200).

\vspace{10mm}


\begin{appendix}


\section{Loop functions in dimensional regularization}
\label{app:loop}
\renewcommand{\theequation}{\thesection.\arabic{equation}}
\setcounter{equation}{0}

Define the basic three-point scalar loop function in $d$-dimension
as
\be%
I(q) \equiv i\int\!\frac{d^dl}{(2\pi)^d}
\frac{1}{\left(l^2-m_1^2+i\epsilon\right)
\left[(P-l)^2-m_2^2+i\epsilon\right]
\left[(l-q)^2-m_3^2+i\epsilon\right] }.
\ee%
Non-relativistically, in the rest frame of the initial particle,
i.e. $P^\mu=\{M,\vec{ 0}\}$, it can be worked out analytically by
performing the contour integration over $l^0$, and then integrating
over the spatial components of the loop momentum using
dimensional regularization,
\ba%
I(q) \al=\al \frac{-i}{8m_1m_2m_3} \int\!\frac{d^dl}{(2\pi)^d}
\frac{1}{ \left(l^0-\frac{\vec{ l}\ ^2}{m_1}+i\epsilon\right)
\left(l^0+b_{12}+\frac{\vec{ l}\ ^2}{m_2}-i\epsilon\right)
\left[l^0+b_{12}-b_{23}-\frac{(\vec{ l}-\vec{
q})^2}{m_3}+i\epsilon\right] } \non\\
\al=\al \frac{\mu_{12}\mu_{23}}{2m_1m_2m_3}
\int\!\frac{d^{d-1}l}{(2\pi)^{d-1}} \frac{1}{\left(\vec{ l}\
^2+c-i\epsilon\right) \left(\vec{ l}\ ^2-\frac{2\mu_{23}}{m_3}\vec{
l}\cdot\vec{
q}+c'-i\epsilon\right)} \non\\
\al=\al \frac{\mu_{12}\mu_{23}}{2m_1m_2m_3} \int_0^1dx
\int\!\frac{d^{d-1}l}{(2\pi)^{d-1}} \frac{1}{\left[\vec{ l}\
^2-ax^2+(c'-c)x+c-i\epsilon\right]^2} \non\\
\al=\al \frac{\mu_{12}\mu_{23}}{16\pi m_1m_2m_3} \frac{1}{\sqrt{a}}
\left[ \tan^{-1}\left(\frac{c'-c}{2\sqrt{ac}}\right) +
\tan^{-1}\left(\frac{2a+c-c'}{2\sqrt{a(c'-a)}}\right) \right],
\ea%
where $\mu_{ij}=m_im_j/(m_i+m_j)$ are the reduced masses, $b_{12} =
m_1+m_2-M$, $b_{23}=m_2+m_3+q^0-M$, and
\ba%
a = \left(\frac{\mu_{23}}{m_3}\right)^2 \vec{ q}\ ^2, \quad c =
2\mu_{12}b_{12}, \quad c'=2\mu_{23}b_{23}+\frac{\mu_{23}}{m_3}\vec{
q}\ ^2.
\ea%
Since there is no pole for $d=4$, in the last step, we have taken
$d=4$.

We also need the vector and tensor loop integrals which are defined
as
\be%
q^i I^{(1)}(q) \equiv i \int\!\frac{d^dl}{(2\pi)^d} \frac{l^i}{
\left(l^2-m_1^2+i\epsilon\right)
\left[(P-l)^2-m_2^2+i\epsilon\right]
\left[(l-q)^2-m_3^2+i\epsilon\right] },
\ee%
and
\be%
\label{Aeq:tensorloop} q^iq^j I^{(2)}_0(q) + \delta^{ij}\vec{ q}\ ^2
I^{(2)}_1(q) \equiv i \int\!\frac{d^dl}{(2\pi)^d} \frac{l^il^j}{
\left(l^2-m_1^2+i\epsilon\right)
\left[(P-l)^2-m_2^2+i\epsilon\right]
\left[(l-q)^2-m_3^2+i\epsilon\right] }.
\ee%
By using the technique of tensor reduction, we get
\ba%
\label{eq:I1}
 I^{(1)}(q) \al=\al \frac{\mu_{23}}{a m_3}
\left[B(c'-a)-B(c) +  \frac{1}{2} (c'-c) I(q)\right],
\\ \label{eq:I20}%
\vec{ q}\ ^2 I^{(2)}_0(q) \al=\al \frac{d-3}{d-2} B(c'-a) +
\frac{c}{d-2} I(q) + \frac{d-1}{d-2} (c'-c) \frac{m_3}{2\mu_{23}}
I^{(1)}(q), \\ \label{eq:I21}%
\vec{ q}\ ^2 I^{(2)}_1(q) \al=\al \frac{1}{d-2} B(c'-a) -
\frac{c}{d-2} I(q) - \frac{1}{d-2} (c'-c) \frac{m_3}{2\mu_{23}}
I^{(1)}(q),
\ea%
where the function $B(c)$ is defined as
\ba%
\frac{4m_1m_2m_3}{\mu_{12}\mu_{23}}B(c) \al\equiv\al
\int\!\frac{d^{d-1}l}{(2\pi)^{d-1}} \frac{1}{\vec{
l}\ ^2+c-i\epsilon} \non\\
\al=\al (4\pi)^{(1-d)/2}\Gamma(\frac{3-d}{2})(c-i\epsilon)^{(d-3)/2} \non\\
\al=\al -\frac{\sqrt{c-i\epsilon}}{4\pi}.
\ea%
In the last step, the dimension $d$ is taken to be 4.

\section{Amplitudes for the charmonium transitions}
\label{app:amp}
\renewcommand{\theequation}{\thesection.\arabic{equation}}
\setcounter{equation}{0}

Here we give the amplitudes for the transitions between charmonia
with the emission of one pion or $\eta$ discussed in the paper. Notice
that the expressions for the amplitudes are only given for the
charged charmed meson loops. The expressions for the neutral or
strange charmed meson loops can be obtained easily by replacing the
charged charmed meson masses by the the corresponding neutral or charmed
ones. To distinguish loops with different $m_1,m_2$ and $m_3$, we write the
loop functions as, for instance, $I^{(1)}(q,M1,M2,M3)$ with $Mi$
denoting the meson with the mass $m_i$ in the following.

\subsection{Transitions between the $S$-wave charmonia}

\begin{itemize}

\item $\psi'\to J/\psi \pi^0 (\eta)$
\ba%
\al\al -i{\cal M}\left(\psi'\to J/\psi\phi\right)_{\pm} \non\\
\al=\al i2\frac{g}{F}g_2g_2'
\epsilon^{ijk}q^i\varepsilon^j(\psi')\varepsilon^k(J/\psi) \vec{q}\
^2
\left[ 2I_1^{(2)}(q,D^\pm,D^\pm,D^{*\pm}) +
2I_1^{(2)}(q,D^{*\pm},D^\pm,D^\pm) \right.\non\\
\al\al + 4I_1^{(2)}(q,D^\pm,D^{*\pm},D^{*\pm}) +
2I_0^{(2)}(q,D^\pm,D^{*\pm},D^{*\pm}) -
I^{(1)}(q,D^\pm,D^{*\pm},D^{*\pm}) \non\\
\al\al + 4I_1^{(2)}(q,D^{*\pm},D^{*\pm},D^\pm) +
2I_0^{(2)}(q,D^{*\pm},D^{*\pm},D^\pm) -
I^{(1)}(q,D^{*\pm},D^{*\pm},D^\pm) \non\\
\al\al - 2I_1^{(2)}(q,D^{*\pm},D^\pm,D^{*\pm}) -
2I_0^{(2)}(q,D^{*\pm},D^\pm,D^{*\pm}) +
I^{(1)}(q,D^{*\pm},D^\pm,D^{*\pm}) \non\\
\al\al \left. - 10I_1^{(2)}(q,D^{*\pm},D^{*\pm},D^{*\pm}) -
2I_0^{(2)}(q,D^{*\pm},D^{*\pm},D^{*\pm}) +
I^{(1)}(q,D^{*\pm},D^{*\pm},D^{*\pm})\right]. \non\\
\label{eq:amppsipsi}
\ea%

\end{itemize}

\subsection{Transitions between the $S$- and $P$-wave charmonia}

\begin{itemize}
\item $\psi'\to h_c\pi^0$
\ba%
\al\al -i{\cal M}\left(\psi'\to h_c\pi^0\right)_{\pm} \non\\
\al=\al -i2\frac{g}{F}g_1g_2' \left\{ \vec{
q}\cdot\vec{\varepsilon}(\psi')\vec{ q}\cdot\vec{\varepsilon}(h_c)
\left[
I^{(1)}(q,D^\pm,D^\pm,D^{*\pm}) - I^{(1)}(q,D^{*\pm},D^\pm,D^{*\pm}) \right.\right. \non\\
\al\al \left.+ I^{(1)}(q,D^\pm,D^{*\pm},D^{*\pm}) -
I^{(1)}(q,D^{*\pm},D^{*\pm},D^{*\pm})
\right] \non\\
\al\al + \vec{ q}\
^2\vec{\varepsilon}(\psi')\cdot\vec{\varepsilon}(h_c) \left[
I^{(1)}(q,D^{*\pm},D^\pm,D^{*\pm}) + I^{(1)}(q,D^{*\pm},D^{*\pm},D^\pm) \right.\non\\
\al\al \left.\left. - I^{(1)}(q,D^\pm,D^{*\pm},D^{*\pm}) -
I^{(1)}(q,D^{*\pm},D^{*\pm},D^{*\pm}) \right] \right\}.
\ea%

\item $\eta_c'\to \chi_{c0}\pi^0$
\ba%
-i{\cal M}\left(\eta_c'\to \chi_{c0}\pi^0\right)_{\pm} \al=\al
-i\frac{2}{\sqrt{3}}\frac{g}{F}g_1g_2'\vec{q}\ ^2 \left[
3I^{(1)}(q,D^{*\pm},D^\pm,D^\pm) \right. \non\\
\al\al \left. - I^{(1)}(q,D^\pm,D^{*\pm},D^{*\pm}) -
2I^{(1)}(q,D^{*\pm},D^{*\pm},D^{*\pm}) \right].
\ea%

\item $h_c\to J/\psi\pi^0$
\ba%
\al\al -i{\cal M}\left(h_c\to J/\psi\pi^0\right)_{\pm} \non\\
\al=\al i\frac{g}{F}g_1g_2 \left\{
\vec{q}\cdot\vec{\varepsilon}(h_c)\vec{q}\cdot\vec{\varepsilon}(J/\psi)
\left[
2I^{(1)}(q,D^{*\pm},D^\pm,D^\pm) - I(q,D^{*\pm},D^\pm,D^\pm)  \right.\right.\non\\
\al\al - 2I^{(1)}(q,D^{*\pm},D^\pm,D^{*\pm}) + I(q,D^{*\pm},D^\pm,D^{*\pm})\non\\
\al\al + 2I^{(1)}(q,D^{*\pm},D^{*\pm},D^\pm) - I(q,D^{*\pm},D^{*\pm},D^\pm)\non\\
\al\al \left.- 2I^{(1)}(q,D^{*\pm},D^{*\pm},D^{*\pm}) +
I(q,D^{*\pm},D^{*\pm},D^{*\pm})
\right] \non\\
\al\al + \vec{ q}\
^2\vec{\varepsilon}(\psi')\cdot\vec{\varepsilon}(h_c) \left[
2I^{(1)}(q,D^{*\pm},D^\pm,D^{*\pm}) - I(q,D^{*\pm},D^\pm,D^{*\pm}) \right.\non\\
\al\al + 2I^{(1)}(q,D^\pm,D^{*\pm},D^{*\pm}) - I(q,D^\pm,D^{*\pm},D^{*\pm}) \non\\
\al\al - 2I^{(1)}(q,D^{*\pm},D^{*\pm},D^\pm) + I(q,D^{*\pm},D^{*\pm},D^\pm)\non\\
\al\al \left.\left. - 2I^{(1)}(q,D^{*\pm},D^{*\pm},D^{*\pm}) +
I(q,D^{*\pm},D^{*\pm},D^{*\pm}) \right] \right\}.
\ea%

\item $\chi_{c0}\to \eta_c\pi^0$
\ba%
-i{\cal M}\left(\chi_{c0}\to \eta_c\pi^0\right)_{\pm} \al=\al
-\frac{i}{\sqrt{3}}\frac{g}{F}g_1g_2\vec{q}\ ^2 \left[
3I(q,D^\pm,D^\pm,D^{*\pm}) - 6I^{(1)}(q,D^\pm,D^\pm,D^{*\pm}) \right. \non\\
\al\al - I(q,D^{*\pm},D^{*\pm},D^\pm) +
2I^{(1)}I(q,D^{*\pm},D^{*\pm},D^\pm)\non\\
\al\al \left.- 2I(q,D^{*\pm},D^{*\pm},D^{*\pm}) +
4I^{(1)}(q,D^{*\pm},D^{*\pm},D^{*\pm}) \right].
\ea%

\end{itemize}

\subsection{Transitions between the $P$-wave charmonia}
\label{app:Pamp}
\begin{itemize}

\item $\chi_{c0}'\to\chi_{c1}\pi^0$
\ba%
-i{\cal M}\left(\chi_{c0}'\to\chi_{c1}\pi^0\right)_{\pm} \al=\al
-i\frac{2}{\sqrt{6}}\frac{g}{F}g_1g_1'
\vec{q}\cdot\vec{\varepsilon}(\chi_{c1}) \non\\
\al\al \times\left[ 3I(q,D^{\pm},D^{\pm},D^{*\pm}) +
I(q,D^{*\pm},D^{*\pm},D^{\pm}) \right].
\ea%

\item $\chi_{c1}'\to\chi_{c0}\pi^0$
\ba%
-i{\cal M}\left(\chi_{c1}'\to\chi_{c0}\pi^0\right)_{\pm} \al=\al
i\frac{2}{\sqrt{6}}\frac{g}{F}g_1g_1'
\vec{q}\cdot\vec{\varepsilon}(\chi_{c1}') \non\\
\al\al \times\left[ 3I(q,D^{*\pm},D^{\pm},D^{\pm}) +
I(q,D^{\pm},D^{*\pm},D^{*\pm}) \right].
\ea%

\item $\chi_{c1}'\to\chi_{c1}\pi^0$
\be%
-i{\cal M}\left(\chi_{c1}'\to\chi_{c1}\pi^0\right)_{\pm} =
-i2\frac{g}{F}g_1g_1'
\epsilon^{ijk}q^i\varepsilon^j(\chi_{c1}')\varepsilon^k(\chi_{c1})
I(q,D^{*\pm},D^{\pm},D^{*\pm}).
\ee%

\item $\chi_{c1}'\to\chi_{c2}\pi^0$
\be%
-i{\cal M}\left(\chi_{c1}'\to\chi_{c2}\pi^0\right)_{\pm} =
i2\sqrt{2}\frac{g}{F}g_1g_1'
\varepsilon^{ij}(\chi_{c2})q^i\varepsilon^j(\chi_{c1}')
I(q,D^{\pm},D^{*\pm},D^{*\pm}),
\ee%
where $\varepsilon^{ij}(\chi_{c2})$ is the polarization tensor of
the $\chi_{c2}$, and it is traceless, i.e.
$\varepsilon^{ii}(\chi_{c2})=0$, and symmetric.

\item $\chi_{c2}'\to\chi_{c1}\pi^0$
\be%
-i{\cal M}\left(\chi_{c2}'\to\chi_{c1}\pi^0\right)_{\pm} =
-i2\sqrt{2}\frac{g}{F}g_1g_1'
\varepsilon^{ij}(\chi_{c2})q^i\varepsilon^j(\chi_{c1}')
I(q,D^{*\pm},D^{*\pm},D^{\pm}).
\ee%

\item $\chi_{c2}'\to\chi_{c2}\pi^0$
\be%
-i{\cal M}\left(\chi_{c2}'\to\chi_{c2}\pi^0\right)_{\pm} =
-i4\frac{g}{F}g_1g_1' \epsilon^{ijk} q^i
\varepsilon^{jl}(\chi_{c2}')\varepsilon^{kl}(\chi_{c2})
I(q,D^{*\pm},D^{*\pm},D^{*\pm}).
\ee%

\item $h_c'\to h_c\pi^0$
\ba%
-i{\cal M}\left(h_c'\to h_c\pi^0\right)_{\pm} \al=\al
-i\frac{g}{F}g_1g_1' \epsilon^{ijk} q^i
\varepsilon^{j}(h_c')\varepsilon^{k}(h_c)\non\\
\al\al \times \left[ I(q,D^{*\pm},D^{\pm},D^{*\pm})  +
I(q,D^{\pm},D^{*\pm},D^{*\pm})\right.\non\\
\al\al \left. - I(q,D^{*\pm},D^{*\pm},D^{\pm}) +
I(q,D^{*\pm},D^{*\pm},D^{*\pm}) \right].
\ea%

\end{itemize}

\section{The Effective Lagrangian Approach (ELA)}
\label{app:ELA}
\subsection{Effective Lagrangians}
Based on the heavy quark
symmetry~\cite{Casalbuoni:1996pg,Colangelo:2003sa}, the relevant
effective Lagrangians used here are:
\begin{eqnarray}
\mathcal{L}_1 &=& i \frac {g_1} {2} Tr[P_{c\bar{c}}^\mu
\bar{H}_{2i}\gamma_\mu
\bar{H}_{1i}] + {\rm H.c.}, \\
\mathcal{L}_2 &=& i \frac {g_2}{2} Tr[R_{c\bar{c}}
\bar{H}_{2i}\gamma^\mu {\stackrel{\leftrightarrow}{\partial}}_\mu
\bar{H}_{1i}] + {\rm H.c.},
\end{eqnarray}
where the spin multiplets for these four $P$-wave and two $S$-wave
charmonium states are expressed as
\begin{eqnarray}
P_{c\bar{c}}^\mu &=& \left( \frac{1+ \rlap{/}{v} }{2} \right)
\left(\chi_{c2}^{\mu\alpha}\gamma_{\alpha} +\frac{1}{\sqrt{2}}
\epsilon^{\mu\nu\alpha\beta}v_{\alpha}\gamma_{\beta}\chi_{c1\nu}
+\frac{1}{\sqrt{3}}(\gamma^\mu -v^\mu)\chi_{c0} +h_c^\mu \gamma_{5}
\right) \left( \frac{1- \rlap{/}{v} }{2} \right), \nonumber\\ &&\\
R_{c\bar{c}}&=&  \left( \frac{1+ \rlap{/}{v} }{2} \right) (\psi^\mu
\gamma_\mu-\eta_c \gamma_5) \left( \frac{1- \rlap{/}{v} }{2}
\right).
\end{eqnarray}
The charmed and anti-charmed meson triplets read
\begin{eqnarray}
H_{1i}&=&\left( \frac{1+ \rlap{/}{v} }{2} \right)
[\mathcal{D}_i^{*\mu}
\gamma_\mu -\mathcal{D}_i\gamma_5], \\
H_{2i}&=& [\bar{\mathcal{D}}_i^{*\mu} \gamma_\mu
-\bar{\mathcal{D}}_i\gamma_5]\left( \frac{1- \rlap{/}{v} }{2}
\right),
\end{eqnarray}
where $\mathcal{D}$ and $\mathcal{D}^*$ denote the pseudoscalar and vector
charmed meson fields, respectively, i.e.
$\mathcal{D}^{(*)}=(D^{0(*)},D^{+(*)},D_s^{+(*)})$. These Lagrangians can be
reduced to the two-component ones used in the NREFT in Section~\ref{sec:loops}.

Consequently, the Lagrangian for $S$-wave $J/\psi$ and $\eta_c$ is
\begin{eqnarray}
\mathcal{L}_S &=& ig_{\psi D^* D^*} (-\psi^\mu
D^{*\nu}\overleftrightarrow{\partial}_\mu {D}_\nu^{*\dagger}+
\psi^\mu D^{*\nu}\partial_\nu{D}^{*\dagger}_{\mu} -
\psi_\mu\partial_\nu D^{*\mu} {D}^{*\nu\dagger})\nonumber\\
&&+ ig_{\psi DD}\psi_\mu(\partial^\mu D{D}^{\dagger}-D\partial^\mu
{D}^{\dagger}) + g_{\psi
DD}\varepsilon^{\mu\nu\alpha\beta}\partial_\mu
\psi_\nu(D^*_\alpha\overleftrightarrow{\partial}_\beta {D}^{\dagger}
- D\overleftrightarrow{\partial}_\beta{D}^{*\dagger}_\alpha)
\nonumber\\
&&+ g_{\eta_c D^* D} D^{*\mu}(\partial_\mu \eta_c D-\eta_c
\partial_\mu D)  + ig_{\eta_c D^* D^*} \varepsilon^{\mu\nu\alpha\beta}
\partial_\mu D^\ast_\nu D^{*\dagger}_\alpha \partial_\beta\eta_c,
\end{eqnarray}
and the Lagrangian for $P$-wave $h_c$ and $\chi_{cJ}$ is
\begin{eqnarray}
\mathcal{L}_P &=& ig_{\chi_{c2}D^* D^*} \chi_{c2}^{\alpha\beta}
D^*_\alpha D^{*\dag}_\beta +g_{\chi_{c1} D^* D}\chi_{c1\mu}D^{*\mu}D
+ig_{\chi_{c0}D^* D^*}D^{*\mu}D^{*\dag}_\mu+ig_{\chi_{c0} DD}DD \nonumber \\
&&+ g_{h_cD^* D} D^{*\mu}h_{c\mu}+ ig_{h_cD^*
D^*}\varepsilon^{\mu\nu\alpha\beta} D^\ast_\mu \partial_\alpha
h_{c\nu} D^*_\beta +{\rm H.c.} ,
\end{eqnarray}
where the couplings are
\begin{eqnarray} 
\label{eq:couplings_ELA}
g_{\psi DD} & = & g_2\sqrt{m_\psi} m_D, \,\,\,\,\,\,\, g_{\psi
D^*D^*} =  -g_2\sqrt{m_\psi} m_{D^*},
\,\,\,\,\,\,\, g_{\psi D^\ast D} = g_2\sqrt{\frac {m_\psi m_D} {m_{D^*}}}, \nonumber\\
g_{\eta_cD^*D}&=& g_2\sqrt{m_{\eta_c}} m_D, \,\,\,\,\, g_{\eta_cD^*D^*}= g_2\sqrt{\frac {m_{\eta_c} m_{D^*}} {m_{D^*}}}, \nonumber\\
g_{h_c D^* D}&=&-g_1\sqrt{m_{h_c} m_D m_{D^*}}, \,\,\,\,\, g_{h_c D^* D^*}= g_1\sqrt{\frac {m_{D^*}^2} {m_{h_c}}}, \nonumber
\ea%
\ba%
g_{\chi_{c0}DD}&=&-\sqrt{3}g_1\sqrt{m_{\chi_{c0}}}m_D, \,\,\,\,\, g_{\chi_{c0}D^*D^*}=-\frac {1} {\sqrt{3}}g_1\sqrt{m_{\chi_{c0}}}m_D^*, \nonumber \\
g_{\chi_{c1}D^*D}&=& \sqrt{2}g_1\sqrt{m_D m_D^*m_{\chi_{c0}}}, \,\,\,\,\,  g_{\chi_{c2}D^*D^*}=g_1m_{D^*}\sqrt{m_{\chi_{c2}}}, \nonumber \\
g_{D^\ast D^\ast \pi} &=& \frac {g_{D^\ast D \pi}} {\sqrt {m_D
m_{D^\ast}}} = \frac {\sqrt{2}} {F} g.
\end{eqnarray}

In the following, we present the transition amplitudes for the intermediate
meson loops listed in Table~\ref{tab:loops} in the framework of the ELA. Some
conventions should be clarified in advance. Notice that the expressions are
similar for the charged, neutral and charmed strange mesons except that
different charmed meson masses are applied. We thus only present the amplitudes
for those charged charmed meson loops. $G_1G_2G_3$ is the product of vertex
couplings for each loop, and the explicit expression can be found in
Eq.~(\ref{eq:couplings_ELA}). ${\cal F}= \prod_i{\cal F}_i(m_i, q_i^2)$ is the
form factor. [$q_1$, $q_3$, $q_2$] are the four-vector momenta for the
intermediate mesons [M1, M3, M2], respectively. $p_i \ (p_f)$, $\varepsilon_i \
(\varepsilon_f)$, and $\phi_i \ (\phi_f)$ are the initial (final) charmonium
four-vector momentum, polarization vector, and polarization tensor,
respectively. $p_2$ is the four-vector momentum of $\pi^0$ or $\eta$.

\subsection{Transitions between the $S$-wave charmonia}

\begin{itemize}

\item $\psi'\to J/\psi \pi^0 (\eta)$

\ba%
\mathcal {M}_{[D D D^*]}&=&-2G_1 G_2 G_3
\varepsilon_{\alpha\beta\mu\nu}\varepsilon_{i\rho} p_f^\alpha
\varepsilon_f^{*\beta}  p_2^\nu
\int \frac {d^4 q_2} {(2\pi)^4} \frac {{\cal F}} {a_1 a_2 a_3}q_1^\rho q_2^\mu  ,\nonumber \\
{\cal M}_{[DD^* D]}&=&-2G_1 G_2 G_3
\varepsilon_{\alpha\beta\mu\nu}p_i^\alpha \varepsilon_i^\beta
\varepsilon_{f\rho}^* p_2^\nu \int \frac {d^4 q_2} {(2\pi)^4} \frac {{\cal
F}} {a_1 a_2 a_3}q_3^\mu q_2^\rho  ,\nonumber\\
 {\cal M}_{[D D^* D^*]}&=&G_1 G_2 G_3 \varepsilon_{\alpha\beta\mu\nu}
\varepsilon_{\rho\sigma\tau\lambda}\varepsilon_{\alpha'\beta'\mu'\nu'}
p_i^\alpha \varepsilon_i^\beta  p_f^\rho \varepsilon_f^{*\sigma} p_2^{\mu'}
g^{\lambda\beta'} g^{\nu\nu'}
\int \frac {d^4 q_2} {(2\pi)^4} \frac {{\cal F}} {a_1 a_2 a_3}q_3^\mu q_2^\tau q_2^{\alpha'} ,\nonumber \\
{\cal M}_{[D^* D D^*]}&=&G_1 G_2 G_3 \varepsilon_{\alpha\beta\mu\nu}
p_i^\alpha \varepsilon_i^\beta  p_{2\rho}  \int \frac {d^4 q_2} {(2\pi)^4}
\frac {{\cal F}} {a_1 a_2 a_3}q_1^\mu \left[-\varepsilon_f^{*\nu} q_{1\sigma} \left(-g^{\rho\sigma} +\frac {q_2^\rho q_2^\sigma } {m_{D^*}^2}\right)\right. \nonumber\\
\left.\right.&& + \left.2\varepsilon_{f\sigma}^* q_1^\sigma
\left(-g^{\nu\rho} +\frac {q_2^\nu q_2^\rho} {m_{D^*}^2}\right)+
\varepsilon_{f\sigma}^* q_2^\nu \left(-g^{\rho\sigma} +\frac {q_2^\rho q_2^\sigma} {m_{D^*}^2}\right)\right] ,\nonumber \\
{\cal M}_{[D^* D^* D]}&=&G_1 G_2 G_3 \varepsilon_{\alpha\beta\mu\nu}
p_f^\alpha \varepsilon_f^{*\beta} p_{2\rho}  \int \frac {d^4 q_2} {(2\pi)^4}
\frac {{\cal F}} {a_1 a_2 a_3}  q_1^\mu \left[-\varepsilon_i^{\nu} q_{1\sigma} \left(-g^{\rho\sigma} +\frac {q_3^\sigma q_3^\rho} {m_{D^*}^2}\right) \right.\nonumber\\
\left.\right.&& +\left.2\varepsilon_{i\sigma} q_1^\sigma
\left(-g^{\nu\rho} +\frac {q_3^\nu q_3^\rho} {m_{D^*}^2}\right)+
\varepsilon_{i\sigma} q_3^\nu \left(-g^{\rho\sigma} +\frac {q_3^\rho q_3^\sigma} {m_{D^*}^2}\right)\right] ,\nonumber \\
{\cal M}_{[D^* D^* D^*]}&=&G_1 G_2 G_3 \varepsilon_{\alpha\beta\mu\nu}  \int
\frac {d^4 q_2} {(2\pi)^4}
\frac {{\cal F}} {a_1 a_2 a_3}  q_3^\alpha q_2^\mu\left[2\varepsilon_i^\nu p_i^\beta\varepsilon_f^*\cdot q_1 -2\varepsilon_i\cdot q_1 \varepsilon_f^\beta p_i^\nu \right.\nonumber \\
\left.\right.&-&\left. p_i^\beta \varepsilon_f^\nu (-\varepsilon_i\cdot q_2+\frac {\varepsilon_i\cdot q_2 q_2^2}{m_{D^*}^2})
-4\varepsilon_i\cdot q_1 \varepsilon_f\cdot q_1 g^{\beta\nu}-2\varepsilon_i\cdot q_1 \varepsilon_f^\nu q_2^\beta \right.\nonumber \\
\left. \right.&+& \left. \varepsilon_i^\beta p_i^\nu (-\varepsilon_f\cdot q_3 + \frac {q_1\cdot q_3 \varepsilon_f\cdot q_1}{m_{D^*}^2}) + 2\varepsilon_i^\beta \varepsilon_f^\nu (-q_2\cdot q_3 + \frac {q_2^2q_2\cdot q_3} {m_{D^*}^2})\right] .
\end{eqnarray}
\end{itemize}

\subsection{Transitions between the $S$- and $P$-wave charmonia}

\begin{itemize}
\item $\psi'\to h_c\pi^0$
\begin{eqnarray}\nonumber
{\cal M}_{[D D^* D]}&=&2G_1 G_2 G_3 \varepsilon_{i\alpha}\varepsilon_{f\beta}^* p_{2\mu} \int \frac {d^4 q_2} {(2\pi)^4}\frac {{\cal F}} {a_1 a_2 a_3}q_1^\alpha\left(-g^{\beta\mu} + \frac{q_3^\beta q_3^\mu} {m_{D^*}^2}\right),\\
{\cal M}_{[D D^* D^*]}&=&2G_1 G_2 G_3
\varepsilon_{\alpha\beta\mu\nu}\varepsilon_{\rho\sigma\tau\lambda}
p_i^\alpha \varepsilon_i^\beta \varepsilon_f^{*\sigma}
g^{\nu\lambda} p_2^\tau \int \frac {d^4 q_2} {(2\pi)^4}
 \frac {{\cal F}} {a_1 a_2 a_3}q_3^\mu q_2^\rho, \nonumber \\
{\cal M}_{[D^* D D^*]}&=&G_1 G_2 G_3 \varepsilon_{\alpha\beta\mu\nu}\varepsilon_{\rho\sigma\tau\lambda}
p_i^\alpha\varepsilon_i^\beta p_f^\rho \varepsilon_f^{*\sigma}  p_{2\xi}(-g^{\mu\tau})
\int \frac {d^4 q_2} {(2\pi)^4} \frac {{\cal F}} {a_1 a_2 a_3}q_1^\mu\left(-g^{ \nu\xi}
+ \frac {q_2^\nu q_2^\xi} {m_{D^*}^2}\right) ,\nonumber\\
{\cal M}_{[D^* D^*D]}&=&G_1 G_2 G_3 \varepsilon_{f\mu}^* p_{2\nu}\varepsilon_{i\alpha}
\int \frac {d^4 q_2} {(2\pi)^4} \frac {{\cal F}} {a_1 a_2 a_3}
\left[q_{1\beta}(-g^{\alpha\mu} + \frac {q_1^\alpha q_1^\mu} {m_{D^*}^2})(-g^{\beta\nu}
+\frac {q_3^\beta q_3^\nu} {m_{D^*}^2})\right. \nonumber \\
\left.\right.& -&\left. 2q_1^\alpha\left(-g^{\beta\mu}+
\frac {q_1^\beta q_1^\mu}{m_{D^*}^2})(-g_\beta^\nu+\frac {q_{3\beta} q_3^\nu}{m_{D^*}^2}\right)\right.\nonumber\\
\left.\right.&-& \left.q_{3\beta}\left(-g^{\alpha\mu}
+\frac {q_3^\alpha q_3^\mu} {m_{D^*}^2}\right)\left(-g^{\beta\nu}
+\frac {q_1^\beta q_1^\nu} {m_{D^*}^2}\right)\right] \nonumber 
\ea%
\ba%
{\cal M}_{[D^* D^*D^*]}&=&-G_1 G_2 G_3 \varepsilon_{\mu\nu\rho\sigma}\varepsilon_{\tau\lambda\xi\eta}p_f^\mu p_2^\xi\varepsilon_f^{*\nu} \varepsilon_{i\alpha}g^{\sigma\lambda}
 \int \frac {d^4 q_2} {(2\pi)^4} \frac {{\cal F}} {a_1 a_2 a_3} q_2^\tau\left[ q_1\eta\left(-g^{\alpha\rho} +\frac{q_1^\alpha q_1^\rho} {m_{D^*}^2}\right)  \right. \nonumber \\
\left. \right.& - &\left.2q_1^\alpha\left(-g^{\eta\rho}
+\frac{q_1^\eta q_1^\rho} {m_{D^*}^2}\right) -
\left(-q_3^\rho +\frac{q_1\cdot q_3 q_1^\rho}
{m_{D^*}^2}\right)g^{\alpha\eta}\right] .
\end{eqnarray}

\item $\eta_c'\to \chi_{c0}\pi^0$
\begin{eqnarray}
{\cal M}_{[D D D^*]}&=&2G_1 G_2 G_3p_i^\alpha p_2^\beta\int \frac
{d^4 q_2} {(2\pi)^4}
\frac {{\cal F} } {a_1 a_2 a_3}\left(-g_{\alpha\beta} + \frac{q_{3\alpha} q_{3\beta}} {m_{D^*}^2}\right), \nonumber \\
{\cal M}_{[D^* D D^*]}&=&2G_1 G_2 G_3p_i^\alpha p_2^\mu\int \frac
{d^4 q_2} {(2\pi)^4}
 \frac {{\cal F} } {a_1 a_2 a_3}\left(-g_\alpha^\beta + \frac{q_{1\alpha} q_1^\beta} {m_{D^*}^2}\right)  \left(-g_{\beta\mu} + \frac{q_{2\beta} q_{2\mu}} {m_{D^*}^2}\right), \nonumber \\
{\cal M}_{[D^* D^* D^*]}&=& 2G_1 G_2
G_3\varepsilon_{\alpha\beta\mu\nu}\varepsilon_{\rho\sigma\tau\lambda}
p_i^\mu  p_2^\tau g^{\beta\sigma} g^{\nu\lambda}\int \frac {d^4 q_2}
{(2\pi)^4}
 \frac {{\cal F}} {a_1 a_2 a_3}q_1^\alpha q_2^\rho .
\end{eqnarray}

\item $h_c\to J/\psi\pi^0$

\begin{eqnarray}\nonumber
{\cal M}_{[D D^* D]}&=&2G_1 G_2 G_3 \varepsilon_{i\alpha}\varepsilon_{f\beta}^* p_{2\mu}
\int \frac {d^4 q_2} {(2\pi)^4}\frac {{\cal F}} {a_1 a_2 a_3}q_1^\beta\left(-g^{\alpha\mu} + \frac{q_3^\alpha q_3^\mu} {m_{D^*}^2}\right),\nonumber\\
{\cal M}_{[D^* D D^*]}&=&G_1 G_2 G_3 \varepsilon_{i\alpha} p_{2\beta}
\int \frac {d^4 q_2} {(2\pi)^4} \frac {{\cal F}} {a_1 a_2 a_3}
\left[\varepsilon_{f\mu}^* q_{1\nu}(-g^{\alpha \mu} +
\frac {q_1^\alpha q_1^\mu} {m_{D^*}^2}) (-g^{\beta \nu} + \frac {q_2^\beta q_2^\nu} {m_{D^*}^2}) \right.\nonumber \\
\left.\right.&& - 2 \varepsilon_{f\mu}^* q_1^\mu(-g^{\alpha \nu} +
\frac {q_1^\alpha q_1^\nu} {m_{D^*}^2}) \left(-g^\beta_\nu +
\frac {q_2^\beta q_{2\nu}} {m_{D^*}^2}\right) \nonumber\\
&& -\left. \varepsilon_{f\mu}^* q_{2\nu}\left(-g^{\alpha \nu} +
\frac {q_1^\alpha q_1^\nu} {m_{D^*}^2}\right) \left(-g^{\beta \mu} +
\frac {q_2^\beta q_2^\mu} {m_{D^*}^2}\right)\right] ,\nonumber 
\ea%
\ba%
{\cal M}_{[D D^* D^*]}&=&2G_1 G_2 G_3
\varepsilon_{\alpha\beta\mu\nu}\varepsilon_{\rho\sigma\tau\lambda}
p_f^\alpha \varepsilon_i^\lambda \varepsilon_f^{*\beta}
g^{\nu\sigma} p_2^\tau \int \frac {d^4 q_2} {(2\pi)^4}
\frac {{\cal F}} {a_1 a_2 a_3}q_2^\mu q_2^\rho,\nonumber \\
{\cal M}_{[D^* D^* D]}&=&-G_1 G_2 G_3 \varepsilon_{\alpha\beta\mu\nu}\varepsilon_{\rho\sigma\tau\lambda} p_i^\alpha\varepsilon_i^\beta p_f^\rho \varepsilon_f^{*\sigma}  p_{2\xi}g^{\mu\lambda}
\int \frac {d^4 q_2} {(2\pi)^4} \frac {{\cal F}} {a_1 a_2 a_3}q_1^\tau \left(-g^{ \nu\xi} + \frac {q_3^\nu q_3^\xi} {m_{D^*}^2}\right) , \nonumber \\
{\cal M}_{[D^* D^* D]}&=&-G_1 G_2 G_3
\varepsilon_{\alpha\beta\mu\nu}\varepsilon_{\rho\sigma\tau\lambda}
p_i^\alpha \varepsilon_i^\beta q_2^\rho p_2^\tau g^{\nu\lambda}
\nonumber\\ &&\times\int \frac {d^4 q_2} {(2\pi)^4}
\frac {{\cal F}} {a_1 a_2 a_3} \left[-q_1^\sigma\left(-\varepsilon_f^\mu + \frac {q_1^\mu q_1\cdot \varepsilon_f^*} {m_{D^*}^2}\right) \right.\nonumber \\
\left.\right. && + \left. 2\varepsilon^*_f\cdot q_1
\left(-g^{\mu\sigma} + \frac {q_1^\mu q_1^\sigma} {m_{D^*}^2}\right)
+\varepsilon_f^{*\sigma} \left(-q_2^\mu + \frac {q_1^\mu q_1\cdot
q_2} {m_{D^*}^2}\right)\right] .
\end{eqnarray}

\item $\chi_{c0}\to \eta_c\pi^0$
\begin{eqnarray}
{\cal M}_{[D D D^*]}&=&G_1 G_2 G_3p_f^\alpha p_2^\beta\int \frac
{d^4 q_2} {(2\pi)^4}
 \frac {{\cal F}} {a_1 a_2 a_3} \left(-g_{\alpha\beta} + \frac{q_{2\alpha} q_{2\beta}} {m_{D^*}^2}\right),\nonumber \\
{\cal M}_{[D^* D^* D]}&=&G_1 G_2 G_3p_f^\beta p_2^\mu\int \frac {d^4
q_2} {(2\pi)^4}
 \frac {{\cal F} } {a_1 a_2 a_3},\left(-g^\alpha_\beta + \frac{q_1^\alpha q_{1\beta}} {m_{D^*}^2}\right)  \left(-g_{\alpha\mu} + \frac{q_{3\alpha} q_{3\mu}} {m_{D^*}^2}\right) \nonumber \\
{\cal M}_{[D^* D^* D^*]}&=&-G_1 G_2
G_3\varepsilon_{\alpha\beta\mu\nu}\varepsilon_{\rho\sigma\tau\lambda}
p_f^\alpha p_2^\tau g^{\beta\lambda} g^{\nu\sigma}\int \frac {d^4
q_2} {(2\pi)^4}
 \frac {{\cal F}} {a_1 a_2 a_3}q_2^\mu q_2^\rho .
\end{eqnarray}
\end{itemize}


\subsection{Transitions between the $P$-wave charmonia}
\begin{itemize}

\item $\chi_{c0}'\to\chi_{c1}\pi^0$

\begin{eqnarray}
{\cal M}_{[D^*D^*D]}&=&G_1 G_2 G_3\varepsilon_{f\beta}^*p_{2\mu}\int
\frac {d^4 q_2} {(2\pi)^4}
 \frac {{\cal F}} {a_1 a_2 a_3} \left(-g_\alpha^\beta + \frac{q_{1\alpha} q_1^\beta} {m_{D^*}^2}\right)  \left(-g^{\alpha\mu} + \frac{q_3^\alpha q_3^\mu} {m_{D^*}^2}\right),\nonumber \\
{\cal M}_{[DDD^*]}&=&G_1 G_2
G_3\varepsilon_{f\alpha}^*p_{2\beta}\int \frac {d^4 q_2} {(2\pi)^4}
 \frac {{\cal F} } {a_1 a_2 a_3}\left(-g^{\alpha\beta} + \frac{q_2^\alpha q_2^\beta} {m_{D^*}^2}\right) .
\end{eqnarray}

\item $\chi_{c1}'\to\chi_{c0}\pi^0$

\begin{eqnarray}
{\cal M}_{[DD^*D]}&=&G_1 G_2 G_3\varepsilon_{i\alpha}p_{2\beta}\int
\frac {d^4 q_2} {(2\pi)^4}
 \frac { {\cal F}} {a_1 a_2 a_3}\left(-g^{\alpha\beta} + \frac{q_3^\alpha q_3^\beta} {m_{D^*}^2}\right),\nonumber \\
{\cal M}_{[D^*DD^*]}&=&G_1 G_2 G_3\varepsilon_{i\alpha}p_{2\mu}\int
\frac {d^4 q_2} {(2\pi)^4}
 \frac {{\cal F}} {a_1 a_2 a_3} \left(-g^\alpha_\beta + \frac{q_1^\alpha
     q_{1\beta}} {m_{D^*}^2}\right) \left(-g^{\beta\mu} + \frac{ q_2^\beta
     q_2^\mu} {m_{D^*}^2}\right) .\nonumber\\
&&
\end{eqnarray}

\item $\chi_{c1}'\to\chi_{c1}\pi^0$

\begin{eqnarray}
{\cal M}_{[DD^*D^*]}=G_1 G_2 G_3\varepsilon_{\alpha\beta\mu\nu}
\varepsilon_i^\beta\varepsilon_f^{*\mu}p_2^\nu\int \frac {d^4 q_2}
{(2\pi)^4}
 \frac {{\cal F} } {a_1 a_2 a_3}q_3^\alpha .
\end{eqnarray}

\item $\chi_{c1}'\to\chi_{c2}\pi^0$
\begin{eqnarray}
{\cal M}_{[D^*DD^*]}&=&G_1 G_2
G_3\phi_{f\alpha\beta}^*\varepsilon_{i\mu} p_{2\nu}\int \frac {d^4
q_2} {(2\pi)^4}
 \frac {{\cal F}} {a_1 a_2 a_3}\left[ \left(-g^{\alpha\mu} + \frac{q_1^\alpha q_1^\mu} {m_{D^*}^2}\right) \left(-g^{\beta\nu} + \frac{ q_2^\beta q_2^\nu} {m_{D^*}^2}\right)\right. \nonumber \\
 \left.\right. && + \left. \left(-g^{\beta\mu} + \frac{q_1^\beta q_1^\mu} {m_{D^*}^2}\right) \left(-g^{\alpha\nu} + \frac{ q_2^\alpha q_2^\nu} {m_{D^*}^2}\right)\right] .
\end{eqnarray}

\item $\chi_{c2}'\to\chi_{c1}\pi^0$
\begin{eqnarray}
{\cal M}_{[D^*D^*D]}&=&G_1 G_2
G_3\phi_{i\alpha\beta}\varepsilon_{f\mu}^* p_{2\nu}\int \frac {d^4
q_2} {(2\pi)^4}
 \frac { {\cal F}} {a_1 a_2 a_3}\left[\left(-g^{\alpha\mu} + \frac{q_1^\alpha q_1^\mu} {m_{D^*}^2}\right) \left(-g^{\beta\nu} + \frac{ q_3^\beta q_3^\nu} {m_{D^*}^2}\right) \right.\nonumber \\
 \left.\right.&& + \left.\left(-g^{\beta\mu} + \frac{q_1^\beta q_1^\mu} {m_{D^*}^2}\right) \left(-g^{\alpha\nu} + \frac{ q_3^\alpha q_3^\nu} {m_{D^*}^2}\right)\right] .
\end{eqnarray}

\item $\chi_{c2}'\to\chi_{c2}\pi^0$
\begin{eqnarray}
{\cal M}_{[D^*D^*D^*]}&=&G_1 G_2 G_3\phi_{i\alpha\beta}\phi_{f\mu\nu}^* \varepsilon_{\rho\sigma\tau\lambda} p_2^\tau \int \frac {d^4 q_2} {(2\pi)^4} \frac {{\cal F}} {a_1 a_2 a_3} q_2^\rho\left[g^{\beta\lambda}g^{\nu\sigma}\left(-g^{\alpha\mu} + \frac{q_1^\alpha q_1^\mu} {m_{D^*}^2}\right) \right.\nonumber \\
\left. \right. && +\left.
g^{\beta\lambda}g^{\mu\sigma}\left(-g^{\alpha\nu} + \frac{q_1^\alpha
q_1^\nu}
{m_{D^*}^2}\right)+g^{\alpha\lambda}g^{\nu\sigma}\left(-g^{\beta\mu}
+ \frac{q_1^\beta q_1^\mu} {m_{D^*}^2}\right) \right.\nonumber\\
&& \left.+g^{\alpha\lambda}g^{\mu\sigma}\left(-g^{\beta\nu} +
\frac{q_1^\beta q_1^\nu} {m_{D^*}^2}\right)\right] .
\end{eqnarray}

\item $h_c'\to h_c\pi^0$

\begin{eqnarray}
{\cal M}_{[DD^*D^*]}&=&G_1 G_2 G_3 \varepsilon_{\alpha\beta\mu\nu}
p_2^\mu\varepsilon_i^\nu \varepsilon_f^{*\beta}\int \frac {d^4 q_2}
{(2\pi)^4}
 \frac {{\cal F}} {a_1 a_2 a_3}q_2^\alpha,\nonumber \\
{\cal M}_{[D^*DD^*]}&=&G_1 G_2 G_3
\varepsilon_{\alpha\beta\mu\nu}p_f^\alpha\varepsilon_{i\rho}
\varepsilon_f^{*\beta} p_{2\sigma} \nonumber\\
&& \times\int \frac {d^4 q_2} {(2\pi)^4}
 \frac {{\cal F}} {a_1 a_2 a_3}\left(-g^{\mu\rho}+\frac {q_1^\mu q_1^\rho}{m_{D^*}^2}\right)    \left(-g^{\nu\sigma}+\frac {q_2^\nu q_2^\sigma}{m_{D^*}^2}\right),\nonumber \\
{\cal M}_{[D^*D^*D^*]}&=&G_1 G_2 G_3
\varepsilon_{\alpha\beta\mu\nu}\varepsilon_{\rho\sigma\tau\lambda}\varepsilon_{\alpha'\beta'\mu'\nu'}p_i^\alpha\varepsilon_i^\beta
p_f^\rho\varepsilon_f^{*\sigma}
p_2^{\mu'}g^{\lambda\beta'}g^{\nu\nu'} \nonumber\\ && \times\int
\frac {d^4 q_2} {(2\pi)^4}
\frac {{\cal F}} {a_1 a_2 a_3}q_2^{\alpha'} \left(-g^{\mu\tau}+ \frac {q_1^\mu q_1^\tau} {m_{D^*}^2}\right),\nonumber \\
{\cal M}_{[D^*D^*D]}&=&G_1 G_2 G_3
\varepsilon_{\alpha\beta\mu\nu}p_i^\alpha\varepsilon_i^\beta
\varepsilon_{f\rho}^* p_{2\sigma}\int \frac {d^4 q_2} {(2\pi)^4}
 \frac {{\cal F}} {a_1 a_2 a_3} \nonumber\\
 && \times\left(-g^{\mu\rho}+\frac {q_1^\mu q_1^\rho}{m_{D^*}^2}\right)    \left(-g^{\nu\sigma}+\frac {q_3^\nu q_3^\sigma}{m_{D^*}^2}\right) .
\end{eqnarray}

\end{itemize}

\end{appendix}

\end{document}